\documentclass[11pt]{article}
\usepackage{amsmath}
\usepackage{amssymb}
\usepackage{cite}
\usepackage{graphicx,bbm,mathrsfs,epstopdf}
\usepackage{nicefrac}
\usepackage{indentfirst} 

\usepackage{geometry}
\usepackage{jheppub}       

\def\ie{{\it i.e.}}
\def\eg{{\it e.g.}}
\def\fbi{~{\rm fb}^{-1}}

\newcommand{\be}{\begin{equation}}  
\newcommand{\ee}{\end{equation}}  
\newcommand{\bea}{\begin{eqnarray}}  
\newcommand{\eea}{\end{eqnarray}}

\newcommand{\WB}{W\hspace{-1.5pt}B}
\newcommand{\WW}{W\hspace{-.5pt}W}
\newcommand{\BB}{B\hspace{-.5pt}B}
\newcommand{\nn}{\nonumber}

\newcommand\lsim{\mathrel{\rlap{\lower4pt\hbox{\hskip1pt$\sim$}}
    \raise1pt\hbox{$<$}}}
\newcommand\gsim{\mathrel{\rlap{\lower4pt\hbox{\hskip1pt$\sim$}}
    \raise1pt\hbox{$>$}}}

\renewcommand{\O}{\mathcal O}

\newcommand{\captionfonts}{\small}

\newcommand{\approptoinn}[2]{\mathrel{\vcenter{
  \offinterlineskip\halign{\hfil$##$\cr
    #1\propto\cr\noalign{\kern2pt}#1\sim\cr\noalign{\kern-2pt}}}}}

\makeatletter  
\long\def\@makecaption#1#2{%
  \vskip\abovecaptionskip
  \sbox\@tempboxa{{\captionfonts #1: #2}}%
  \ifdim \wd\@tempboxa >\hsize
    {\captionfonts #1: #2\par}
  \else
    \hbox to\hsize{\hfil\box\@tempboxa\hfil}%
  \fi
  \vskip\belowcaptionskip}
\makeatother   

\begin{document}

\rightline{CPHT-RR025.0413}
\rightline{ICTP-SAIFR/2013-005}
\rightline{LPSC13097}

\vspace*{1.2cm}

\begin{center}

\thispagestyle{empty}

{\Large\bf A Bayesian view of the Higgs sector with higher dimensional operators }\\[10mm]

\renewcommand{\thefootnote}{\fnsymbol{footnote}}

{\large B\'eranger Dumont$^{\,a}$, Sylvain~Fichet$^{\,b}$, Gero von Gersdorff$^{\,c,d\,}$\footnote{sylvain.fichet@lpsc.in2p3.fr, dumont@lpsc.in2p3.fr, gersdorff@gmail.com}}\\[10mm]

\addtocounter{footnote}{-1} 

{\it
$^a\,$Laboratoire de Physique Subatomique et de Cosmologie, UJF Grenoble 1,
CNRS/IN2P3, INPG, 53 Avenue des Martyrs, F-38026 Grenoble, France\\
$^{b}$~International Institute of Physics, UFRN, 
Av. Odilon Gomes de Lima, 1722 - Capim~Macio - 59078-400 - Natal-RN, Brazil \\
$^{c}$ Centre de Physique Th\'eorique, \'Ecole Polytechnique and CNRS, 
F-91128 Palaiseau, France\\
$^{d}$ ICTP South American Institute for Fundamental Research,
Instituto de Fisica Teorica, Sao Paulo State University, Brazil\\
}

\vspace*{12mm}

{  \bf  Abstract }
\end{center}

\noindent

We investigate the possibilities of New Physics affecting the Standard Model (SM) Higgs sector. An effective Lagrangian with dimension-six operators is used to capture the effect of New Physics.
We carry out a global Bayesian inference analysis, considering the recent LHC data set including all available correlations, as well as results from Tevatron. Trilinear gauge boson couplings and electroweak precision observables are also taken into account. The case of weak bosons tensorial couplings is closely examined and NLO QCD corrections are taken into account in the deviations we predict.
We consider two scenarios, one where the coefficients of all the dimension-six operators are essentially unconstrained, and one where a certain subset is loop suppressed.
In both scenarios, we find that large deviations from some of the SM Higgs couplings can still be present, assuming New Physics arising at 3 TeV.
In particular, we find that a significantly reduced coupling of the Higgs to the top quark is possible and slightly favored by searches on Higgs production in association with top quark pairs. The total width of the Higgs boson is only weakly constrained and can vary between  0.7 and 2.7 times the Standard Model value within 95\% Bayesian credible interval (BCI).
We also observe sizeable effects induced by New Physics contributions to tensorial couplings. In particular, the Higgs boson decay width into $Z\gamma$ can be enhanced by up to a factor 12 within 95\% BCI.

\clearpage
%

\section{Introduction}

The existence of a scalar resonance compatible with the SM Higgs boson has now been firmly established at the LHC by the ATLAS and CMS experiments~\cite{atlas_higgs,cms_higgs}. Based on the latest results presented at the Rencontres de Moriond 2013, the combined significance of the excess around 125 GeV reaches more than $7~\sigma$ in both experiments. The analyses are based on up to 5 fb$^{-1}$ at 7 TeV and 21~fb$^{-1}$ at 8 TeV, collected in 2011 and 2012.

Following this fundamental breakthrough, new questions need to be addressed. Some of them may find an answer during the LHC era, through the analysis of  the properties of the new boson. A first question is whether or not this resonance is a Higgs boson, \ie\  the manifestation of a field involved in the electroweak symmetry breaking (EWSB) and unitarization of $WW$ scattering. Second, if it is indeed a Higgs, one may wonder to what extent it is compatible with the Standard Model Higgs. Indeed, the properties of the Higgs boson could be dramatically modified with the theoretical prejudice that New Physics has to emerge near the electroweak scale.
On the other hand, direct searches for new states beyond the SM have so far turned out to be unsuccessful, and indirect constraints from electroweak precision measurements at LEP push the limits on masses of new particles somewhat further above the electroweak scale.
If New Physics is indeed present and is somehow separated from the electroweak scale, the couplings of the Higgs boson will be close to those of the SM and will only be modified by the effect of a few higher dimensional operators.
In this paper, we will explore an effective field theory (EFT) with only relatively few new parameters.

Many aspects of these fundamental questions have already been investigated in several works (see \eg~\cite{Espinosa:2012im, Plehn:2012iz, Bonnet:2011yx,Carmi:2012in,Giardino:2013bma,Azatov:2012bz,Corbett:2012ja,
Cacciapaglia:2012wb,Falkowski:2013dza,Alanne:2013dra,Djouadi:2013qya,Corbett:2013pja} for studies based on the 8 TeV results).
For example, the chiral EW Lagrangian with a non-linear realization of the $SU(2)_L \times U(1)_Y$ symmetry is a quite general framework in order to study the properties of electroweak symmetry breaking. Many scenarios producing possibly large deviations with respect to the SM Higgs properties are successfully captured in such EFT approach.

We are going to consider that new states appear at a typical scale $\Lambda$ substantially larger than the electroweak scale. For physical processes involving an energy scale smaller than $\Lambda$, New Physics can be integrated out. As a consequence of this hypothesis, the resulting low-energy effective theory consists in the Standard Model, supplemented by infinite series of local operators with higher dimension, which involves negative powers of $\Lambda$, 
\be
\mathcal{L}_{\rm eff}=\mathcal{L}_{\rm SM}+\sum_{i} \frac{\alpha_{i}}{\Lambda^{n_i}} \mathcal{O}_{i}\,. \label{L_eff}
\ee
The effects of such higher dimensional operators (HDOs) have been investigated in many contexts such as flavor physics, or the study of the properties of the electroweak gauge bosons through LEP precision measurements. The purpose of this paper is to study the electroweak sector again, which now includes new Higgs observables. For our analysis, we only have to consider the leading HDOs. The only operator with $n_i=1$ is the one giving a Majorana mass to the neutrino, and is not relevant for our study. We will thus be exclusively interested  in the $n=2$ terms, \ie~dimension-$6$ operators.

We adopt the statistical framework of Bayesian inference in this work, which allows us to assign probabilies to our parameters and to deal with partially constrained problems.
Another interesting property is that the unnatural (\ie~fine-tuned) character of precise cancellations which may occur between HDO contributions is built-in in this framework. Indeed, regions of the parameter space in which precise cancellations occur have by construction a weak statistical weight (see also \cite{Fichet:2012sn}). 
 The results we will present can thus be considered as generic, \ie~free of improbable cancellations. Our Bayesian analysis will rely on Markov chain Monte Carlo (MCMC) techniques, which allows us to easily sample the posterior probability distribution function.

The outline of the paper is as follows. In Section \ref{se_HDOs}, we lay out the formalism for higher-dimension operators in the  electroweak sector. In Section \ref{se_data}, we present the dataset used for the analysis and the measurements entering the likelihood functions. The peculiar case of observables sensitive to tensorial couplings relating Higgs and weak bosons is investigated in Section \ref{se_tensorials}. In Section \ref{se_deviations}, we derive the observable deviations from the SM induced by the higher dimensional operators, taking into account leading NLO QCD effects. Section \ref{se_bayesian_setup} presents the setup of our Bayesian analysis. Section \ref{se_results} is devoted to our results. Our conclusion is given in Section \ref{se_conclusion}.

\section{Electroweak higher-dimension operators} \label{se_HDOs}

In this section, we define the basis of dimension-6 operators  supplementing the renormalizable electroweak sector of  the SM Lagrangian.
We refer to  \cite{Buchmuller:1985jz, Grzadkowski:2010es} for further details on the Standard Model HDOs. 

A basis of CP-even operators not involving fermions can be chosen as~\footnote{The operator $\mathcal O_6$ plays no role in what follows and is listed here only for completeness.}
\be
\O_6=|H|^6\,,\qquad \O_{D^2}=|H|^2|D_\mu H|^2\,,\qquad \O'_{D^2}=|H^\dagger D_\mu H|^2\,, \label{HDO_O_6}
\ee
\be
\O_{\WW}=H^\dagger H\,(W_{\mu\nu}^a)^2\,,\qquad \label{HDO_O_WW}
\O_{\BB}=H^\dagger H\,(B_{\mu\nu})^2\,,\qquad
\O_{\WB}=H^\dagger\,W_{\mu\nu} H\,B_{\mu\nu}\,,
\ee
\be
\O_{GG}=H^\dagger H\,(G_{\mu\nu}^a)^2\,. \label{HDO_O_GG}
\ee
Any other operator can be reduced to these via integration by parts and the use of the SM equations of motion for the Higgs and gauge fields, possibly generating operators involving fermions. 
Amongst the latter, only a limited set will be relevant for our purpose. 
Operators of the form $J_H\cdot J_{f}$, where $J_H$ and $J_{f}$ are $SU(2)$ or $U(1)_Y$ currents involving Higgs field and fermion $f$  respectively, will in general contribute to FCNC as well as electroweak non-oblique corrections (\eg, non-universal couplings of fermions to gauge bosons).~\footnote{The non-universal corrections to the weak bosons couplings of the top quark are only very mildly constrained by EW data, and it is a priori not justified to set them to zero. 
However the only effect to Higgs observables at leading order is a modification of the top loop contribution to the $h\to Z\gamma$ decay due to the anomalous $Ztt$ vertex. The top contribution is however about one order of magnitude  smaller than the leading contribution from the $W$ loop \cite{Djouadi:2005gi}.  
We will therefore only consider universal (oblique) corrections to EW data.}
 However, the operators
\be
\O_D=J^a_{H\,\mu}\,J^a_{\mu}\,,\qquad
\O_{D}'=J^Y_{H\,\mu}\,J^Y_{\mu}\,,
\ee
where $J=\sum_f J_{f}$ are the SM fermion currents coupling to $B_\mu$ and $W_\mu$,
are flavor diagonal and only result in universal corrections to gauge couplings and should hence be viewed as contributing to $S$ and $T$. 
\footnote{In fact this is the way how contributions to $S$ and $T$ can arise in theories with new spin-1 states, such as in warped extra dimensions \cite{Davoudiasl:2009cd,Cabrer:2011fb}.}
We will also need to consider Yukawa corrections of the form
\be
\O_f= 2 y_f\,|H|^2 \, H\bar f_L f_R \,, \label{HDO_O_f}
\ee
where $f_R=t_R,\ b_R,\ \tau_R$ and $f_L$ the corresponding doublet ($\bar f^a_L=\epsilon^{ab} \bar q^{b,3}_L,\ \bar q_L^{a,3},\bar\ell_L^{a,3}$) and $y_f$ the Yukawa coupling.

Note that the operators $\O_D$ and $\O_D'$ could be traded for the operators 
\be
\O_{W}=(D_\mu H)^\dagger W_{\mu\nu}D_\nu H\,,\qquad \O_B=(D_\mu H)^\dagger D_\nu H\,B_{\mu\nu}\,,
\ee
by use of the SM equations of motion for $B$ and $W$. While $\O_D$ and $\O_D'$ contribute to $S$ and $T$ but not to the modified Higgs couplings, for $\O_B$ and $\O_W$ it is the other way around. Both choices of basis are physically equivalent.  
Before passing from a general redundant set of operators to a convenient irreducible basis via the equations of motion, it is useful to first identify the operators that cannot be generated at tree-level. \footnote{A detailed study about perturbative generation of HDOs can be found in \cite{Arzt:1994gp}. } This is valuable information and we would like to avoid it to be lost in the course of the reduction. However this is what would happen if we eliminated $\O_D$ and $\O_D'$ in favor of $\O_W$ and $\O_B$. Indeed, this would cause the coefficient of \eg~$\O_{\WB}$ (which cannot be generated at tree-level) to be shifted by the coefficient of $\O_D$  (which can be generated at the tree-level via exchange of spin-one states).  This is why we choose this basis.

The only remaining two-fermion operators are of the dipole type. These operators are tightly constrained by FCNC as well as by their contributions to electric and magnetic dipole moments. Moreover, they are necessarily generated at the loop-level, and only affect Higgs couplings to gauge bosons by modifying existing SM loops. They will not have any impact on our results, therefore we can neglect them entierely.

We do not take into account CP-violating HDOs. These operators are constrained  by observables such as electric dipole moments. If we choosed to include these CP-odd HDOs in our analysis, we would also need to consider the whole set of data sensitive to CP violation. Although there is no fundamental problem with such extended analysis, that is beyond the scope of the present work. Moreover, the effects induced by CP-violating HDOs are often subleading with respect to the effects of CP-even operators, unless the latter are sufficiently suppressed. This is the case for Higgs decays, because CP-violating amplitudes do not interfere with SM amplitudes, whereas CP-conserving amplitudes do interfere with SM amplitudes \cite{Manohar:2006gz}. In the following we will derive observable deviations from the Standard Model using the full set of HDOs, and perform the analysis presented in Section~\ref{se_bayesian_setup} taking into account only operators that respect custodial symmetry.

We could also rigorously take into account the running of the HDO coefficients $\alpha_i$ from the scale $\Lambda$ to the low scale ($m_h$ or $\sqrt{s}$, depending on the process considered), see for example \cite{Grojean:2013kd}.  However, the consequences of this running are rather mild so we will neglect them in this study. Notice that the strong effect of the operator $O_{\WB}$ on the $h\gamma\gamma$ vertex found in \cite{Grojean:2013kd} requires large enhancement of $\alpha_{\WB}$ with respect to $ s_w^2\,\alpha_{\WW}+ c_w^2\, \alpha_{\BB}-\frac{1}{2} s_w c_w\,\alpha_{\WB}$ (the coefficient of $h\,F_{\mu\nu}F^{\mu\nu}$, see Eq.~(\ref{lambda_gamma_tens})). As is evident in our basis, this cannot be explained by a relative loop factor as none of these operators receive contributions at tree-level. Moreover, it has been shown in Ref.~\cite{Elias-Miro:2013gya} that operators that can be generated at tree-level (such as $\mathcal O_{D}$) do not mix with the loop-suppressed operators such as $\mathcal O_{VV}$ in the renormalization group flow.
In the absence of large hierarchies in the couplings of New Physics states, we conclude that operator mixing does not lead to a large enhancement of the $h \rightarrow \gamma\gamma$ rate. 

\subsection{Effective Lagrangian}

In this section we will present the  effect of the HDOs on the SM tree-level couplings. Loops involving SM particles are considered in Sec.~\ref{loops}. We define the physical Higgs field $h$ as
\be
H=\left(\begin{array}{c}0\\ \frac{1}{\sqrt{2}}\,(\tilde v+h)\end{array}\right)\,,
\ee
and parametrize the couplings of $h$ to gauge bosons and fermions as~\footnote{For the $hVV$ couplings with different tensor structure see below.}  
\be
\mathcal L^{\rm tree}_{v,f}=\lambda_{Z}\,h\,(Z_\mu)^2+\lambda_{W}\, h\,W_{\mu}^+W_{\mu}^-  +\sum_f\lambda_f\, h\,\bar f_Lf_R \,. \label{L_vect}
\ee
The SM tree-level predictions for these quantities are given in terms of the SM input parameters $\tilde g$, $\tilde v$ and $\tilde s_w^2\equiv \tilde g'^2/(\tilde g^2+\tilde g'^2)$:
\be
\lambda_{Z}=\frac{\tilde g^2\,\tilde v}{4\,\tilde c_w^2}\equiv\frac{\tilde m_{Z}^2}{\tilde v}\,,\qquad\lambda_{W}=\frac{\tilde g^2\,\tilde v}{2}\equiv\frac{2\,\tilde m_{W}^2}{\tilde v}\,,\qquad 
\lambda_{f}=-\frac{\tilde y_{f}}{\sqrt{2}}\equiv-\frac{\tilde m_{f}}{\tilde v} \,, \label{lambda_vect}
\ee
where the quantities with a tilde are the ones that appear in the SM part of the Lagrangian. For instance, $\tilde g$ and $\tilde g'$ are the couplings appearing in the covariant derivatives. However, these couplings do not take the same values as in the SM, since there are corrections from HDOs. 
There are distinct effects, as follows (see Ref.~\cite{Burgess:1993vc} for an analogous discussion on fermion couplings).
\begin{itemize}
\item
Operators such as $\O_{D^2}$ correct directly the tree-level SM vertices.
\item
Some operators (\eg~$\O_{D^2}$, $\O_{\WW}$) modify the kinetic terms of Higgs and gauge fields and thus indirectly lead to the rescaling of some couplings.
\item
Finally, there can be indirect effects from input parameters. They are taken to be the fine-structure constant $\alpha$, the $Z$ boson mass $m_Z$ and the Fermi constant $G_F$, as well as the physical fermion masses $m_f$ and the strong coupling constant $\alpha_s$. These quantities receive corrections from HDOs but must be held fixed in the analysis. Yet, this causes the SM parameters $\tilde g$, $\tilde v$ and $\tilde s_w$ to become functions of the HDO coefficients.
\end{itemize}

The last point is sometimes not taken into account in the literature. Let us focus on it and {\em define} the quantities $v$, $g$ and $s_w$ via
\be
4\pi\alpha \equiv s_w^2 g^2\,,\qquad
m_Z^2\equiv\frac{ v^2\, g^2}{4\,  c_w^2}\,,\qquad 
G_F\equiv\frac{1}{\sqrt{2}\, v^2} \,.
\label{hat}
\ee
These quantities can be viewed as the ``familiar'' numbers from the SM (\eg~$v=246$~GeV). Like the input parameters they stay fixed in our analysis. On the other hand, the parameters $\tilde g$, $\tilde s_w$ and $\tilde v$ are the gauge couplings appearing in the covariant derivatives and the vacuum expectation value (vev) of the Higgs field, and must be expressed in terms of the HDO coefficients. We present the details of this procedure in Appendix~\ref{inputs}. Taking into account all the above effects, we obtain
\be
\lambda_{Z}=a_Z\,\frac{m_Z^2}{ v}\,,\qquad
\lambda_{W}=a_W\,\frac{2\,m_W^2}{ v}\,,\qquad
\lambda_{f}=-c_f\,\frac{m_f}{ v} \,,
\ee
where $m_f$ and $m_W$ are the {\em physical} masses. In particular, $m_W$ is given by~\footnote{Unlike $m_Z$ and $m_f$, which are input parameters, the $W$ mass is a prediction in terms of input parameters and HDO coefficients.}
\bea
m_W^2&=&\frac{ g^2\, v^2}{4}\left(1+\left(\frac{1}{2}\alpha_D-\frac{ c_w^2}{2( c_w^2- s_w^2)}[\alpha_{D^2}'+\alpha_D]-
\frac{ c_w s_w}{ c_w^2- s_w^2}\alpha_{\WB}
\right) \frac{ v^2}{\Lambda^2} \right)\nn\\
&=&\frac{ g^2\, v^2}{4}\left(1-\frac{\alpha S}{2( c_w^2- s_w^2)}
+\frac{ c_w^2\,\alpha T}{ c_w^2- s_w^2}
\right) \, . \label{m_W_ST}
\eea
In the last row we have used Eqs.~(\ref{S_tree}) and (\ref{T_tree}) in order to compare our derivation of $m_W$ with the one in \cite{Burgess:1993vc}. The SM prediction of $m_W$ is thus only corrected by the oblique parameters. 
In this parametrization, the rescaling factors $a_Z$, $a_W$ and $c_f$ are given by
\bea
a_Z&=&1+\left(\frac{1}{2}\alpha_{D^2}\, -\frac{1}{4}\alpha_D\,  +\frac{1}{4}\alpha_{D^2}'\, \right) \frac{ v^2}{\Lambda^2} \, , \nn\\
a_W&=&1+\left( \frac{1}{2}\alpha_{D^2}\,-\frac{1}{4}\alpha_D\,-\frac{1}{4}\alpha'_{D^2}\, 
\right) \frac{ v^2}{\Lambda^2} \, , \nn\\
c_f&=&1-\left(\frac{1}{4}\alpha'_{D^2}\,-\frac{1}{4}\alpha_D\, -\alpha_f\right) \frac{ v^2}{\Lambda^2} \, .
\label{anomalous1}
\eea
As a nontrivial consistency check, note that the vector anomalous couplings are rescaled in a custodially symmetric way $(a_Z=a_W)$ once the custodial-symmetry violating operator $\O_{D^2}'$ is turned off.

To conclude this subsection we compute the direct tree-level HDO contribution to the tensor couplings,
\be
\mathcal L^{\rm tree}_{t}=
\zeta_\gamma\,h\,(F_{\mu\nu})^2+\zeta_g\,h\,(G_{\mu\nu})^2
+\zeta_{Z \gamma}\,h\,F_{\mu\nu}Z_{\mu\nu}\,+\zeta_{Z}\,h\,(Z_{\mu\nu})^2+\zeta_{W}\, h\,W_{\mu\nu}^+W_{\mu\nu}^-\,, \label{L_tens}
\ee
which are all zero in the SM at tree-level. One finds
\be
\zeta_\gamma=\left( s_w^2\,\alpha_{\WW}+ c_w^2\, \alpha_{\BB}-\frac{1}{2} s_w c_w\,\alpha_{\WB}\right)\,\frac{ v}{\Lambda^2}\,,\qquad
\zeta_g=\alpha_{GG}\, \frac{ v}{\Lambda^2} \, , \label{lambda_gamma_tens}
\ee
\be
\zeta_{Z \gamma}=\left(2 c_w s_w\,\alpha_{\WW}-2 c_w s_w\,\alpha_{\BB}-\frac{1}{2}( c_w^2- s_w^2)\,\alpha_{\WB}\right) \,\frac{ v}{\Lambda^2} \, , \label{lambda_Zgamma_tens}
\ee
\be
\zeta_{Z}=  \left( c_w^2\,\alpha_{\WW} +  s_w^2\,\alpha_{\BB} + \frac{1}{2} c_w s_w\, \alpha_{\WB}\right)\,\frac{ v}{\Lambda^2}\,, \label{lambda_WZ_tens}
\qquad
\zeta_{W}=  2\,\alpha_{\WW}\,\frac{ v}{\Lambda^2}\,. 
\ee
The first two quantities constitute important corrections to the production and decay of the Higgs boson.
The last two corrections modify the tensorial structure of the SM Higgs--weak bosons coupling in a non-trivial way, which is discussed in detail in Sec. \ref{se_tensorials}.

\subsection{Standard Model loop-induced HDOs}
\label{loops}

In this section we compute the  Standard Model loop-induced operators relevant for Higgs physics. These operators contain indirect modifications due to couplings modified by the HDOs considered in the previous subsection. We want to make sure that we do not double-count possible New Physics contribution to the Higgs couplings. 
In order to have a well-defined HDO framework at loop-level, we should consider that the HDOs we present in Eqs.~\eqref{HDO_O_6}--\eqref{HDO_O_f} are generated exclusively through New Physics states at leading order, and enclose higher-order SM corrections only from irreducible loops.~\footnote{This last point is important for NLO QCD corrections, see Sec. \ref{se_deviations}.} Hence, the modified SM loops are not included in the tree-level contributions computed in the previous subsection. Our strategy is thus to compute the one-loop corrections to $\mathcal L^{\rm tree}$ using the couplings shown in Eq.~(\ref{anomalous1}). 

The one-loop Lagrangian is parametrized as
\be
\mathcal L^{1-{\rm loop}}=\lambda_\gamma\,h\,(F_{\mu\nu})^2+\lambda_g\,h\,(G_{\mu\nu})^2
+\lambda_{Z \gamma}\,h\,F_{\mu\nu}Z_{\mu\nu} \,.
\ee
Let us decompose these couplings according to the particle in the loop, $\lambda_i=\sum_X\lambda_i^X$. We find~\footnote{Note that in Eqs.~(\ref{lambda_loop_gammaW})--(\ref{lambda_loop_Zgammaf}) only the quantites with a tilde appear. Besides the modified Higgs couplings, the HDOs we consider only affect the couplings of the fermions to the $W$ and $Z$ bosons, precisely via the oblique parameters $S$ and $T$. The latter would in fact only show up in $\lambda_{Z \gamma}^f$. However these corrections are subleading and rather small (few percents at most), so that it is safe to neglect them.}
\be
\lambda_\gamma^W=a_W\,\lambda_\gamma^{W, {\rm SM}}= \frac{7}{2}\, \frac{ g^2\, s_w^2}{16\pi^2} \, \frac{a_W}{ v} A_v(\tau_W) \,, \label{lambda_loop_gammaW}
\ee
\be
\lambda_\gamma^f=c_f\,\lambda_\gamma^{f, {\rm SM}}= -\frac{2}{3}\, N^c_{f}\, e_f^2 \,\frac{ g^2  s_w^2}{16\pi^2} \, \frac{c_f}{ v}\, A_f(\tau_f)\,,\qquad
\lambda_g^f=c_f\,\lambda_g^{f,{\rm SM}}=-\frac{1}{3}\,\frac{ g^2_s}{16\pi^2}\, \frac{c_f}{ v} \,A_f(\tau_f) \,,\label{lambda_loop_gf}
\ee
\begin{multline}
\lambda_{Z \gamma}^W=a_W\,\lambda_{Z \gamma}^{W, {\rm SM}}= \frac{ e^2}{16\pi^2} \, \frac{a_W}{ v}\,
 t_w^{-1} \left( 2\left[ t^2_w -3\right] A_{Z \gamma}(\tau_W,\kappa_W) \phantom{\int} \right. \\ \left. + \left[
\frac{5- t_w^2}{2}+\frac{1- t_w^2}{\tau_W}
\right]B_{Z \gamma}(\tau_W,\kappa_W) \right)\,,
\end{multline}
\be
\lambda_{Z \gamma}^f=-c_f\,\lambda_{Z \gamma}^{f, {\rm SM}}= \frac{ e^2}{16\pi^2} \, \frac{c_f}{ v}\,
N^c_{f}\,\frac{e_f^2\, (T^{3L}_f-2e_f  s^2_w)}{ s_w  c_w} \bigl(B_{Z \gamma}(\tau_f,\kappa_f) - A_{Z \gamma}(\tau_f,\kappa_f)\bigr)\,.
\label{lambda_loop_Zgammaf}
\ee
where $N^c_f$ and $e_f$ are the number of colors and the fraction of electric charge of the fermion running in the loop, respectively.
We define $\tau_i=4m^2_i/m_h^2$, $\kappa_i=4m^2_i/m_Z^2$ . The form factors $A_i, B$ are given in Appendix  \ref{App_SMloops}. They are defined so that in the decoupling limit, $A_{f,v}\rightarrow 1$ when $\tau\rightarrow \infty$, and $A_{Z \gamma}\rightarrow 1$, $B_{Z \gamma}\rightarrow 0$  when $\tau, \kappa\rightarrow \infty$.

\subsection{Trilinear gauge boson vertices}

The higher dimensional operators that we are considering also affect charged triple gauge boson vertices (TGV).
In the parametrization of Ref.~\cite{Hagiwara:1986vm},
\be
\mathcal L_{\rm TGV}=
-i\, e\, \kappa_\gamma F_{\mu\nu}\,W_\mu^-W_\nu^+
-i\, g\, c_w\,\kappa_Z Z_{\mu\nu}\,W_\mu^-W_\nu^+
-i\, g\, c_w\,g_1^Z\left[ W_{\mu\nu}^+W_\nu^-- W_{\mu\nu}^-W_\nu^+\right]
 Z_{\mu} \,,
\ee
the deviations from the Standard Model can be expressed in terms of the HDO coefficients as follows:
\bea
\kappa_\gamma&=&1+\frac{\alpha_{\WB} }{2 t_w} \frac{v^2}{\Lambda^2} \,, \nn\\
\kappa_Z&=&1-\left(\frac{ s_w c_w}{( c_w^2- s_w^2)}\alpha_{\WB} +\frac{1}{4( c_w^2- s_w^2)}[\alpha'_{D^2}+\alpha_D] \right) \frac{v^2}{\Lambda^2} \,, \nn\\
g_1^Z&=& 1-\left(\frac{ s_w}{2 c_w( c_w^2- s_w^2)}\alpha_{\WB} +\frac{1}{4( c_w^2- s_w^2)}[\alpha'_{D^2}+\alpha_D] \right) \frac{v^2}{\Lambda^2} \,,
\label{eq:tgv_hdo}
\eea
where again some indirect effects from fixing input parameters were taken into account.
Gauge invariance implies the relation $\kappa_Z=g_1^Z-(\kappa_\gamma-1) t_w^2$  and one can check that it is indeed fulfilled. We then choose $\kappa_\gamma$ and $g_1^Z$ as independent couplings.

\section{Data treatment} \label{se_data}

We exploit the results from Higgs searches at the LHC and at Tevatron as well as electroweak precision observables and trilinear gauge couplings. The results from Higgs searches are given in terms of signal strengths $\mu(X,Y)$, the ratio of the observed rate for some process $X\to h \to Y$ relative to the prediction for the SM Higgs. An experimental channel is defined by its final state ($\gamma\gamma$, $ZZ$, $Z\gamma$, $WW$, $b\bar{b}$, $\tau\tau$) and is often divided into subchannels having different sensitivity to the various production processes. The accessible production mechanisms at the LHC are {\it i)} gluon-gluon fusion (ggF), {\it ii)} vector boson
fusion (VBF), {\it iii)} associated production with an electroweak gauge
boson $V=W,Z$ (VH), and {\it iv)} associated production with a $t\bar{t}$ pair (ttH).

As higher dimensional operators modify not only the Higgs decays but also its production (see Section~\ref{se_deviations}), care has to be taken in extracting the information given by the experiments. In particular, when available, we use the results given in the plane $\left(\mu({\rm ggF+ttH},Y), \mu({\rm VBF+VH},Y)\right)$, which (partly) account for the correlations between the subchannels.

The values for the signal strengths in the various (sub)channels as reported by the experiments and used in this analysis, together with the estimated decompositions into production channels are given in Tables~\ref{ATLASresults}--\ref{Tevatronresults}. Some of the decompositions into production channels are taken from~\cite{Belanger:2012gc}. In case of missing information, we take the relative ratios of production cross sections for a SM Higgs as a reasonable approximation, \ie~we assume that the experimental search is fully inclusive and compute the signal strength modified by HDOs accordingly. To this end, we use the latest predictions of the cross sections at the LHC~\cite{HXSWG} and at Tevatron~\cite{tevatron:2012zzl}. 

\begin{table}\centering
\begin{tabular}{|c|c|c|ccccc|}
\hline
Channel & Signal strength $\mu$ & $m_h$ (GeV) & \multicolumn{5}{c|}{Production mode} \\
& & & ggF & VBF & WH & ZH & ttH \\
\hline
\multicolumn{8}{|c|}{$h \rightarrow \gamma\gamma$ (4.8 fb$^{-1}$ at 7 TeV + 20.7 fb$^{-1}$ at 8 TeV)~\cite{ATLAS-CONF-2013-012,ATLAS-CONF-2013-014}} \\
\hline
$\mu({\rm ggF+ttH},\gamma\gamma)$ & $1.60 \pm 0.41$ & 125.5 & 100\% & -- & -- & -- & -- \\
$\mu({\rm VBF+VH} ,\gamma\gamma)$ & $1.94 \pm 0.82$ & 125.5 & -- & 60\% & 26\% & 14\% & -- \\
\hline
\multicolumn{8}{|c|}{$h \rightarrow ZZ$ (4.6 fb$^{-1}$ at 7 TeV + 20.7 fb$^{-1}$ at 8 TeV)~\cite{ATLAS-CONF-2013-013,ATLAS-CONF-2013-014}} \\
\hline
$\mu({\rm ggF+ttH},ZZ)$ & $1.51 \pm 0.52$ & 125.5 & 100\% & -- & -- & -- & -- \\
$\mu({\rm VBF+VH} ,ZZ)$ & $1.99 \pm 2.12$ & 125.5 & -- & 60\% & 26\% & 14\% & -- \\
\hline
\multicolumn{8}{|c|}{$h \rightarrow WW$ (4.6 fb$^{-1}$ at 7 TeV + 20.7 fb$^{-1}$ at 8 TeV)~\cite{ATLAS-CONF-2013-030,ATLAS-CONF-2013-034}} \\
\hline
$\mu({\rm ggF+ttH},WW)$ & $0.79 \pm 0.35$ & 125.5 & 100\% & -- & -- & -- & -- \\
$\mu({\rm VBF+VH} ,WW)$ & $1.71 \pm 0.76$ & 125.5 & -- & 60\% & 26\% & 14\% & -- \\
\hline
\multicolumn{8}{|c|}{$h \rightarrow b\bar{b}$ (4.7 fb$^{-1}$ at 7 TeV + 13.0 fb$^{-1}$ at 8 TeV)~\cite{ATLAS-CONF-2012-161,ATLAS-CONF-2013-014}} \\
\hline
VH tag & $-0.39 \pm 1.02$ & 125.5 & -- & -- & 64\% & 36\% & -- \\
\hline
\multicolumn{8}{|c|}{$h \rightarrow \tau\tau$ (4.6 fb$^{-1}$ at 7 TeV + 13.0 fb$^{-1}$ at 8 TeV)~\cite{ATLAS-CONF-2013-014}} \\
\hline
$\mu({\rm ggF+ttH},\tau\tau)$ & $2.31 \pm 1.61$ & 125.5 & 100\% & -- & -- & -- & -- \\
$\mu(\mathrm{VBF}+\mathrm{VH},\tau\tau)$ & $-0.20 \pm 1.06$ & 125.5 & -- & 60\% & 26\% & 14\% & -- \\
\hline
\end{tabular}
\caption{ATLAS results, as employed in this analysis. The following correlations are included in the fit: $\rho_{\gamma\gamma} = -0.27$, $\rho_{ZZ} = -0.50$, $\rho_{\WW} = -0.18$, $\rho_{\tau\tau} = -0.49$.}
\label{ATLASresults}
\end{table}

\begin{table}\centering
\begin{tabular}{|c|c|c|ccccc|}
\hline
Channel & Signal strength $\mu$ & $m_h$ (GeV) & \multicolumn{5}{c|}{Production mode} \\
& & & ggF & VBF & WH & ZH & ttH \\
\hline
\multicolumn{8}{|c|}{$h \rightarrow \gamma\gamma$ (5.1 fb$^{-1}$ at 7 TeV + 19.6 fb$^{-1}$ at 8 TeV)~\cite{CMS-PAS-HIG-13-001}} \\
\hline
$\mu({\rm ggF+ttH},\gamma\gamma)$ & $0.49 \pm 0.39$ & 125 & 100\% & -- & -- & -- & -- \\
$\mu({\rm VBF+VH},\gamma\gamma)$ & $1.65 \pm 0.87$ & 125 & -- & 60\% & 26\% & 14\% & --\\
\hline
\multicolumn{8}{|c|}{$h \rightarrow ZZ$ (5.1 fb$^{-1}$ at 7 TeV + 19.6 fb$^{-1}$ at 8 TeV)~\cite{CMS-PAS-HIG-13-002}} \\
\hline
$\mu({\rm ggF+ttH},ZZ)$ & $0.99 \pm 0.46$ & 125.8 & 100\% & -- & -- & -- & -- \\
$\mu({\rm VBF+VH} ,ZZ)$ & $1.05 \pm 2.38$ & 125.8 & -- & 60\% & 26\% & 14\% & -- \\
\hline
\multicolumn{8}{|c|}{$h \rightarrow WW$ (up to 4.9 fb$^{-1}$ at 7 TeV + 19.5 fb$^{-1}$ at 8 TeV)~\cite{CMS-PAS-HIG-13-003, CMS-PAS-HIG-12-039,CMS-PAS-HIG-12-042,CMS-PAS-HIG-12-045}} \\
\hline
0/1 jet & $0.76 \pm 0.21$ & 125 & 97\% & 3\% & -- & -- & -- \\
VBF tag & $-0.05^{+0.74}_{-0.55}$ & 125.8 & 17\% & 83\% & -- & -- & -- \\
VH tag & $-0.31^{+2.22}_{-1.94}$ & 125.8 & -- & -- & 64\% & 36\% & -- \\
\hline
\multicolumn{8}{|c|}{$h \rightarrow b\bar{b}$ (up to 5.0 fb$^{-1}$ at 7 TeV + 12.1 fb$^{-1}$ at 8 TeV)~\cite{CMS-PAS-HIG-12-044,CMS-PAS-HIG-12-025,CMS-PAS-HIG-12-045}} \\
\hline
${\rm Z}(\ell^-\ell^+){\rm h}$ & $1.55_{-1.07}^{+1.20}$ & 125 & -- & -- & -- & 100\% & -- \\
${\rm Z}(\nu\bar{\nu}){\rm h}$ & $1.79_{-1.02}^{+1.11}$ & 125 & -- & -- & -- & 100\% & -- \\
${\rm W}(\ell\nu){\rm h}$ & $0.69_{-0.88}^{+0.91}$ & 125 & -- & -- & 100\% & -- & -- \\
ttH tag & $-0.80^{+2.10}_{-1.84}$ & 125.8 & -- & -- & -- & -- & 100\% \\
\hline
\multicolumn{8}{|c|}{$h \rightarrow \tau\tau$ (4.9 fb$^{-1}$ at 7 TeV + 19.4 fb$^{-1}$ at 8 TeV)~\cite{CMS-PAS-HIG-13-004}} \\
\hline
0/1 jet & $0.76_{-0.52}^{+0.49}$ & 125 & 76\% & 16\% & 4\% & 3\% & 1\% \\
VBF tag & $1.40_{-0.57}^{+0.60}$ & 125 & 19\% & 81\% & -- & -- & -- \\
VH tag & $0.77_{-1.43}^{+1.48}$ & 125 & -- & -- & 64\% & 36\% & -- \\
\hline
\multicolumn{8}{|c|}{$h \rightarrow Z\gamma$ (5.0 fb$^{-1}$ at 7 TeV + 19.6 fb$^{-1}$ at 8 TeV)~\cite{CMS-PAS-HIG-13-006}} \\
\hline
Inclusive  &  $<9.3$ at 95\% CL   &  125.5   & 87\%  &7\%  & 3\% & 2\% & 1\% \\
\hline
\end{tabular}
\caption{CMS results, as employed in this analysis. The following correlations are included in the fit: $\rho_{\gamma\gamma} = -0.50$, $\rho_{ZZ} = -0.73$.}
\label{CMSresults}
\end{table}

\begin{table}\centering
\begin{tabular}{|c|c|c|ccccc|}
\hline
Channel & Signal strength $\mu$ & $m_h$ (GeV) & \multicolumn{5}{c|}{Production mode} \\
& & & ggF & VBF & WH & ZH & ttH \\
\hline
\multicolumn{8}{|c|}{$h \rightarrow \gamma\gamma$~\cite{HCPHiggsTevatron}} \\
\hline
Combined & $6.14^{+3.25}_{-3.19}$ & 125 & 78\% & 5\% & 11\% & 6\% & -- \\
\hline
\multicolumn{8}{|c|}{$h \rightarrow WW$~\cite{HCPHiggsTevatron}} \\
\hline
Combined & $0.85^{+0.88}_{-0.81}$ & 125 & 78\% & 5\% & 11\% & 6\% & -- \\
\hline
\multicolumn{8}{|c|}{$h \rightarrow b\bar{b}$~\cite{HCPtevBBtalk}} \\
\hline
VH tag & $1.56^{+0.72}_{-0.73}$ & 125 & -- & -- & 62\% & 38\% & -- \\
\hline
\end{tabular}
\caption{Tevatron results for up to $10\fbi$ at $\sqrt{s} = 1.96$~TeV, as employed in this analysis.}
\label{Tevatronresults}
\end{table}

In both ATLAS and CMS, the Higgs mass is estimated from the two ``high-resolution'' channels: $ZZ$ and $\gamma\gamma$. In our analysis, the Higgs mass is set to $m_h = 125.5$~GeV (close to the combined mass measurement from the two experiments) since it is not yet possible the take it as a nuisance parameter without losing the correlations between production channels. We consider experimental measurements of the signal strengths as close as possibe to this value.

We take into account the electroweak precision observables using the Peskin--Takeuchi $S$ and $T$ parameters~\cite{Peskin:1990zt,Peskin:1991sw}. 
Beyond $S$ and $T$, the $W$ and $Y$ parameters~\cite{Barbieri:2004qk} should be used in the HDO framework. However we find that constraints arising from these parameters are by far subleading with respect to our other constraints. 
 Experimental values of $S$ and $T$ are taken from the latest electroweak fit of the SM done by the Gfitter Group~\cite{Baak:2012kk}: $S=0.05 \pm 0.09$ and $T= 0.08 \pm 0.07$ with a correlation coefficient of 0.91.
Regarding constraints on TGV, we take into account the LEP measurements~\cite{TGV}:
\bea
\kappa_\gamma&=&0.973^{+0.044}_{-0.045} \,, \nn \\
g_1^Z&=&0.984^{+0.022}_{-0.019} \,.
\label{eq:lep_tgv}
\eea

The global likelihood function is defined as the product of the likelihoods associated to the various observables,
\be
L = L_{\rm Higgs} \times L_{S,T} \times L_{\rm TGV} \,,
\label{eq:likelihood}
\ee
where $L_{\rm Higgs}$ is the product of the likelihoods associated to each of the experimental (sub)categories, including available correlations.
The likelihood associated to the measurement of an observable $\hat O$, given as a central value $O$ and a symmetric uncertainty $\sigma$, is modeled by a normal law,
\be
L_{O}\propto e^{-(O-\hat O)^2/2\sigma^2}\,.
\ee
When uncertainties are asymmetric, we use the positive error bar if $(\hat{O} - O) > 0$, whereas we use the negative error bar if $(\hat{O} - O) < 0$.
Finally, the CMS bound on the decay channel $h \rightarrow Z\gamma$ is implemented as a step function, 
\be
L_{\mu_{Z\gamma}}\propto
\begin{cases}
  1~  &\textrm{if} ~~ \hat\mu_{Z\gamma}<9.3 \,,\\
  0 ~  &\textrm{otherwise}\,.
\end{cases} 
\ee 

We will now derive the deviations induced by the HDOs to the observables presented in Sec. \ref{se_data}. We first discuss the particular treatment of tensorial couplings. All formulas are given in the following section.

\section{On weak bosons tensorial couplings} \label{se_tensorials}

Because of electroweak symmetry breaking, the $W,Z\equiv V$ bosons generally couple to the Higgs through two  different Lorentz structures. The coupling can be vectorial, $\propto g^{\mu\nu}$, or it can be tensorial with a vertex $\propto (g^{\mu\nu}-\frac{q_1^\mu q_2^\nu}{q_1.q_2})$, where $q_1$, $q_2$ are the momenta of the two gauge bosons. The leading SM couplings $\lambda_{W},\lambda_{Z}$ given in Eqs.~\eqref{L_vect} and \eqref{lambda_vect} are vectorial. Tensorial couplings are generated only at one-loop and are $\mathcal{O}(\alpha) \sim 10^{-2}$. 

Once HDOs are taken into account, the relative importance of the vectorial and tensorial terms is modified. On one hand vectorial couplings are rescaled by the coefficients $a_{W,Z}$. On the other hand new tensorial contributions $\zeta_{W}$, $\zeta_{Z}$ are generated following Eq.~\eqref{lambda_WZ_tens}. The amplitude associated to a $hVV$ vertex (with the $V$'s possibly off-shell) is in general
\be
\mathcal{M}(h VV)^{\lambda_1,\lambda_2}=e^{\mu (*)}_{\lambda_1} e^{\nu (*)}_{\lambda_2} \left( ia_V \lambda_V^{\rm SM}g^{\mu\nu} -i2\zeta_{V}q_1.q_2\left[g^{\mu\nu}-\frac{q_1^\mu q_2^\nu}{q_1.q_2}\right]  \right)\,,
\label{M_hvv}
\ee
where $\mathcal{M}^{0,0}$ and $\mathcal{M}^{\pm,\pm}$ are the longitudinal and transverse helicities amplitudes, respectively. Interferences among helicity amplitudes then determine angular distributions (see \eg~\cite{Hagiwara:2009wt}). 
In this work, we consider that the SM contribution to the tensorial coupling is small with respect to the one induced by New Physics.
The relative magnitude of the longitudinal and transverse amplitudes in case of a vectorial coupling is given by
\be
r_v=\left|\frac{\mathcal{M}_v^{0,0}}{\mathcal{M}_v^{\pm,\pm}}\right|=\frac{\left|m_h^2 - q_1^2-q_2^2\right|}{2|q_1||q_2|}\,,
\ee
while it is the inverse in case of a tensorial coupling,
\be
r_t=\left|\frac{\mathcal{M}_t^{0,0}}{\mathcal{M}_t^{\pm,\pm}}\right|=\frac{2|q_1||q_2|}{\left|m_h^2 - q_1^2-q_2^2\right|}\,.
\ee
The two vector bosons can be off-shell in the above expression, while the Higgs is on-shell.

As $r_v\neq r_t$, the two Lorentz structures imply generally different angular distributions. Moreover, even for unpolarized processes, the energy dependence in  Eq. \eqref{M_hvv} is different for both contributions, such that also energy distributions are modified.
Because of this different energy dependence, kinematic cuts prepared for the SM are generally unadapted to such a non-trivial modification. 
That is, in Eq. \eqref{signal_strength_modif}, $\varepsilon_{\rm SM+HDO }\neq\varepsilon_{\rm SM}$.  The consequences may be an incorrect estimation of the signal strength and of the Higgs mass.  To perform an exact analysis, one should redo the fits to LHC data taking into account the modified Lorentz structure in the expected signal. Such work is clearly beyond the scope of our present study.
Instead we will show that under reasonable  approximations we can use  $\varepsilon_{\rm SM+HDO }=\varepsilon_{\rm SM}$ in the present analysis.

There are three processes sensitive to the $\zeta_{V}$ tensorial couplings in the context of the searches for the Higgs boson at around 125 GeV: the leading decay to weak bosons $h\rightarrow VV^*$, and the VBF and VH production modes.  We now discuss how we treat these three tensorial contributions.

\subsection{$h\rightarrow VV^*$}

In the case of a light Higgs boson, the leading decay occurs with one of the $V$ off the mass shell. The weak bosons then decay into fermions. For massless fermions, the kinematic bounds on the on-shell boson energy  $E_V$ are $m_V<E_V<(m_h^2+m_V^2 ) /2 m_h$ in the rest frame of the Higgs. Because of the $V^*$ propagator, the lower bound $E_V=m_V$ is favorized, implying that both weak bosons are preferentially produced at rest. Longitudinal and transverse amplitudes are then equally populated, $r_v=1$. Therefore, one has $r_t=1$ as well, such that one can see qualitatively that a tensorial contribution  cannot radically modify  angular distributions. This is confirmed with the exact  angular and invariant mass distributions among leptons induced  by pure vectorial and pure tensorial couplings \cite{Gao:2010qx, Stolarski:2012ps}.~\footnote{
Overall, the situation is much less striking than for a CP-violating contribution, which forbids the decay to the longitudinal polarization state. } 
In our study, the tensorial contributions are constrained to be subleading with respect to the vectorial contributions, such that the deviations induced on angular and invariant mass distributions can be easily be smaller than the current statistical uncertainty. In addition, they could also be misidentified with the background. For example, in $h\rightarrow VV^*$, the distribution of the most discriminant observable, ``lepton-opposite $Z$ momentum angle'', is very similar to the distribution of the irreducible background  $q\bar q\rightarrow ZZ^*$ (see Fig.~3 in \cite{Gainer:2011xz}).

Following what discussed above, we can reasonably assume that angular and invariant mass distributions are not affected by the presence of tensorial couplings given the current level of precision.
Polarization of the on-shell $V$ can thus be averaged, and we are left with  a matrix element scaling as 
\be
|\mathcal{M}|^2=|\mathcal{M}_v+\mathcal{M}_t|^2\propto\left| a_V\lambda_{V}^{\rm SM}-2\zeta_{V}q_1.q_2 \right|^2  \,, \label{amplitude_vect_tens}
\ee
where $q_1$, $q_2$ are the momenta of the two vector bosons.  In the Higgs rest frame, one has $q_1.q_2=m_h E_V-m_V^2$, which is bounded as
\be
m_V(m_h-m_V)< q_1.q_2< \frac{m_h^2-m^2_V}{2} \,. \label{eq_bound_EV}
\ee 
 The exact tensorial contributions to the total decay widths are given in App.~\ref{App_tens}. We introduce the dimensionless positive quantity
\be\nu_{VV}=q_1.q_2/m_h^2\,,\ee
with $V\equiv W,Z$.
Defining
\be
\langle \nu_{VV} \rangle= \frac{ \int \nu_{VV} \mathcal{M}_v \mathcal{M}_t^* dPS}{\int \mathcal{M}_v \mathcal{M}_t^* dPS}\,,\,\,\,\,\,\,
\langle \nu_{VV}^2\rangle = \frac{ \int \nu^2_{VV} |\mathcal{M}_t|^2dPS}{\int  |\mathcal{M}_t|^2 dPS} \,, \label{nu_VV}
\ee
the vector-tensor interference term will be $\propto \zeta_{V} \langle \nu_{VV}\rangle$ and the pure tensor contribution will be $\propto|\zeta_{V}|^2 \langle \nu^2_{VV}\rangle $. 
For $m_h=125.5\,\textrm{GeV}$, $m_Z=91\,\textrm{GeV}$, $m_W=80\,\textrm{GeV}$,  one gets $\langle \nu_{ZZ}\rangle=0.2209$, $\langle \nu^2_{ZZ}\rangle^\frac{1}{2}=0.2211$, $\langle \nu_{\WW}\rangle=0.2653$, $\langle\nu^2_{\WW}\rangle^\frac{1}{2}=0.2659$. In the following we will make the approximation $\langle \nu_{VV}^2\rangle\approx \langle \nu_{VV}\rangle^2$.

\subsection{VBF production mode} \label{se_vbf}

For the VBF process, both ATLAS and CMS apply hard cuts on the outgoing jets rapidities and their difference. The rapidity distributions of the two jets are similar in presence of a tensorial coupling, just like in the decay into two photons or in the production via gluon-gluon fusion, such that one can assume that cut efficiency is the same. The crucial change lies in the azimuthal angle $\phi_{jj}$ between the two tagging jets  (see \eg~\cite{Hagiwara:2009wt} and references therein).
Indeed, both weak bosons are space-like, with virtualities considerably smaller than $m_h^2$. Such values are favorized to balance the space-like $V$ and the outgoing jets virtualities. As a result one has typically $r_v\gg 1$, $r_t\ll 1$ \ie~vectorial and tensorial amplitudes are mostly longitudinal and transverse, respectively.  Consequently, the $\phi_{jj}$ distribution is almost flat for a pure vectorial coupling, and strongly peaked at $\pi/2$ for a pure tensorial coupling. For a large enough HDO contribution to the tensorial coupling, an anomalous $\phi_{jj}$ distribution could thus be observed.
 However, this variable is not used for the selection of the events in the experimental analyses we consider. Therefore, the selection efficiencies are also suitable in the case of large tensorial contributions, and one has $\varepsilon_{\rm SM}=\varepsilon_{\rm SM+HDO}$. One can average over the polarizations, and the squared amplitude is then simply rescaled by a factor 
$\left| a_V\lambda_{VBF}^{\rm SM}-2\zeta_{V}q_1.q_2 \right|^2 $.

We still have to determine the magnitude of the tensorial contribution.
In this process, the scalar product of the weak boson momenta $q_1.q_2$ is related to the incoming and outgoing quarks as $q_1.q_2=m_h^2/2+p_1.p_3+p_2.p_4$. The outgoing quarks are highly energetic with respect to the amount of $p_T$ they receive from the $V$ fusion, such that one has $\bf{\left|p_1\right|\simeq\left|p_3\right|}$ and $\bf{\left|p_2\right|\simeq\left|p_4\right|}$. In terms of the $p_T$ and rapidities of the outgoing quarks we have then 
\be
q_1.q_2=\frac{m_h^2}{2}+ |p_{T,3}|^2\frac{1+e^{-\eta_{3}}}{2}+ |p_{T,4}|^2\frac{1+e^{-\eta_{4}}}{2}\,.
\ee 
Without the tensorial contribution, the $p_T$ distribution peaks typically at values smaller than $m_V$.  The tails of the $p_T$ distributions drop quickly for higher energies \cite{Aad:2009wy}, with typically one jet at a time getting a large $p_T$ \cite{Djouadi:2005gi}. One can thus assume $q_1.q_2\approx m_h^2/2$  to a good approximation.
Once the tensorial coupling is taken into account, a deviation from the expected SM distributions might be present in the high-$p_T$ tails, as $q_1.q_2$ is enhanced at large $p_T$.
However, as long as one counts the total number of events, \ie~the integral of the distribution, this enhancement of $q_1.q_2$ has a small weight and can be safely neglected.
Finally, defining the dimensionless positive quantity \be\nu_{\rm VBF}=q_1.q_2/m_h^2\,,\ee
with $V=W,Z$, we have thus $\nu_{\rm VBF}\approx 1/2$ after phase space integration.

\subsection{VH production mode}

In the case of the associated production with an electroweak gauge boson, the scalar product of the momenta of the weak bosons is given by
\be
q_1.q_2=\frac{s+m_V^2-m_h^2}{2}\,,
\ee
where $\sqrt{s}$ is the partonic center-of-mass energy, which is typically  $s/m_V^2=\mathcal{O}(100)$   at the LHC. 
Therefore, contrary to the two other processes, the product $q_1.q_2$ is large because it contains the partonic center of mass energy $\sqrt{s}$. The tensorial contribution is then substantially enhanced in this process.
Besides, we have $r_v \neq r_t$, as $r_v=r_t^{-1} \approx \sqrt{s}/2m_V$, such that the angular distributions may in principle be substantially modified by the presence of the tensorial coupling.

However, it turns out that for both polar and angular distributions, the angular effects can be neglected. We refer to \cite{Kilian:1996wu} and references therein for the expressions. Although results are given for $e^+ e^-$ collisions, they can be trivially generalized in the case of the LHC. For the distribution of the polar angle of the vector boson in the laboratory frame, it is the longitudinal component of $V$ which enters mainly, such that the tensorial contribution to the distribution is suppressed by an additional factor $\mathcal{O}(m_V^2/s)$. 
For the azimuthal distributions, the tensorial contributions can be sizeable, but the whole distribution tends to be flat for $s\gg m_V^2$, with non-flat terms suppressed by powers of $m_V/\sqrt{s}$. 
As a result, although various pieces of angular information are used in event selection for this  mode of production, we can safely neglect the angular effects of the tensorial coupling. 

Concerning the magnitude of the tensorial contribution, it appears that it reduces to a simple rescaling $\propto \lambda_V^{\rm SM}+ 12\zeta_V m_V^2$  in the limit $s\gg m_V^2$. The rescaling is exact up to a subleading term $\mathcal{O}(12 m_V^2/s)\approx 0.1$. To include the subleading $s$-dependent terms, an integration over the partonic density functions would be necessary.

\section{Deviations caused by New Physics} \label{se_deviations}

\subsection{Higgs signal strengths}

Theoretical signal strengths for Higgs searches can be expressed as 
\be
\hat\mu(X,Y)=\frac {\left[ \sigma(X\rightarrow h) \, \mathcal{B}(h\rightarrow Y ) \, \varepsilon^{XY}\right]_{\textrm{SM+HDO} }}  {\left[ \sigma(X\rightarrow h) \, \mathcal{B}(h\rightarrow Y ) \, \varepsilon^{XY}\right]_{\textrm{SM}}}\, \label{signal_strength_modif}
\ee
where $\mathcal B$ is the branching ratio of the decay and the coefficient $\varepsilon^{XY}\in[0,1]$ characterizes the efficiency of event selection for a given subcategory. In all generality, efficiencies in the SM with and without HDOs are not necessarily the same, \ie~$\varepsilon_{\textrm{SM+HDO}} \neq \varepsilon_{\textrm{SM}}$, because kinematic distributions can be modified in a non-trivial way by HDOs.  The selection criteria calibrated on the SM expectations are then unadapted in such situation and complicates the interpretation of the signal strengths. 

However, we have seen in  Sec.~\ref{se_tensorials} that one can safely ignore these possibilities of HDOs affecting the kinematic distributions, given the current precision of the experimental searches. It is therefore a good approximation to set $\varepsilon_{\textrm{SM+HDO}} =\varepsilon_{\textrm{SM}}$. Thus, for each signal strength, one can simply incorporate the contributions coming from the tensorial couplings in the rescaling of the Standard Model signal strength.

The gluon-gluon fusion process is modified both by the tree-level HDO contribution $\zeta_g$ and the anomalous Higgs--fermion couplings $c_f$. Keeping only the third generation, we get
\be
\sigma_{\rm ggF}=\sigma^{\rm SM}_{\rm ggF} \left|\frac{c_t\lambda_g^{t, {\rm SM}}+c_b\lambda_g^{b, {\rm SM}}+\zeta_g}{\lambda_g^{t, {\rm SM}}+\lambda_g^{b, {\rm SM}}} \right|^2\,.
\ee
Vector boson fusion is modified by the anomalous vectorial couplings $a_{W,Z}$ and by $\zeta_{W,Z}$. Denoting by $\lambda_{\rm VBF}^{\rm SM}$ the effective SM couplings, one has
\be
\sigma_{\rm VBF}=\sigma_{\rm VBF}^{\rm SM} \left| \frac{a_W\lambda_{W}^{\rm SM} + a_Z\lambda_{Z}^{\rm SM} - 2 \, \nu_{\rm VBF} \, m_h^2 \, (\zeta_{W}+\zeta_{Z})}{\lambda_{W}^{\rm SM}+\lambda_{Z}^{\rm SM}} \right|^2\,.
\ee
The parameter $\nu_{\rm VBF}$ is defined in Sec.~\ref{se_vbf}. We take $\nu_{\rm VBF} = 1/2$.
The associated production with an electroweak gauge boson is modified as 
\be
\sigma_{\rm VH}=\sigma_{\rm VH}^{\rm SM}\left|\frac{a_V\lambda_V^{\rm SM}+12\zeta_V m_V^2}{\lambda_V^{\rm SM}}   \right|^2\,,
\label{eq:VH}
\ee
where $V=W,Z$. Finally, the associated production with a $t\bar{t}$ pair is rescaled as  
\be
\sigma_{\rm ttH} = |c_t|^2 \sigma_{\rm ttH}^{\rm SM}\,.
\ee
The decays of the Higgs boson into fermions are modified as
\be
\Gamma_{ff}=|c_f|^2\Gamma^{\rm SM}_{ff}\,.
\ee
The tree-level decays to vector bosons are modified as
\be
 \Gamma_{VV}=\left|\frac{a_V\,\lambda_{V}^{\rm SM}-2\,\zeta_{V} m_h^2 \,\langle\nu_{VV}\rangle }{\lambda_{V}^{\rm SM}
 }\right|^2 \Gamma^{\rm SM}_{VV}\,,
\ee
where the  parameter $\langle \nu_{VV}\rangle$,
 defined in Eq.~\eqref{nu_VV}, encodes the modification of phase space integrals.
Loop-induced decays are sensitive to more deviations, 
\be
 \Gamma_{\gamma\gamma}= \Gamma_{\gamma\gamma}^{\rm SM} \left|\frac{a_W\lambda_\gamma^{W, {\rm SM}}+c_t\lambda_\gamma^{t, {\rm SM}}+c_b\lambda_\gamma^{b, {\rm SM}}+c_\tau\lambda_\gamma^{\tau, {\rm SM}}+\zeta_\gamma}{\lambda_\gamma^{W, {\rm SM}}+\lambda_\gamma^{t, {\rm SM}}+\lambda_\gamma^{b, {\rm SM}}+\lambda_\gamma^{\tau, {\rm SM}}} \right|^2\,,
\ee
\be
 \Gamma_{Z \gamma}= \Gamma_{Z \gamma}^{\rm SM} \left|\frac{a_W\lambda_{Z \gamma}^{W, {\rm SM}}+c_t\lambda_{Z \gamma}^{t, {\rm SM}}+c_b\lambda_{Z \gamma}^{b, {\rm SM}}+c_\tau\lambda_{Z \gamma}^{\tau, {\rm SM}}+\zeta_{Z \gamma}}{\lambda_{Z \gamma}^{W, {\rm SM}}+\lambda_{Z \gamma}^{t, {\rm SM}}+\lambda_{Z \gamma}^{b, {\rm SM}}+\lambda_{Z \gamma}^{\tau, {\rm SM}}} \right|^2\,.
\ee
In such cases, the tensorial couplings can compete with the SM effective couplings.

\subsection{QCD radiative corrections}
 
Many of the above described processes receive leading radiative corrections from QCD loops. For all the tree-level processes, the structure of loop diagrams is not modified by the insertion of HDOs, including the tensorial couplings, such that radiative corrections factorize up to higher order corrections. It is thus straightforward to take them into account, simply using the NLO predictions of $\sigma^{\rm SM}$ and $\Gamma^{\rm SM}$.

The situation is more involved in the case of the loop-induced processes ($h \rightarrow \gamma \gamma$, $h \rightarrow Z \gamma$, and $gg \rightarrow h$) because this time the tensorial coupling is competing with the SM loops. Hence the effects of the $\zeta$'s may be very large in these processes, such that it is important to properly take into account the radiative corrections.
As stated in Sec.~\ref{se_HDOs}, the HDOs implicitely contain higher-order corrections from irreducible SM loops. These contributions therefore have to be taken into account for the SM effective couplings and not for the $\zeta$ couplings.~\footnote{We are grateful to M. Spira for enlightening discussion on this subject.} 
 
The processes $h \rightarrow \gamma \gamma$ and $h\rightarrow Z \gamma$ only receive virtual NLO QCD corrections. For $h \rightarrow \gamma \gamma$, we take into account the exact values of the correction factor to the quark effective couplings 
 \be
 \lambda_\gamma^{q, {\rm SM}}=\lambda_\gamma^{q, {\rm SM}}|_{\rm LO} \left( 1+\frac{\alpha_s}{\pi}C_H(\tau_q) \right)\,,
 \ee   
 where the $C_H$ function can be found in \cite{Djouadi:2005gi}. For $h\rightarrow Z \gamma$, one can take the correction in the heavy top limit as a good approximation~\cite{Djouadi:2005gi},
 \be
 \lambda_\gamma^{t, {\rm SM}}=\lambda_\gamma^{t, {\rm SM}}|_{\rm LO} \left( 1-\frac{\alpha_s}{\pi} \right)\,.
 \ee   

The situation  is more subtle for the ggF process, because of the presence of important NLO real corrections. Introducing the tensorial coupling leads generally to non-trivial modifications of the integrals over parton densities for real emissions. However, in the heavy-top limit and neglecting the small bottom quark contribution, the QCD corrections to the SM loop and to the tensorial coupling $\zeta_g$ become similar and factorize. Adopting this fairly good approximation, the SM effective coupling are rescaled as 
 \be
 \lambda_g^{t, {\rm SM}}=\lambda_g^{t, {\rm SM}}|_{\rm LO} \left( 1+\frac{11}{4}\frac{\alpha_s}{\pi} \right)\,.
 \ee

\subsection{$S$ and $T$ parameters}

The electroweak precision observables are affected in the presence of the HDOs. At tree-level the $S$ and $T$ parameters are related to the HDO coefficients as follows:
\bea
\alpha\,S&=&\biggr(2\, s_w c_w\,\alpha_{\WB}+ s_w^2\,\alpha_D+ c_w^2\,\alpha_{D}' \biggr)\frac{ v^2}{\Lambda^2}\,, \label{S_tree}\\
\alpha\,T&=&\left(-\frac{1}{2}\,\alpha_{D^2}'+\frac{1}{2}\,\alpha_{D}' \right)\frac{ v^2}{\Lambda^2}\,. \label{T_tree}
\eea
Moreover, the SM loops are modified by the HDOs. The $T$ parameter receives  new divergent contributions from the modified SM couplings $a_Z$ and $a_W$ in Eq.~(\ref{anomalous1}).  A quadratic divergence,
\be
\alpha \Delta T=-\frac{\Lambda^2}{16 \,\pi^2\, v^2} \frac{\alpha_{D^2}' v^2}{\Lambda^2} \,,
\ee
arises from custodial breaking \cite{Farina:2012ea}. Dropping other terms that are proportional to $\alpha_{D}'$ and $\alpha_{D^2}'$ (that already appear at tree-level) the two couplings coincide and we can take the result from Ref.~\cite{Espinosa:2012im},
\be
\alpha \Delta T=-\frac{3\, e^2}{32\, \pi^2\, c_w^2} \left(\alpha_{D^2}-\frac{1}{2}\alpha_D\right) \frac{ v^2}{\Lambda^2}
\log\left(\frac{m_h}{\Lambda}\right) \,.
\ee 
Similarly, the $S$ parameter receives corrections due to the modified Higgs coupling $\alpha_Z$ \cite{Espinosa:2012im}, 
hence it is expected to get new contributions proportional to $\alpha_{D^2}$ and $\alpha_{D^2}'$. 
Finally, the tensor couplings $\zeta_{V}$ can also generate new SM loop contributions which have been given in Ref.~\cite{Alam:1997nk}, 
\be
\alpha \Delta S=
\frac{ e^2}{24 \pi^2}\left(\alpha_{D^2}+\frac{1}{2}\alpha_{D^2}'\right)\frac{ v^2}{\Lambda^2}\log\left(\frac{m_h}{\Lambda}\right)
+\frac{ e^2}{2\pi^2}(\alpha_{\BB}+\alpha_{\WW})\frac{ v^2}{\Lambda^2}
\log\left(\frac{m_h}{\Lambda}\right)
\,.
\label{eq:DeltaS}
\ee
Finally we neglect the contraints coming from the $W$ and $Y$ parameters \cite{Barbieri:2004qk} as they are expected to have a small impact on our results.

\section{Bayesian setup and low-$\Lambda$ scenario} \label{se_bayesian_setup}

\subsection{Bayesian inference}

We are working in the framework of Bayesian statistics (see \cite{Trotta:2008qt} for an introduction). In this approach, a probability is interpreted as a measure of the degree of belief about a proposition.
Our study lies in the domain Bayesian inference, which is based on the relation
\be
p(\theta |d,\mathcal{M})\propto p(d|\theta,\mathcal{M})p(\theta|\mathcal{M})\,,
\ee
where $\theta\equiv\{\theta_{1\ldots n}\}$ are the parameters of the model $\mathcal{M}$, and $d$ denotes the experimental data. The distribution $p(\theta |d,\mathcal{M})$ is the so-called posterior probability density function (PDF), $p(d|\theta,\mathcal{M})\equiv L(\theta)$ is the likelihood function enclosing experimental data, and  $p(\theta|\mathcal{M})$ is the prior PDF, which represents our a priori degree of belief on the parameters. The model $\mathcal{M}$ is in our case the Standard Model extended with higher dimensional operators. The likelihood is defined in Sec.~\ref{se_data} (see Eq.~(\ref{eq:likelihood})) and the theoretical expressions for the HDO modified signal strengths are given in Sec.~\ref{se_deviations}.
The prior PDF is discussed in the next subsection.

The posterior PDF is the core of our results. Integrating the posterior over a subset $\lambda$ of the parameter set $\theta\equiv \{\psi,\lambda\}$,
\be p(\psi|d,\mathcal{M})\propto\int d\lambda\, p(\psi,\lambda|\mathcal{M})L(\psi,\lambda)\,,\ee
leads to inference on the parameters $\psi$.

Also the notion of naturalness and fine-tuning are built-in \cite{Fichet:2012sn} in the Bayesian approach. This is relevant for our study, in which precise (``fine-tuned'') cancellations between various HDO contributions can happen. Intrinsically, the regions of parameter space in which precise cancellations occur have a weak statistical weight, such that they are flushed away after integration. The results we will present can thus be considered as generic, \ie~free of improbable cancellations. 

We will consider uniform (flat) priors for the quantities
\be
\beta_i\equiv \alpha_i\frac{v^2}{\Lambda^2}
\ee
and demand $|\beta_i|<1$.
Moreover, we will fix the cutoff scale to be $\Lambda=4\pi v$. In the following we will justify these choices and argue that it ensures in particular convergence of the HDO expansion as well as perturbativity of the UV theory, and minimizes the dependence on the choice of the HDO basis.

\subsection{Priors and low-$\Lambda$ scenarios}

The prior distributions associated to our parameters is a key feature of Bayesian inference. We follow the ``principle of indifference'' \cite{press,jaynes} that maximizes the objectiveness of the priors. Once a transformation law $\gamma =f(\theta)$ irrelevant  for a given problem is identified, this principle let us find the most objective prior by identifying $p_\Theta\equiv p_\Gamma$ in the relation $p_\Theta(\theta)d\theta=p_\Gamma(\gamma)d\gamma$.

The cutoff scale $\Lambda$ is given a logarithmically uniform PDF,
\be p(\Lambda)\propto \frac{1}{\Lambda}\,. \label{prior_lambda}\ee
By doing so, all order of magnitudes are given the same probability density. 
Regarding the dimensionless coefficients $\alpha$, note that the choice of the HDO basis should be irrelevant for the conclusions of our study. Given that coefficients in different basis are related through linear transformations, the most objective prior to associate to each $\alpha_i$ is the uniform PDF,~\footnote{Here the principle of indifference sets the shape of the PDFs but does not set the bounds. One can see that ranges on $\alpha$'s are not conserved from one basis to another. In the scenario of democratic HDOs, this issue will be automatically solved, as one relies only on  perturbativity of the HDO expansion to set the bounds on $\alpha$'s. In the scenario of loop suppressed $\mathcal{O}_{FF}$'s, one takes advantage of a particular choice of basis, so the same argument does not apply in that case.  
 } 
\be
p(\alpha_i)\propto 1\,.\label{prior_ci}
\ee 
This choice of prior is well justified, however, one should keep in mind that other possibilities still exist.

Let us emphasize that in our general framework, the following hypotheses need to be scrutinized.
\begin{itemize}
\item Perturbativity of the HDO expansion, $|\alpha_i|/\Lambda^2 < \mathcal{O}( 1/v^2)$,
\item Perturbativity of the couplings expansions in the UV theory, $|\alpha_i| < \mathcal{O}(16\pi^2)$,
\item HDO generation by loops, 
\item Custodial symmetry.
\end{itemize}

In the present work, we investigate scenarios of low-scale New Physics, with values of $\Lambda$ going up to  ${\cal O}(4\pi v)$. We take  custodial symmetry  to be an exact symmetry of the theory.
This forbids the presence of the operators $\mathcal O'_{D^2}$ and $\mathcal O'_{D}$.  As a consequence, one has $a_W=a_Z\equiv a_V$ and some contributions to the EW precision observables are suppressed including the potentially large quadratic divergence in $T$. Recall that $\mathcal O_{\WW}$, $\mathcal O_{\WB}$, and $\mathcal O_{\BB}$ are all independently custodially symmetric. This generally implies that processes involving the $W$ and $Z$ are not identically rescaled, for instance
\be
\frac{\sigma_{\rm WH}}{\sigma_{\rm WH}^{\rm SM}} \neq \frac{\sigma_{\rm ZH}}{\sigma_{\rm ZH}^{\rm SM}} \,.
\ee
Our approach goes therefore beyond the fits involving pure rescalings induced by anomalous couplings.

Over this range of $\Lambda$, perturbativity of the HDO expansion is the dominant constraint as it requires $|\alpha_i|<\Lambda^2/v^2$ which automatically implies $|\alpha_i|<16 \pi^2$ and hence perturbativity of the couplings expansions in the UV theory.

When the HDOs are generated within a perturbative UV theory, none of the field strength--Higgs operators ${\cal O}_{FF}\equiv {\cal O}_{\WW,\,\WB,\,\BB,\,GG}$ (see Eqs. \eqref{HDO_O_WW} and \eqref{HDO_O_GG}) can be generated at tree-level. Because of our appropriate choice of basis, these loop-generated HDOs are exactly the ones associated with the tensorial couplings $\zeta_{g,\gamma,Z\gamma}$. We will therefore distinguish between two scenarios, depending on whether or not the $\mathcal O_{FF}$'s are loop suppressed with respect to the other HDOs. Given that tensorial couplings can play an important role, this distinction is particularly crucial. The two scenarios, denoted by I and II, are respectively dubbed ``democratic HDOs'' and ``loop-suppressed $\mathcal O_{FF}$'s''. The main features are summarized in Table \ref{tab_scenarios}.
These two scenarios are generic, in the sense that they encompass all known UV models in addition to the ones not yet thought of. This implies that features predicted only by specific UV models--\textit{e.g.} suppression of HDOs or precise cancellations between HDOs--will get a small statistical weight, as we consider the whole set of UV realizations. Finally, we emphasize that the interpretation of $\Lambda$ as a true New Physics scale also depends at which order  the whole set of HDOs is generated. For instance, in the $R$-parity conserving MSSM, the whole set of HDOs is generated only at one-loop order, such that the actual NP scale should be $\mathcal{O}(4\pi \Lambda)$.

A parameterization particularly adapted to low-$\Lambda$ scenarios is as follows. Defining the parameters
\be
\beta_i=\alpha_i \frac{v^2}{\Lambda^2}\,,
\ee
it follows that the $\beta$'s and $\Lambda$ are independent, \ie~$p(\alpha_i,\Lambda)=p(\Lambda)p(\beta_i)$. 
The $\beta$'s prior is the uniform PDF over $[-1;1]$, noted $U(\beta_i)$. The prior of $\Lambda$ is $p(\Lambda)\propto\Lambda^{2n-1}$, where $n$ is the number of  $\beta$'s. In our case, $n=9$ is large enough such that this prior is essentially peaked at $\Lambda_{\rm max}$, $p(\Lambda)\approx\delta(\Lambda-\Lambda_{\rm max})$. We have therefore 
\be
p(\alpha_i,\Lambda)=\delta(\Lambda-\Lambda_{\rm max})U(\beta_1)\ldots U(\beta_n)\,.
\ee

This factorization allows us to marginalize over $\Lambda$, and to present our results in terms of $\beta$'s, which contain all the relevant information. A mild dependence on $\Lambda$ will remain through loop-level $\mathcal{O}(\log \Lambda)$ terms in the $S$ and $T$ parameters, that will be discussed below. The fact that $\beta$'s prior is  uniform and spans a constant range is essential  to facilitate interpretation of the posterior PDFs. The fact that $\Lambda\approx \Lambda_{\rm max}$ is also useful, as it renders straightforward the evaluation of the few $\Lambda$-dependent terms.

\begin{table}
\center
\begin{tabular}{|c|c|c|}
\hline
& I) Democratic HDOs & II) Loop-suppressed $\mathcal O_{FF}$'s \\
\hline
$\Lambda$   &  $4\pi v$ &  $4\pi v$ \\
\hline
$\beta_{FF}$   &  $[-1,1]$ &  $[-1/16\pi^2,1/16\pi^2]$ \\
\hline
Other $\beta$   &  $[-1,1]$ & $[-1,1]$ \\
\hline
\end{tabular}
\caption{Summary of the setup of the scan in the two scenarios we consider. The $\beta_{FF} \equiv \alpha_{FF} \, v^2 / \Lambda^2$ coefficients (where $FF = \WW,\,WB,\,BB,\,GG$) correspond to the field-strength--Higgs operators. In both cases we take custodial symmetry to be an unbroken symmetry.} \label{tab_scenarios}
\end{table}

This parameterization turns out to be convenient in order to extract information about HDOs  in a scale independent way, up to a mild ${\cal O}(\log\Lambda_{\rm max})$ dependence. For example, for a given $\Lambda$, one can directly read the values of $\alpha$'s on the $\beta$'s plot. Similarly, for given $\alpha$'s, one can deduce the allowed $\Lambda$ values from the plots. 
This parameterization is appropriate at low $\Lambda$, up to $\Lambda={\cal O}(4\pi v)$. Beyond this scale, the bound from HDO perturbative expansion competes with the bound from the perturbative expansion of the couplings. Once the latter dominates, the features of factorization no longer hold.

\subsection{The MCMC setup}

We evaluate posterior PDFs by means of a Markov Chain Monte Carlo (MCMC) method. The basic idea of a MCMC is setting a random walk in the parameter space such that the density of points asymptotically reproduces the posterior PDF. Any marginalisation is then reduced to a summation over the points of the Markov chain. We refer to~\cite{Allanach:2005kz,Trotta:2008qt} for details on MCMCs and Bayesian inference. Our MCMC method uses the Metropolis-Hastings algorithm with a symmetric, Gaussian proposal function.
We run respectively 50 and 15 chains with $\mathcal{O}(10^8)$ iterations each for the democratic HDOs case and the loop-suppressed $\mathcal{O}_{FF}$'s case.
Finally, we check the convergence of our chains using an improved Gelman and Rubin test with multiple chains~\cite{Gelman92}. The first $10^4$ iterations are discarded (burn-in).

\section{Inference on HDOs} \label{se_results}

In this section we present and analyze the posterior PDFs arising in our scenarios of democratic HDOs and loop-suppressed $\mathcal{O}_{FF}$'s, denoted by I and II, respectively. Our results will be shown in terms of the $\beta_i \equiv \alpha_i v^2 / \Lambda^2$ parameters, which encode information about the fundamental parameters. Recall that the $\beta$'s prior PDF is uniform, and that the $\beta$ PDFs we show are valid for any value of the cutoff scale $\Lambda<\mathcal{O}(4\pi v)$, up to a mild $\log\Lambda$ dependence.
Moreover, this parameterization sets $\Lambda\approx \Lambda_{\rm max}$.
The posterior PDF we present is computed for $\Lambda=4\pi v\approx 3\,$TeV. For smaller $\Lambda$, we expect the $\propto \log\Lambda$ constraints from $\Delta S$ and $\Delta T$ to mildly relax. 
We will comment below on this effect.

We will also discuss deviations from the SM cross sections and decay widths, defining
\be
R_X = \frac{\sigma_X}{\sigma^{\rm SM}_X}\,, \quad R_Y=\frac{\Gamma_{h \rightarrow Y}}{\Gamma^{\rm SM}_{h \rightarrow Y}}\,, \quad R_{\rm width}=\frac{\Gamma_h}{\Gamma^{\rm SM}_h}\,,
\label{eq:Rs}
\ee
where $X = {\rm ggF}, {\rm VBF}, {\rm WH}, {\rm ZH}, {\rm ttH}$, and $Y = \gamma\gamma$, $ZZ$, $Z\gamma$, $WW$, $b\bar{b}$, $\tau\tau$. Note that the observables are the signal strengths $\hat \mu(X,Y)$ rather than the individual $R_X$ and $R_Y$. Furthermore, the total width of the Higgs boson is about 4 MeV in the SM and cannot be probed directly currently at the LHC. The signal strengths, associated with a production mechanism $X$ and a decay $Y$, can be expressed as
\be
\hat \mu(X,Y)=R_{X} \frac{R_{Y}}{R_{\rm width} }\,.
\label{eq:Rmu}
\ee

\begin{figure}
	\centering
	\includegraphics[width=4.5cm]{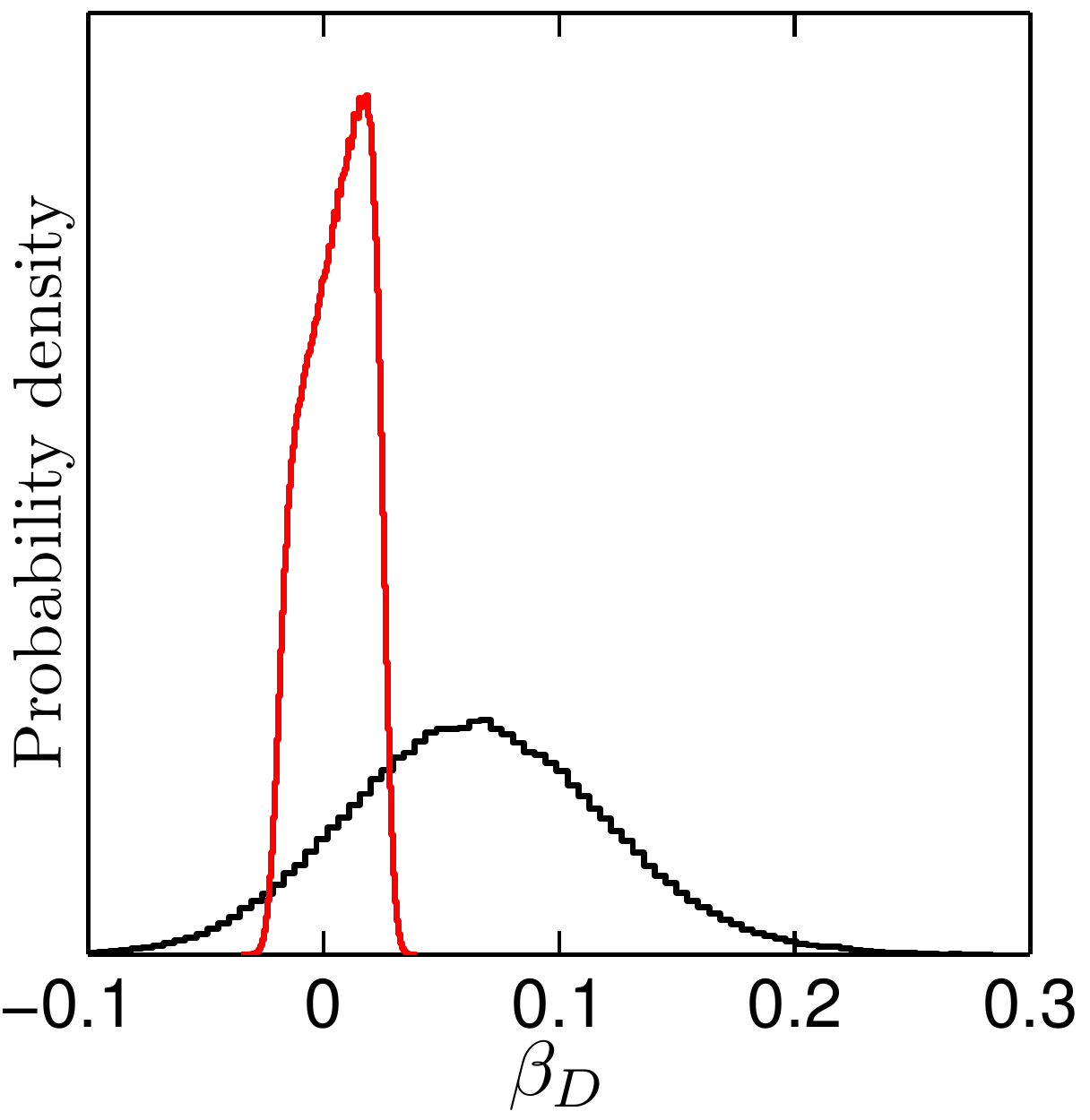}
    \includegraphics[width=4.5cm]{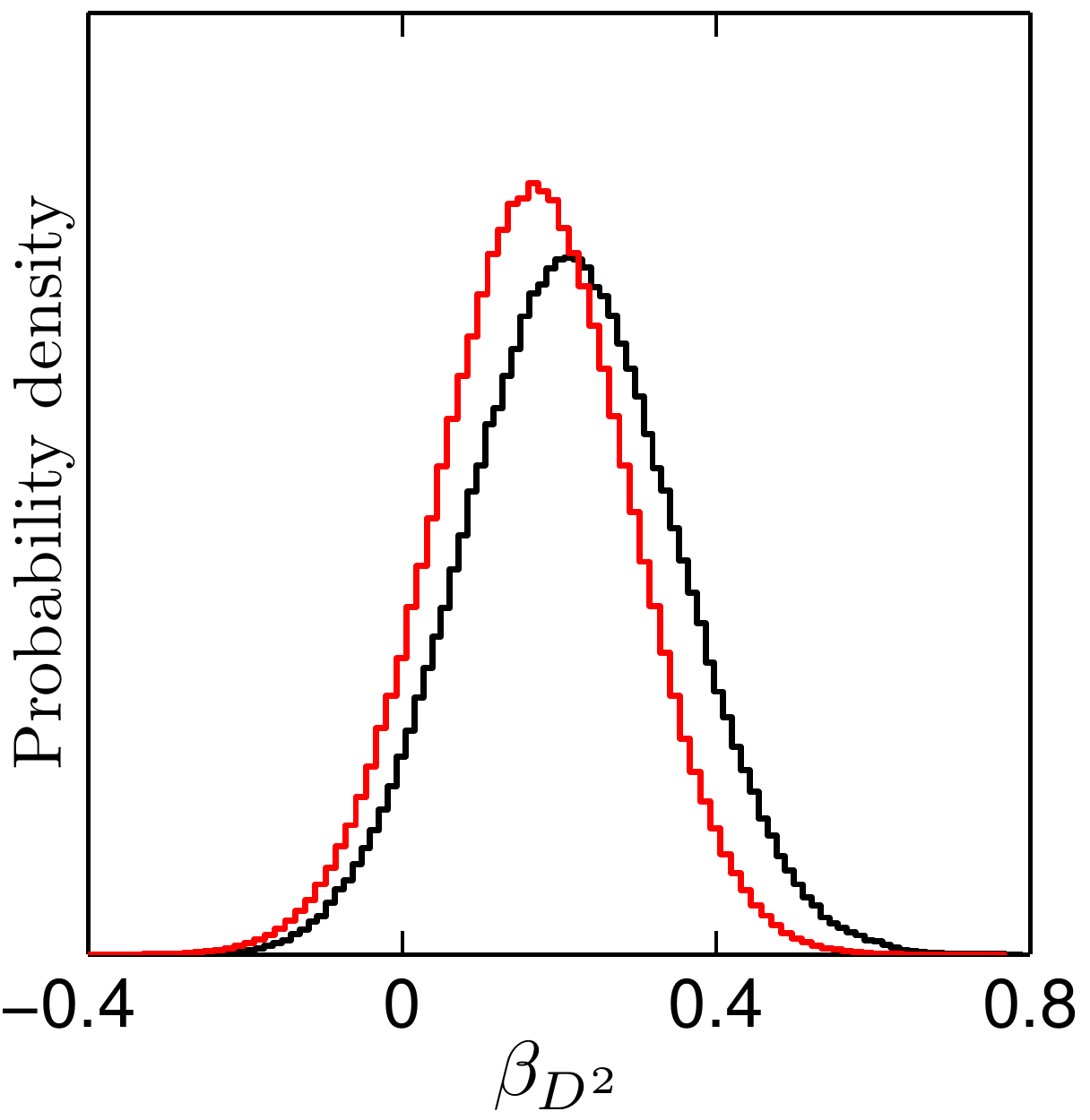}						
	\includegraphics[width=4.4cm]{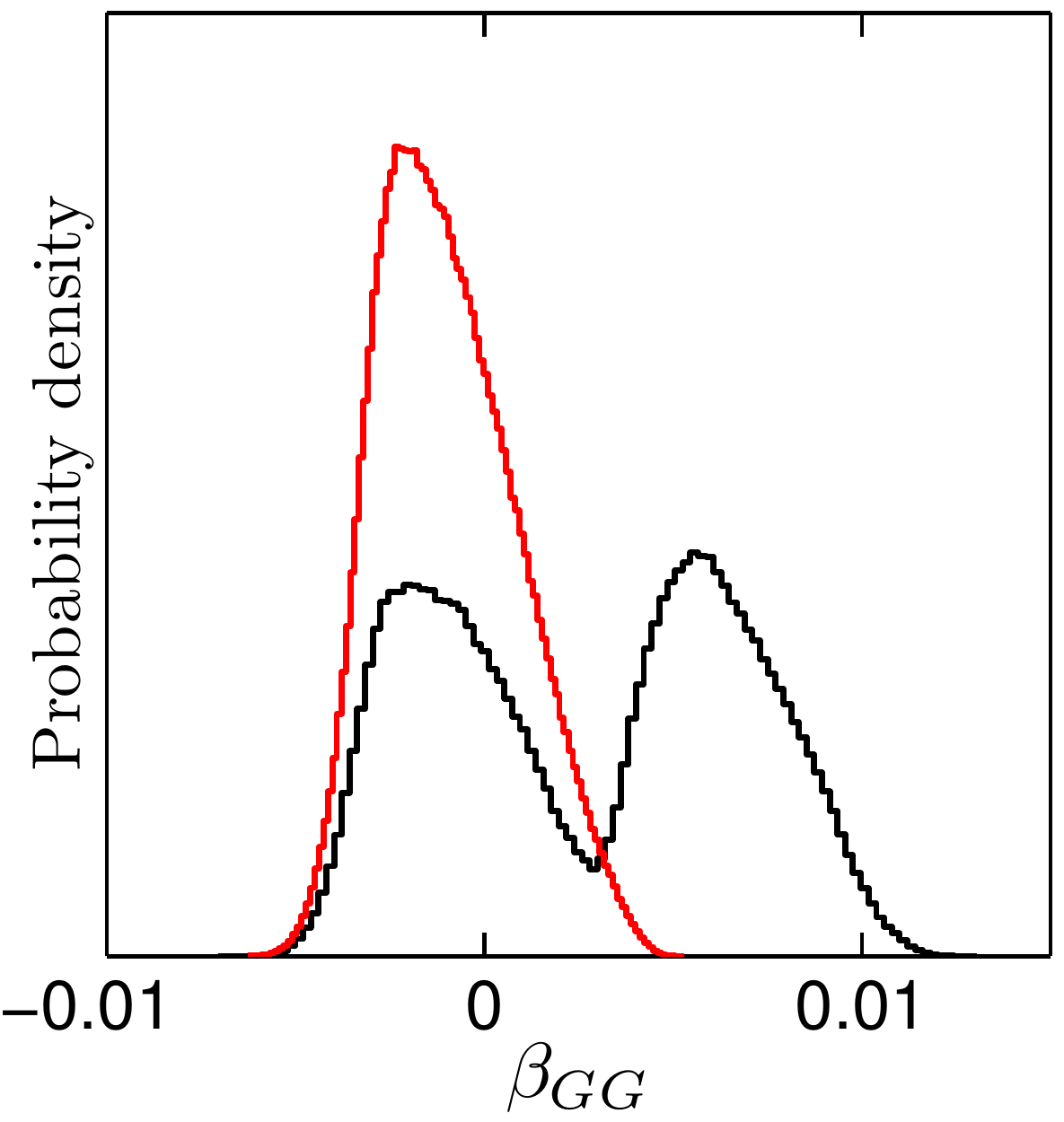}
	\includegraphics[width=4.5cm]{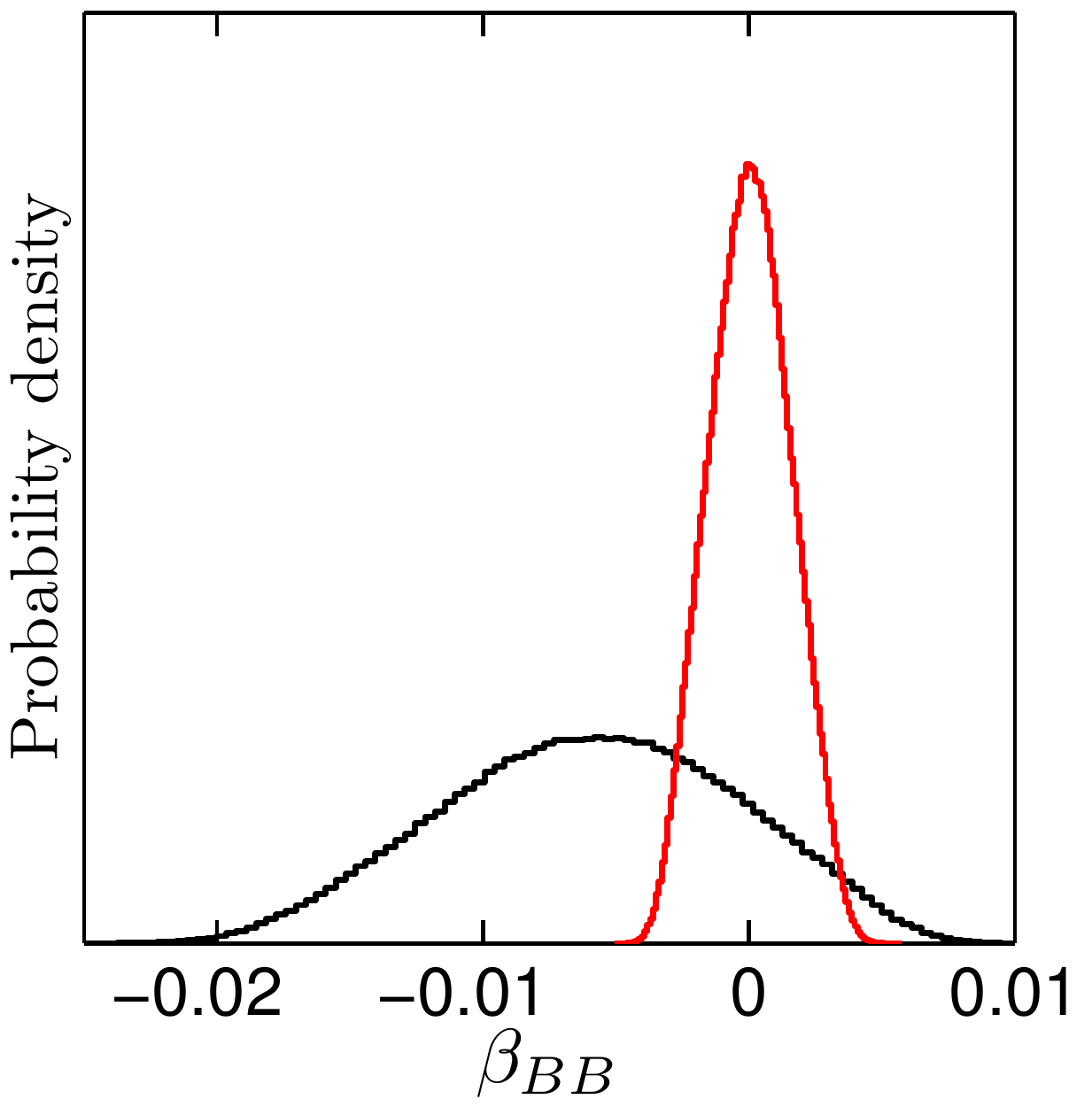}
	\includegraphics[width=4.6cm]{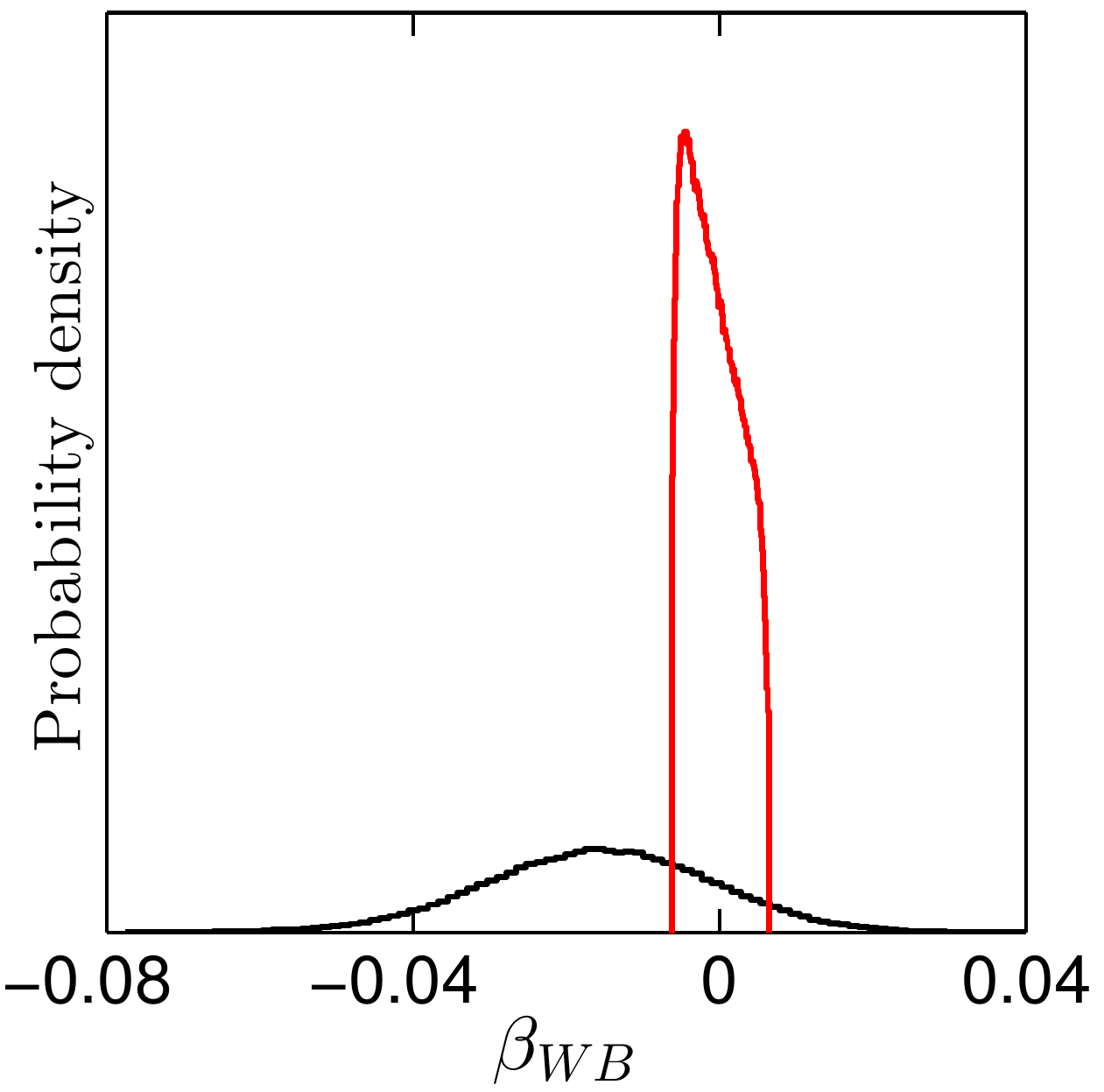}
	\includegraphics[width=4.5cm]{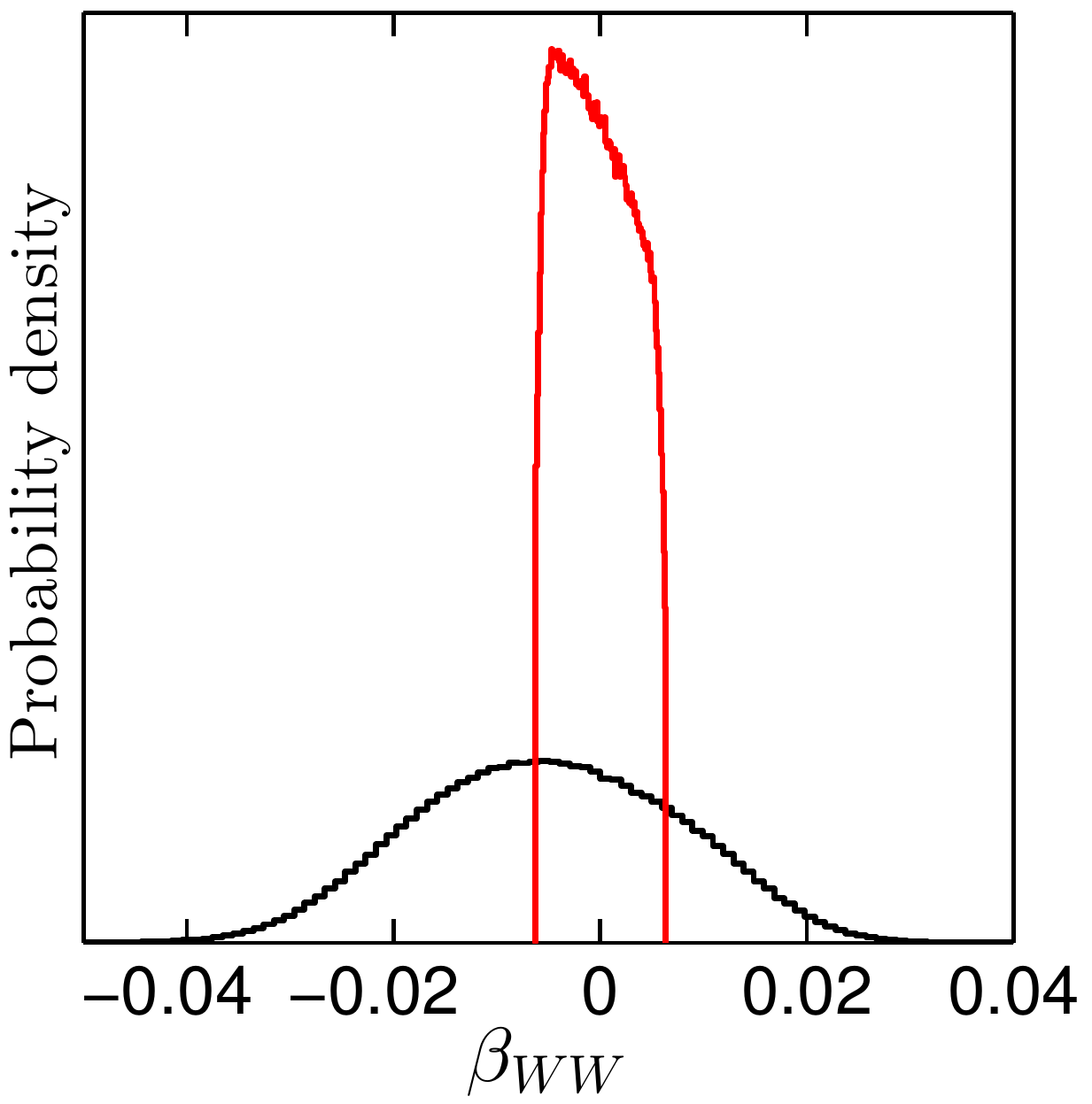}
	\includegraphics[width=4.5cm]{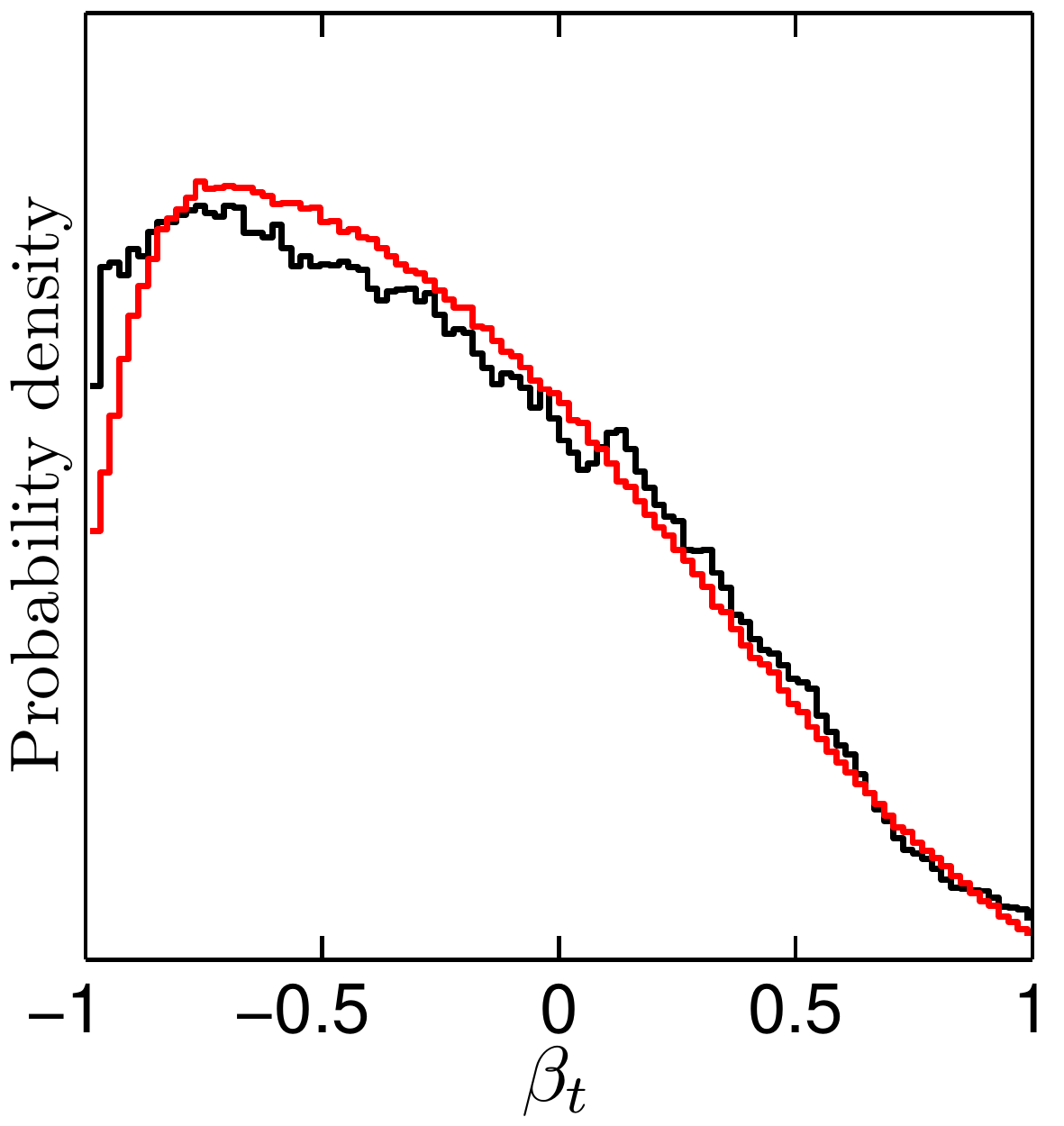}
	\includegraphics[width=4.5cm]{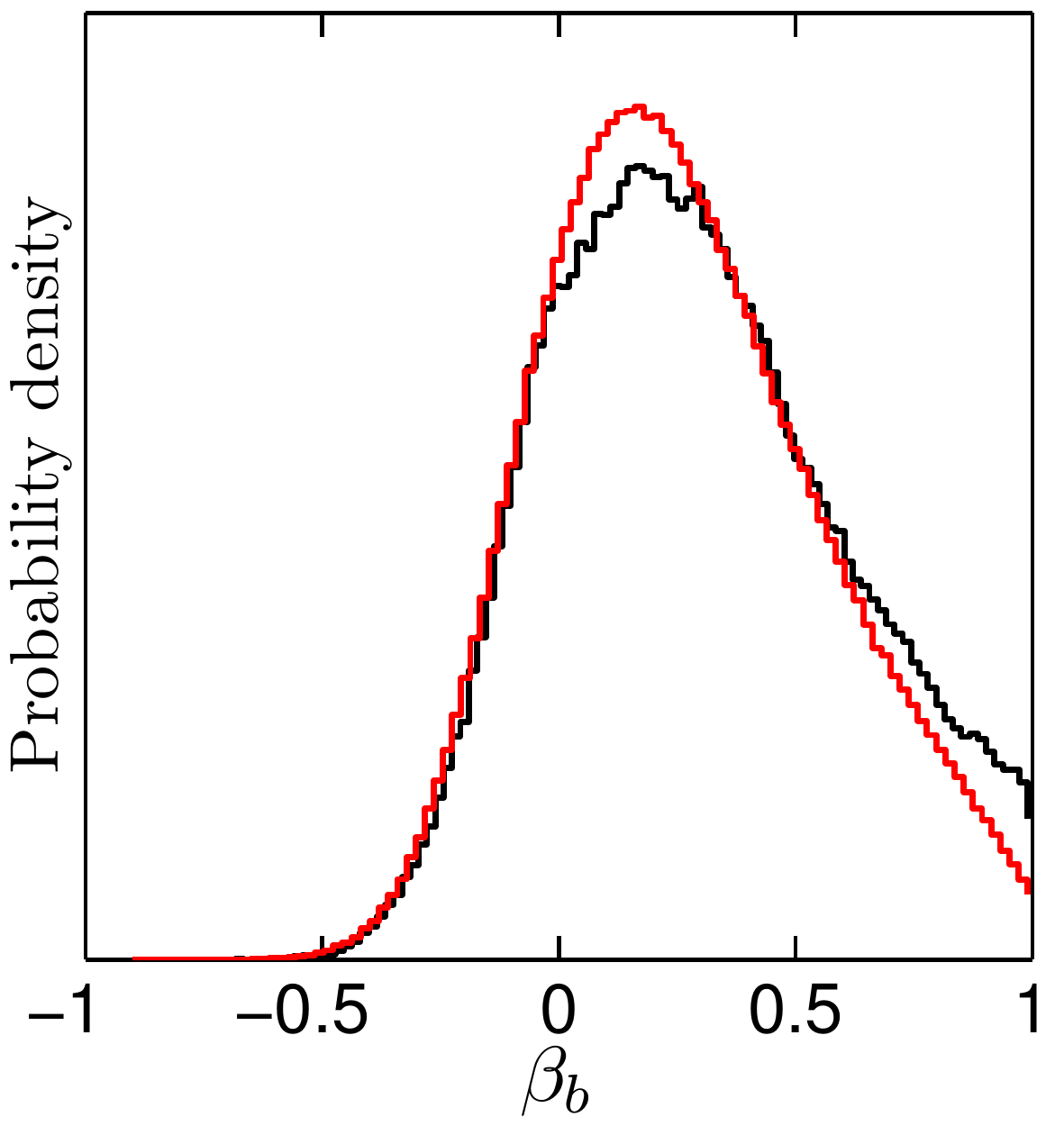}
	\includegraphics[width=4.5cm]{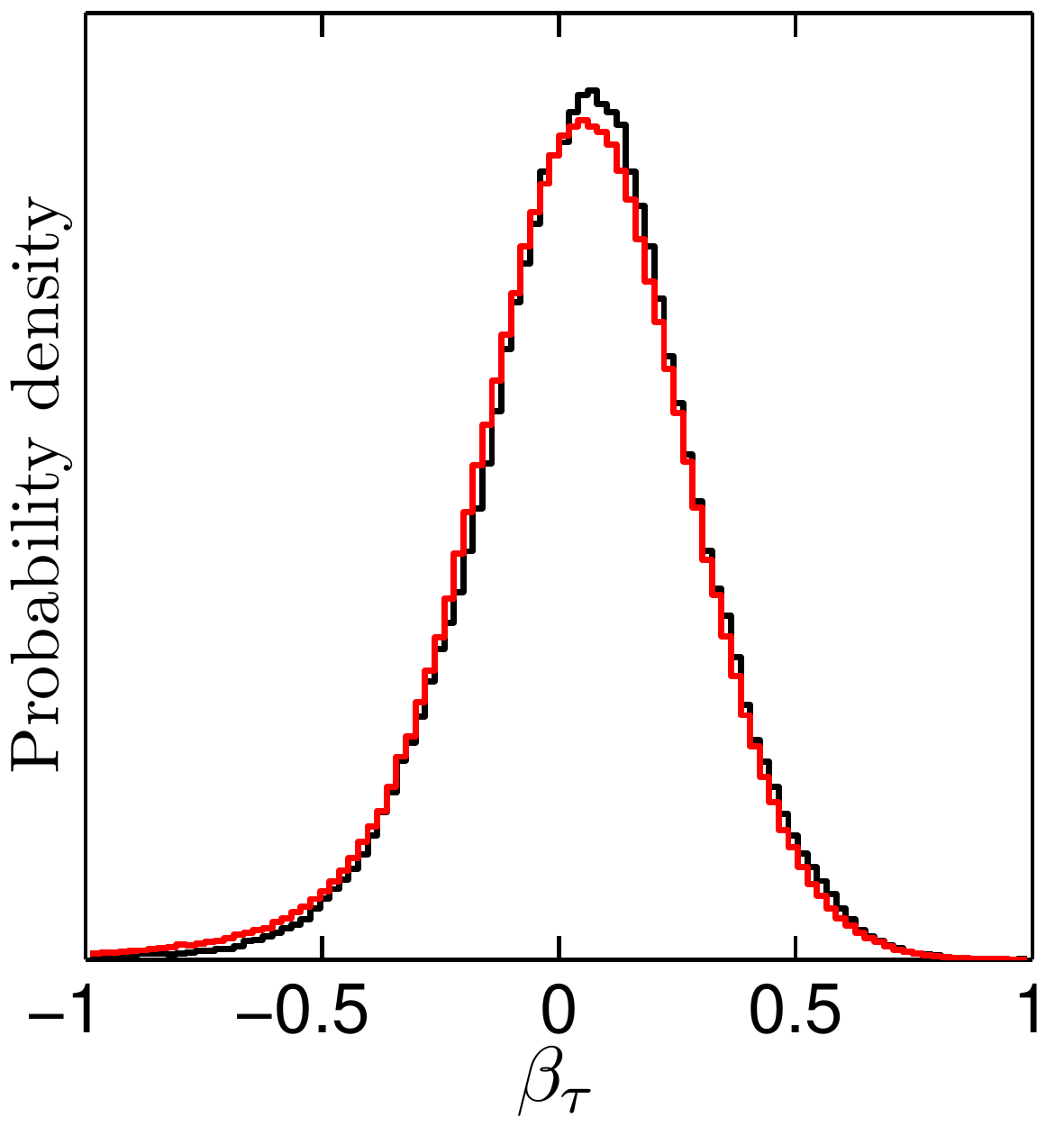}	
	\caption{Posterior PDFs of the 9 fundamental parameters, $\beta_i \equiv \alpha_i v^2/\Lambda^2$, in scenario I (black) and scenario II (red). \label{fig:betas}}
\end{figure}

\begin{figure}
	\centering
     	\includegraphics[width=5cm]{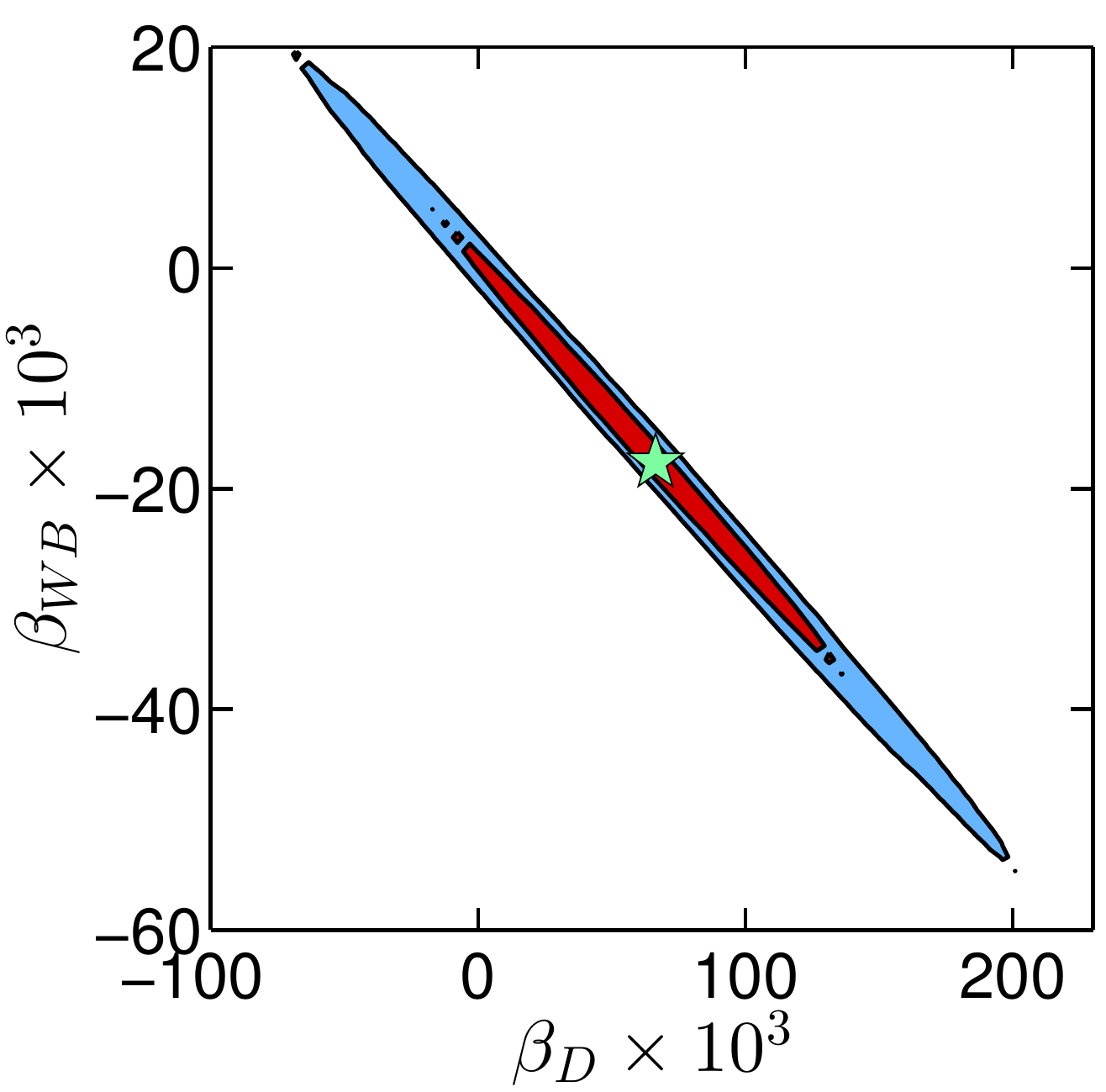}	
		\includegraphics[width=5cm]{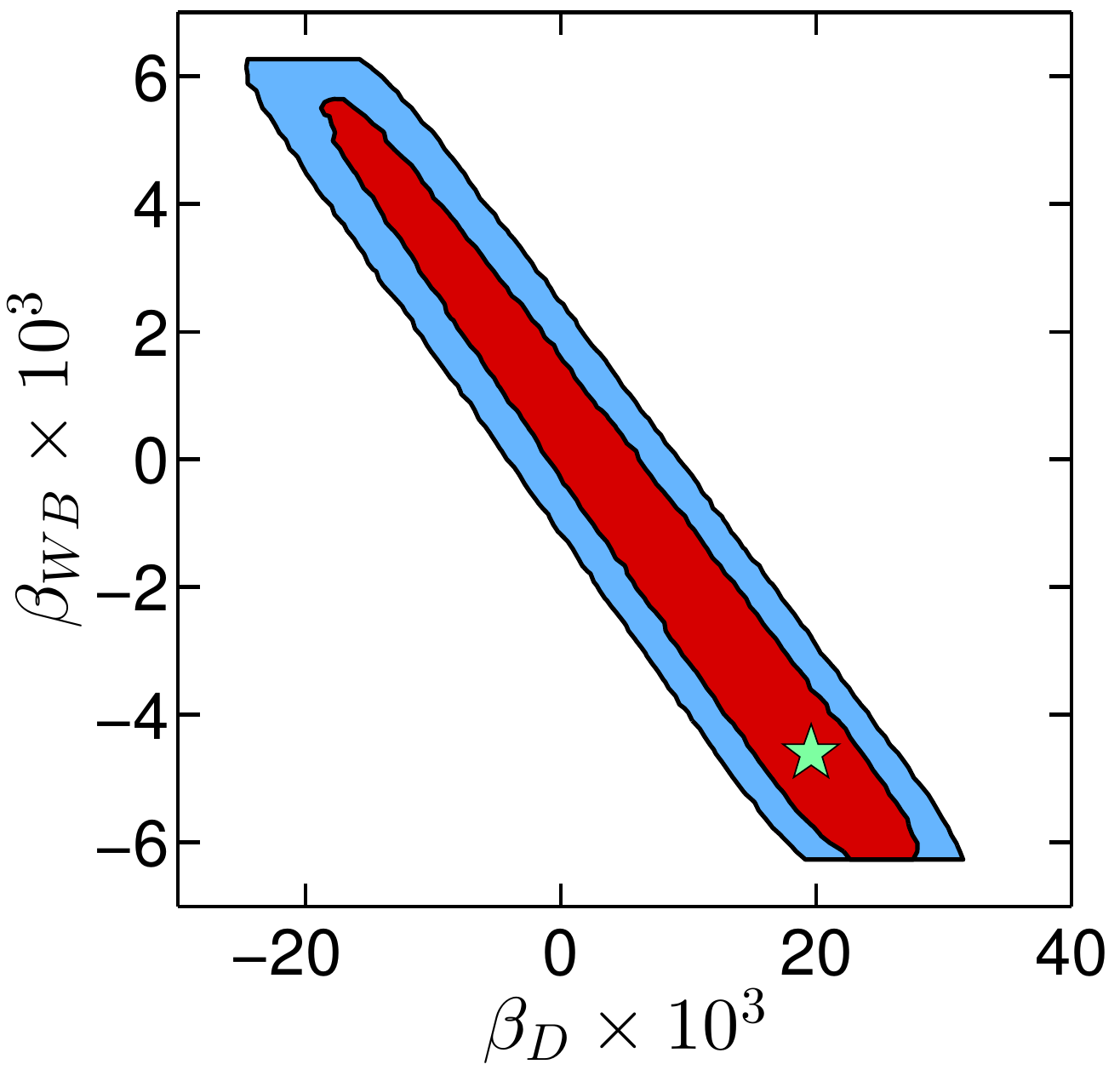}
	\caption{Posterior PDFs of $\beta_{\WB}$ versus $\beta_{D}$ in scenario I (left) and scenario II (right). The red and blue regions correspond to the 68\% and 95\% Bayesian credible regions (BCRs). The green star indicates the maximum of our posterior PDF. \label{fig:corr_D_WB}}
\end{figure}

We present one-dimensional PDFs of the fundamental parameters $\beta_i$ for both scenarios in Fig.~\ref{fig:betas}.
Moreover, in Table~\ref{tab:BCRs_7} we report the $68\%$ and $95\%$ Bayesian credible intervals (BCIs) for these quantities.  We also present the BCIs for the other, dependent quantities, \ie~the anomalous couplings $a_V$ and $c_f$, the tensorial couplings $\zeta_i$, and the various $R$'s. 

\begin{table}[h]
\center
\begin{tabular}{|c|c|c|c|c|}
\hline
& \multicolumn{2}{|c|}{scenario I} & \multicolumn{2}{|c|}{scenario II} \\
& 68\% BCI & 95\% BCI & 68\% BCI & 95\% BCI \\
\hline
$\beta_D \times 10^3$ & $[10,120]$ & $[-50,180]$ & $[-6, 23]$ & $[-19,26]$ \\
$\beta_{D^2} \times 10^3$ & $[70,350]$ & $[-50,480]$ & $[40,290]$ & $[-90,400]$ \\
$\beta_t \times 10^3$ & $[-1000,110]$ & $[-1000,610]$ & $[-930,10]$ & $[-1000,590]$ \\
$\beta_b \times 10^3$ & $[-10,530]$ & $[-220,930]$ & $[-110,500]$ & $[-280,860]$ \\
$\beta_\tau \times 10^3$ & $[-170,300]$ & $[-420,510]$ & $[-190,270]$ & $[-450,510]$ \\
$\beta_{GG} \times 10^3$ & $[-3.2,8.0]$ & $[-4.0,9.6]$ & $[-3.3,0.6]$ & $[-4.2,2.7]$ \\
$\beta_{\WW} \times 10^3$ & $[-19,7]$ & $[-30,18]$ & $[-5.6,2.3]$ & $[-6.0,5.6]$ \\
$\beta_{\WB} \times 10^3$ & $[-32,1]$ & $[-49,13]$ & $[-6.0,1.6]$ & $[-6.3,5.3]$ \\
$\beta_{\BB} \times 10^3$ & $[-12,0]$ & $[-17,4]$ & $[-1.7,1.6]$ & $[-2.9,3.0]$ \\
\hline
$a_{V}$ & $[1.02, 1.15]$ & $[0.96,1.21]$ & $[1.02,1.14]$ & $[0.96,1.20]$ \\
$c_t$ & $[0.05,1.14]$ & $[0.03, 1.63]$ & $[0.06,1.01]$ & $[0.04,1.60]$ \\
$c_b$ & $[0.90,1.54]$ & $[0.79,1.96]$ & $[0.89,1.50]$ & $[0.72,1.86]$ \\
$c_\tau$ & $[0.84,1.31]$ & $[0.58, 1.53]$ & $[0.81,1.27]$ & $[0.55,1.51]$ \\
\hline

$\zeta_g \, v \times 10^3$ & $[-3.2,8.0]$ & $[-4.0,9.6]$ & $[-3.3,0.6]$ & $[-4.2,2.7]$ \\
$\zeta_\gamma\, v \times 10^3$ & $[-5.5,0.5]$ & $[-6.1,0.9]$ & $[-0.33,0.46]$ & $[-0.69,0.86]$ \\
$\zeta_{Z\gamma}\, v \times 10^3$ & $[-13,18]$ & $[-18,30]$ & $[-4.9,4.4]$ & $[-7.6,7.9]$ \\
$\zeta_Z\, v \times 10^3$ & $[-20,2]$ & $[-31,11]$ & $[-3.4,2.3]$ & $[-5.1,4.4]$ \\
$\zeta_W\, v \times 10^3$ & $[-39,15]$ & $[-59,37]$ & $[-11,5]$ & $[-12,11]$ \\
\hline
$R_{\rm ggF}$ & $[0.6,1.3]$ & $[0.5,2.0]$ & $[0.6,1.3]$ & $[0.4,2.0]$ \\
$R_{\rm VBF}$ & $[1.0,1.4]$ & $[0.9,1.6]$ & $[1.0,1.3]$ & $[0.9,1.4]$ \\
$R_{\rm WH}$ & $[0.7,1.3]$ & $[0.5,1.7]$ & $[1.0,1.3]$ & $[0.9,1.4]$ \\
$R_{\rm ZH}$ & $[0.7,1.2]$ & $[0.5,1.5]$ & $[1.0,1.3]$ & $[0.9,1.4]$ \\
$R_{\rm ttH}$ & $[0.02,1.0]$ & $[0.02,2.6]$ & $[0,0.9]$ & $[0,2.5]$ \\
\hline
$R_{\gamma\gamma}$ & $[1.1,1.9]$ & $[0.8,2.5]$ & $[1.1,1.8]$ & $[0.8,2.3]$ \\
$R_{Z\gamma}$ & $[0,5.2]$ & $[0,12.0]$ & $[0,2.2]$ & $[0,4.3]$ \\
$R_{ZZ}$ & $[1.0,1.3]$ & $[0.9,1.5]$ & $[1.0,1.3]$ & $[0.9,1.4]$ \\
$R_{\WW}$ & $[1.0,1.3]$ & $[0.9,1.5]$ & $[1.0,1.3]$ & $[0.9,1.4]$ \\
$R_{b\bar{b}}$ & $[0.7,2.2]$ & $[0.5,3.6]$ & $[0.7,2.1]$ & $[0.4,3.3]$ \\
$R_{\tau\tau}$ & $[0.6,1.6]$ & $[0.3,2.2]$ & $[0.6,1.5]$ & $[0.2,2.1]$ \\
\hline
$R_{\rm width}$ & $[0.8, 1.9]$ & $[0.7,2.7]$ & $[0.8,1.8]$ & $[0.6,2.5]$ \\
\hline
\end{tabular}
\caption{$68\%$ and $95\%$ Bayesian credible intervals (BCIs) for the democratic HDOs case (scenario I) and for the loop-suppressed $\mathcal{O}_{FF}$'s case (scenario II).}
\label{tab:BCRs_7}
\end{table}

One can first remark that all of our HDO coefficients except $\beta_t$ and $\beta_b$ are constrained enough to stay within the bound $|\beta_i| <1$, as required for the convergence of the HDO expansion. Furthermore, the $\beta_{FF} \equiv \beta_{\WW,\,WB,\,BB,\,GG}$ coefficients are ${\cal O}(0.01)$ in both scenarios.
$\beta_D$ and $\beta_{\WB}$ are strongly correlated in both scenarios as they appear in the $S$ parameter at tree-level, see Eq.~\eqref{S_tree} (we recall that we fix $\alpha'_{D} = \alpha'_{D^2} = 0$ in order to preserve custodial symmetry). We thus have $2\,c_w \, \beta_{\WB}\approx-s_w\, \beta_{D}$ as can be seen in Fig.~\ref{fig:corr_D_WB}.
The TGV observables also involve $\beta_{D}$ and $\beta_{\WB}$ (see Eq.~\eqref{eq:tgv_hdo}), and thus provide an independent constraint on $\beta_{D}$ (or equivalently $\beta_{\WB}$). The slight deficit in $\kappa_{\gamma}$ and $g_1^Z$ as measured by LEP, see Eq.~\eqref{eq:lep_tgv}, tend to favors positive (negative) $\beta_{\WB}$ ($\beta_D$). Finally, note that in scenario II the PDF of $\beta_{\WB}$ is limited to the $[-1/16\pi^2,1/16\pi^2]$ range since we consider that the operator ${\cal O}_{\WB}$ is loop-suppressed. This in turn fixes the allowed range for $\beta_{D}$.

The $\beta_{D^2}$ coefficient is allowed to deviate significantly from 0 as it only appears in loop contributions to $S$ and $T$ and in $a_V$. The probability of having $\beta_{D^2} > 0$ is 94\% (90\%)  in scenario I (II) and comes from $T$, as well as VBF and VH production modes and $h\to VV$ decays. A value for $a_V > 1$ leads to a positive contribution to $T$, as well as an enhancement of the VBF and VH production processes, the $h \rightarrow VV^*$ decays, and also to the loop-induced decay rates, $h \rightarrow \gamma\gamma$ and $h \rightarrow Z\gamma$.

$\beta_{\WW}$ and $\beta_{\BB}$ are mainly constrained by the searches for $h \rightarrow \gamma\gamma$ and $h \rightarrow Z\gamma$  as they contribute to the tensorial couplings $\zeta_\gamma$ and $\zeta_{Z\gamma}$, see Eq.~\eqref{lambda_gamma_tens} and \eqref{lambda_Zgamma_tens}. Given the large allowed range for $\beta_{\WB}$, a cancellation has to occur with $\beta_{\WW}$ and $\beta_{\BB}$ in order to achieve a $h \rightarrow \gamma\gamma$ rate compatible with experiment.
As a result, negative values of $\beta_{\WW}$ and $\beta_{\BB}$ are favored. 
Moreover, $\beta_{\WW}$ is also constrained from VH production processes via the quantities $\zeta_V$. In contrast $\beta_{\BB}$ and $\beta_{\WB}$ play no role for these measurements, as they are more strongly constrained by the other effects mentioned above. 
The contributions of $\beta_{\WW}$ and $\beta_{\BB}$ to $S$ are up to $\mathcal O(0.03)$ and do not impact the PDFs. This effect is even smaller in scenario II, and is also smaller if we take $\Lambda < 4 \pi v$ due to the $\log(m_h/\Lambda)$ factor in Eq.~(\ref{eq:DeltaS}). 
We note that in scenario II the PDFs for $\beta_{\WW}$ and $\beta_{\WB}$ can easily reach the bounds set by the priors, while $\beta_{\BB}$ is more strongly constrained by the data. This is due to the fact that $\beta_{\BB}$ enters in $\zeta_\gamma$ with a coefficient roughly four times larger than the other two.

\begin{figure}
	\centering
		\includegraphics[width=4.5cm]{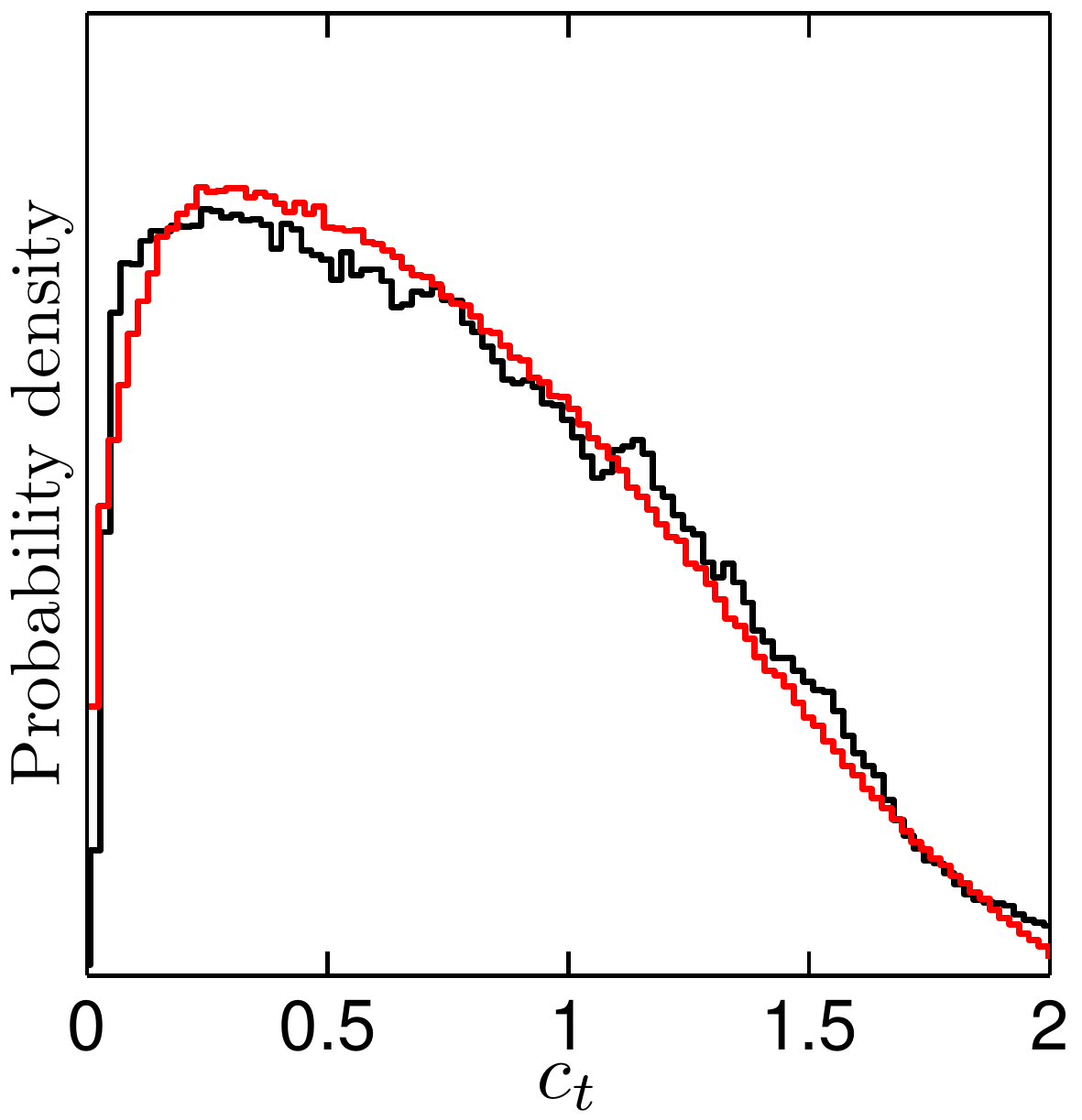}		
		\includegraphics[width=4.4cm]{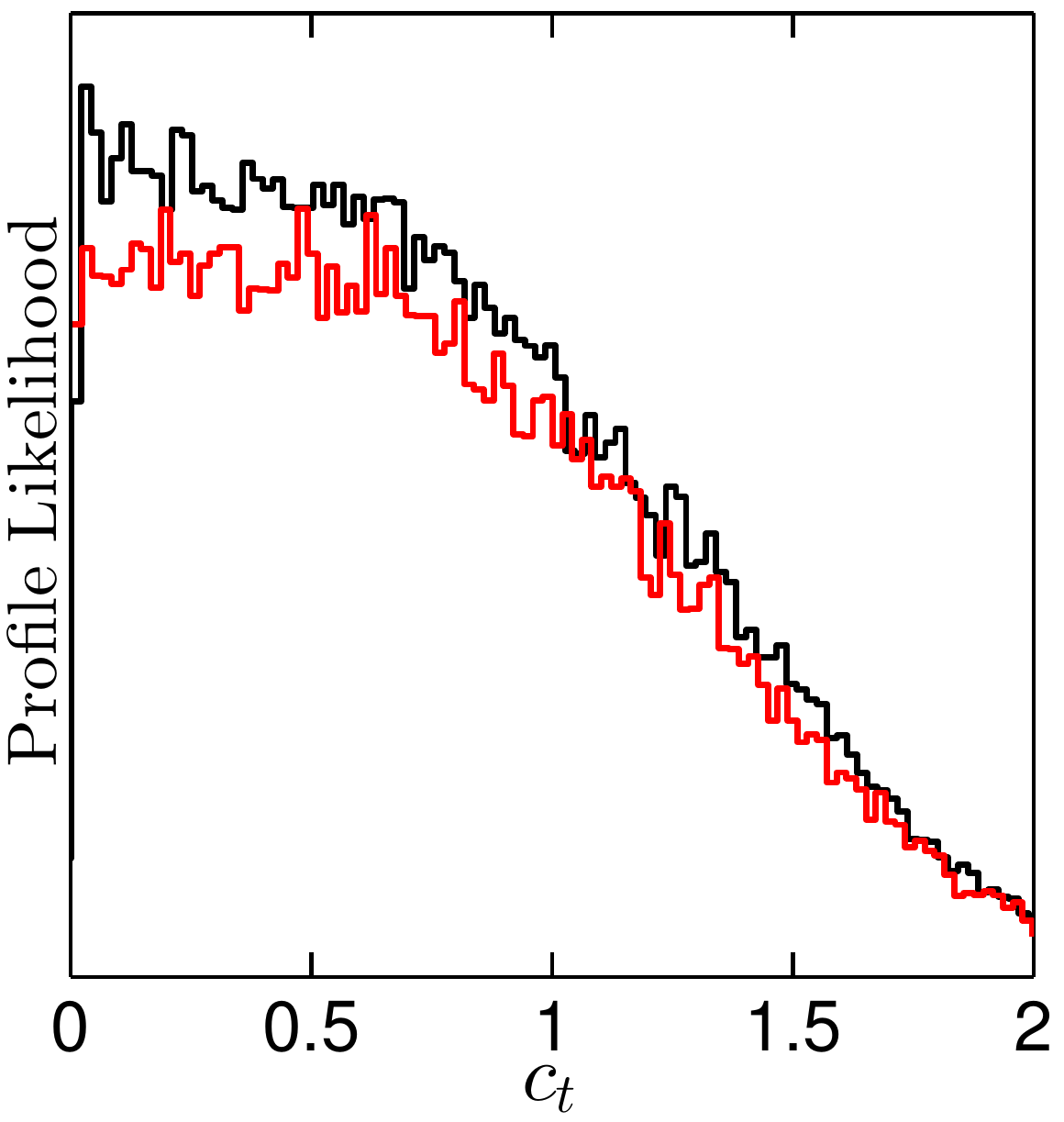}
	\caption{On the left, posterior PDF of $c_t$ in scenario I (black) and scenario II (red). On the right, profile likelihood along the $c_t$ axis in scenario I and scenario II (same color code). \label{ct_profile}}
\end{figure}

Finally, the Yukawa corrections parametrized by $\beta_f$ ($f = t,b,\tau$) are much less constrained as they only contribute to the rescaling factors $c_f$, but account for most of the deviations of $c_f$ from 1, such that we often have $|\beta_f| \gg |\beta_D/4|$ and thus $c_f \approx 1 + \beta_f$.
It is worth noting that $\beta_t$ has a fairly large probability of being close to $-1$, which leads to small or vanishing $c_t$. The posterior PDF of $c_t$ is shown on the left pannel of Fig.~\ref{ct_profile}.

In such case, one may wonder whether or not the preference for small $c_t$ is due to a volume effect caused by the process of marginalization. To this end, we display on the right panel of Fig.~\ref{ct_profile} the profile likelihood for the parameter $c_t$, \ie~the likelihood for given $c_t$, maximized over all the other parameters. We conclude that in both scenarios, the preference for small $c_t$ originates from the likelihood and not from a volume effect.~\footnote{We have also checked that no volume effects appear for any of the other posterior PDFs either.}

\begin{figure}
	\centering
	\includegraphics[width=5.1cm]{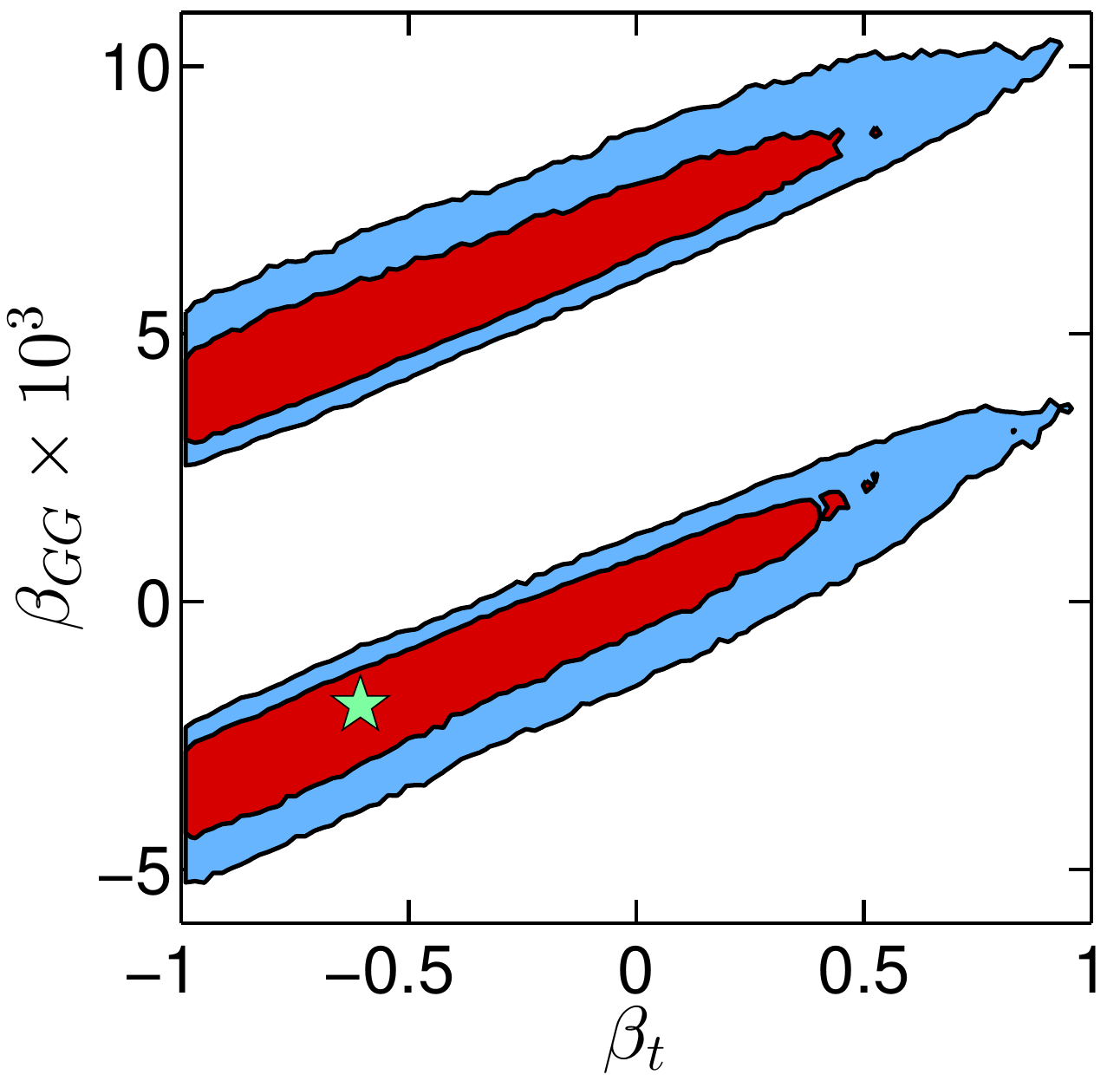}	
	\includegraphics[width=5.1cm]{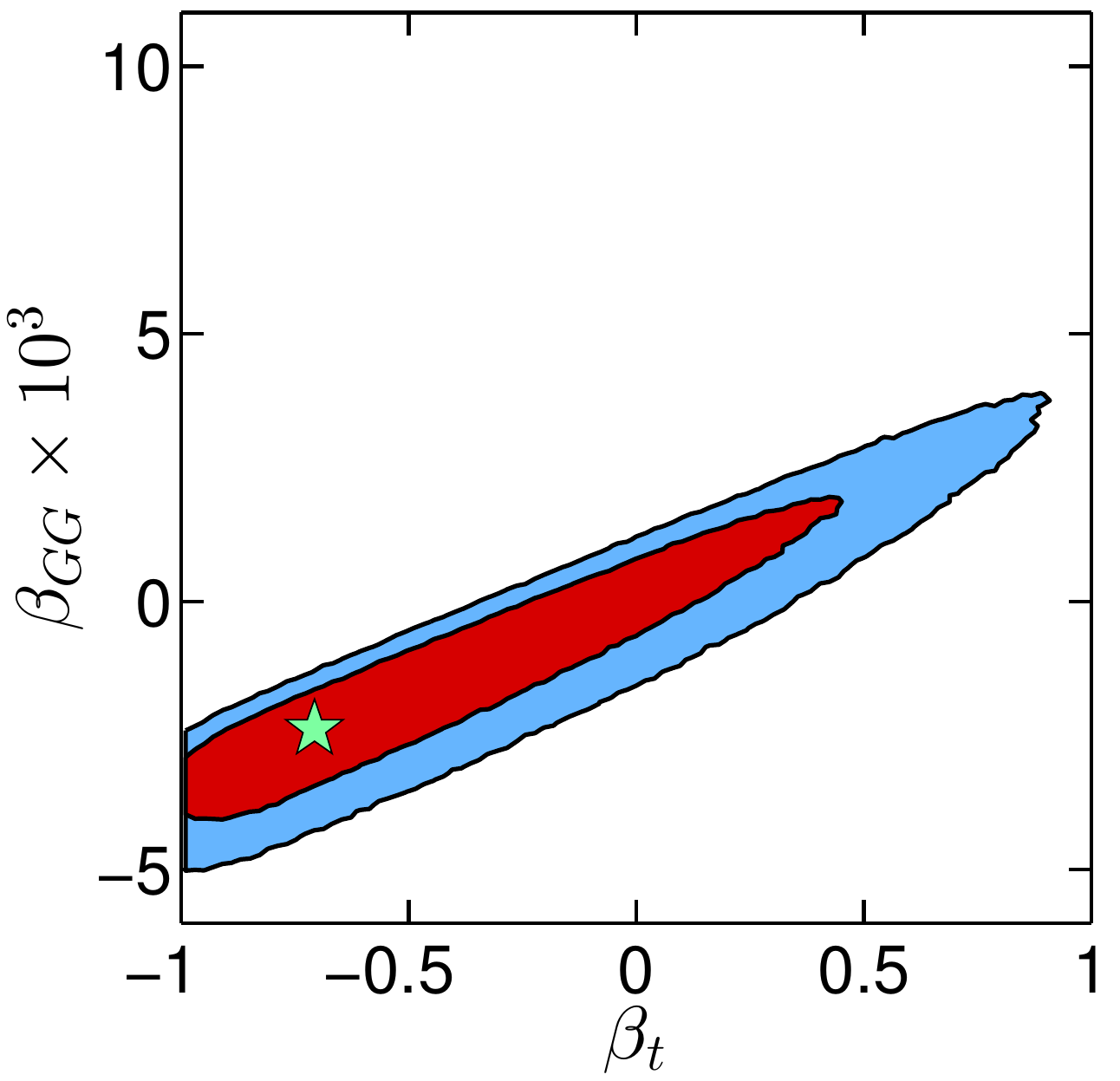} 							
	\caption{Posterior PDF of $\beta_{GG}$ versus $\beta_t$ in scenario I (left) and scenario II (right). Color code as in
Fig.~\ref{fig:corr_D_WB}.
  \label{fig:beta_t_beta_GG}}
\end{figure}

\begin{figure}
	\centering
     	\includegraphics[width=4.45cm]{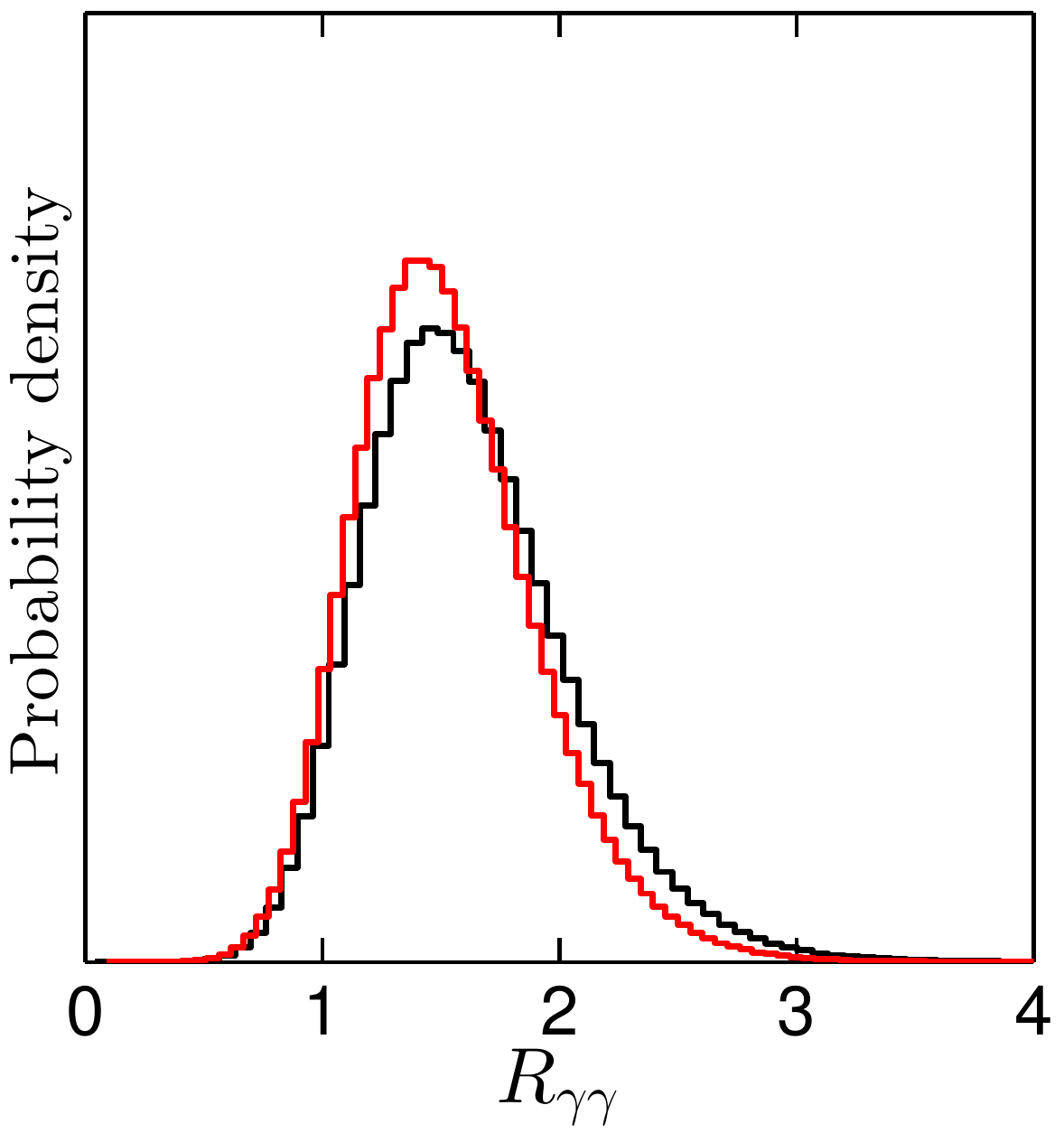}
     	\includegraphics[width=4.9cm]{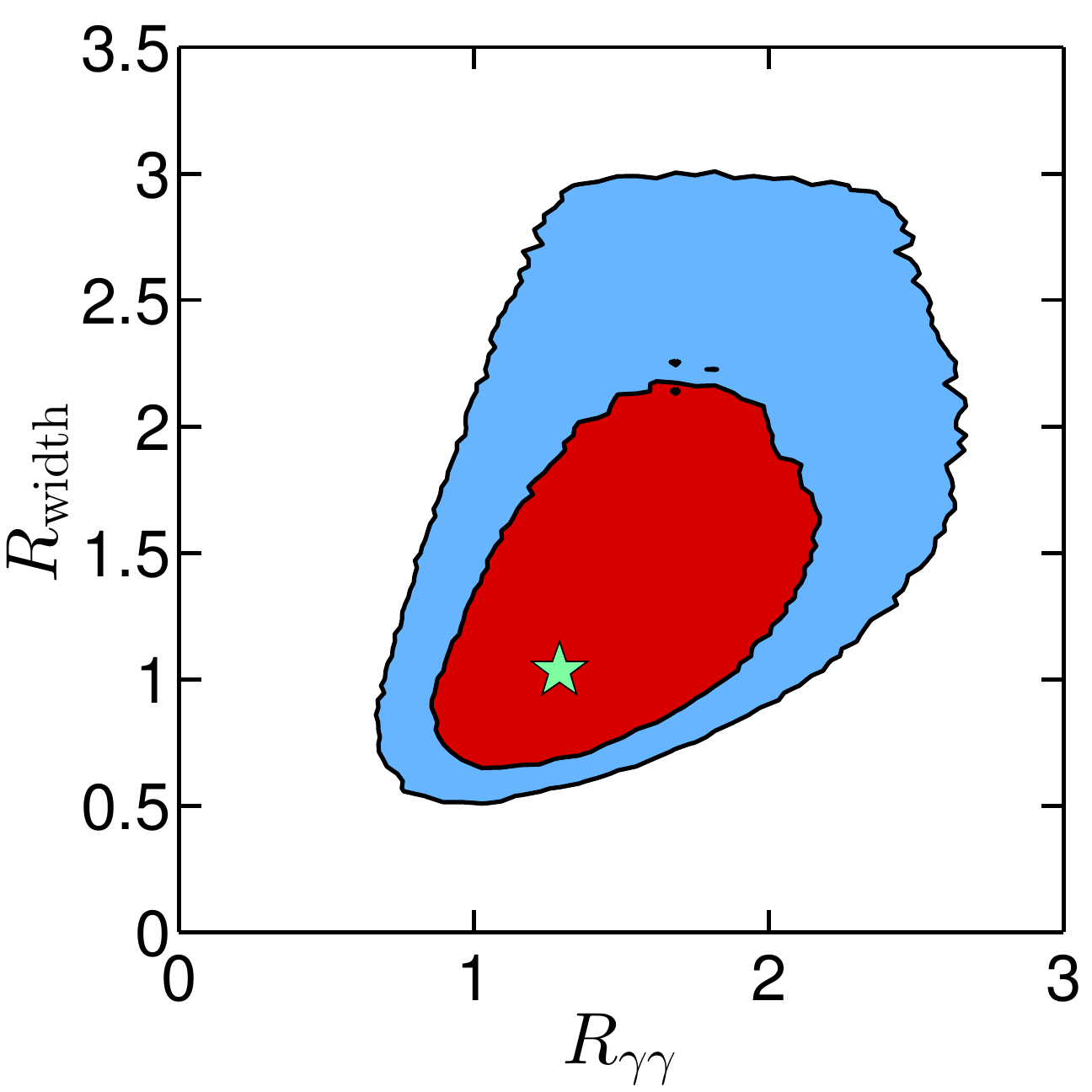}
     	\includegraphics[width=4.9cm]{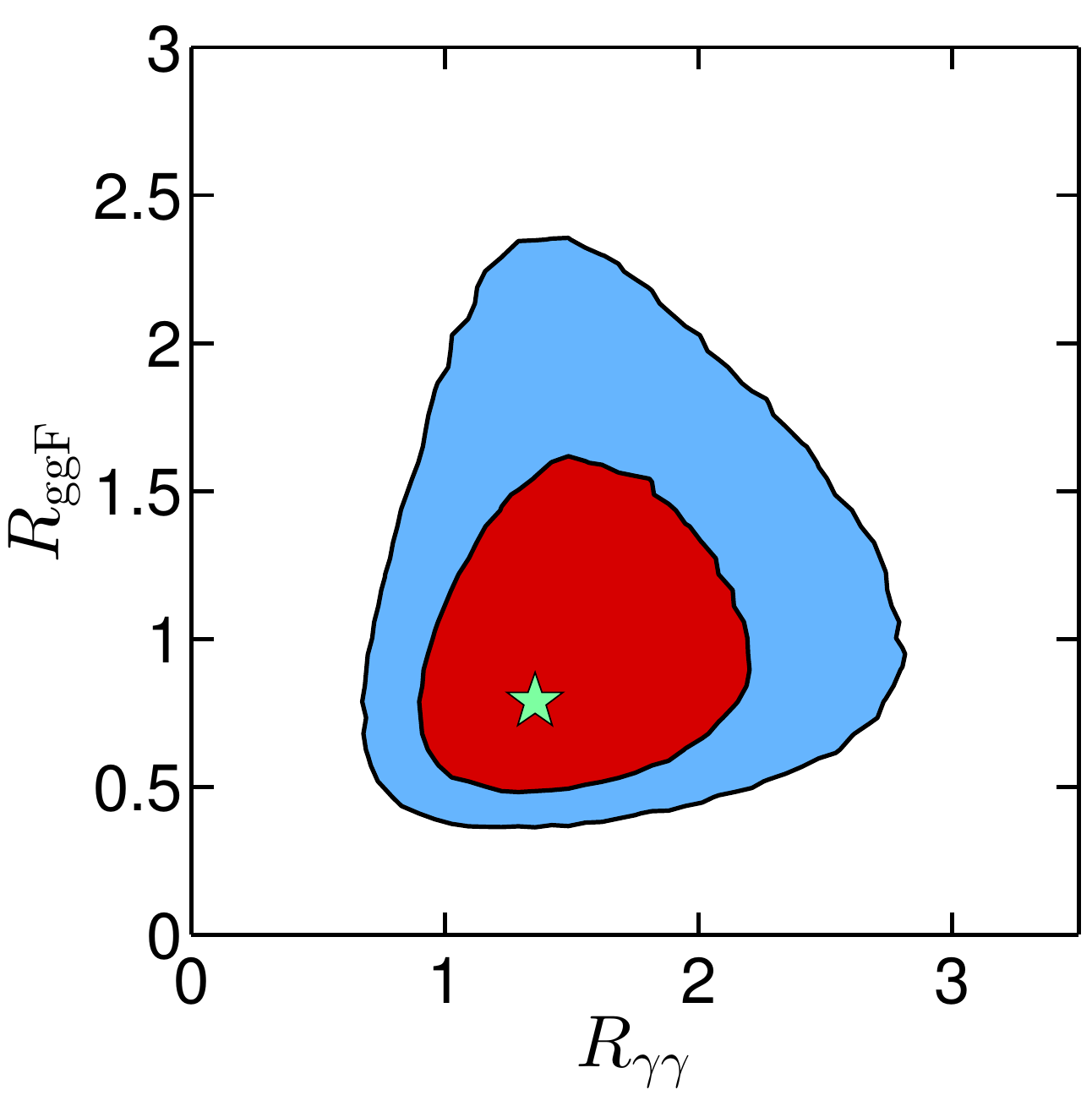}
	\caption{On the left, posterior PDF of $R_{\gamma\gamma}$ in scenario I (black) and scenario II (red). Also shown are the 2D posterior PDFs of $R_{\rm width}$ versus $R_{\gamma\gamma}$ (middle) and $R_{\rm ggF}$ versus $R_{\gamma\gamma}$ (right) in scenario I. Color code as in the previous figure. \label{fig:R_2gamma}}
\end{figure}

The shapes of the PDF and profile likelihood for $c_t$ in Fig.~\ref{ct_profile} are in fact a direct consequence of the signal strength measurement $\mu({\rm ttH},b\bar b)$ by CMS \cite{CMS-PAS-HIG-12-025}, see Table~\ref{CMSresults}. Notice that the latter is so far the only analysis sensitive to the ttH production mode.
In spite of its large error, the low central value drives $c_t$ efficiently to small values because of the relation $R_{\rm ttH}=c_t^2$. Although small $c_t$ decreases (increases) the value of $R_{\rm ggF}$  ($R_{\rm \gamma\gamma}$), these changes can be compensated for  without decreasing the likelihood.
In the case $c_t \approx 0$, the gluon-gluon fusion (ggF) process is mainly driven by the tensorial coupling $\zeta_g \equiv \beta_{GG}/v$. We show in Fig. \ref{fig:beta_t_beta_GG} the correlation between $\beta_{GG}$ and $\beta_t$, which is needed to reproduce the observed ggF rate.
For the decay $h\to\gamma\gamma$, we observe an increased rate $R_{\gamma\gamma} > 1$, which can be seen in Fig.~\ref{fig:R_2gamma}. Indeed, in the SM the $h \rightarrow \gamma\gamma$ process is dominated by the $W$ loops, and there is a destructive interference between the $t$ and $W$ contributions. Therefore, the suppression of $c_t$ helps increasing $R_{\gamma\gamma}$. To better understand this enhanced rate, notice that naively combining the data in Table~\ref{ATLASresults} -- \ref{Tevatronresults} one obtains $\mu({\rm ggF+ttH},\gamma \gamma)=1.05\pm 0.28$ and $\mu({\rm VBF+VH},\gamma\gamma)=1.8\pm 0.6$. It turns out that these different values are then realized with a slighly reduced $R_{\rm ggF}$ and an increased $R_{\gamma\gamma}$.


\begin{figure}
	\centering
	\includegraphics[width=4.5cm]{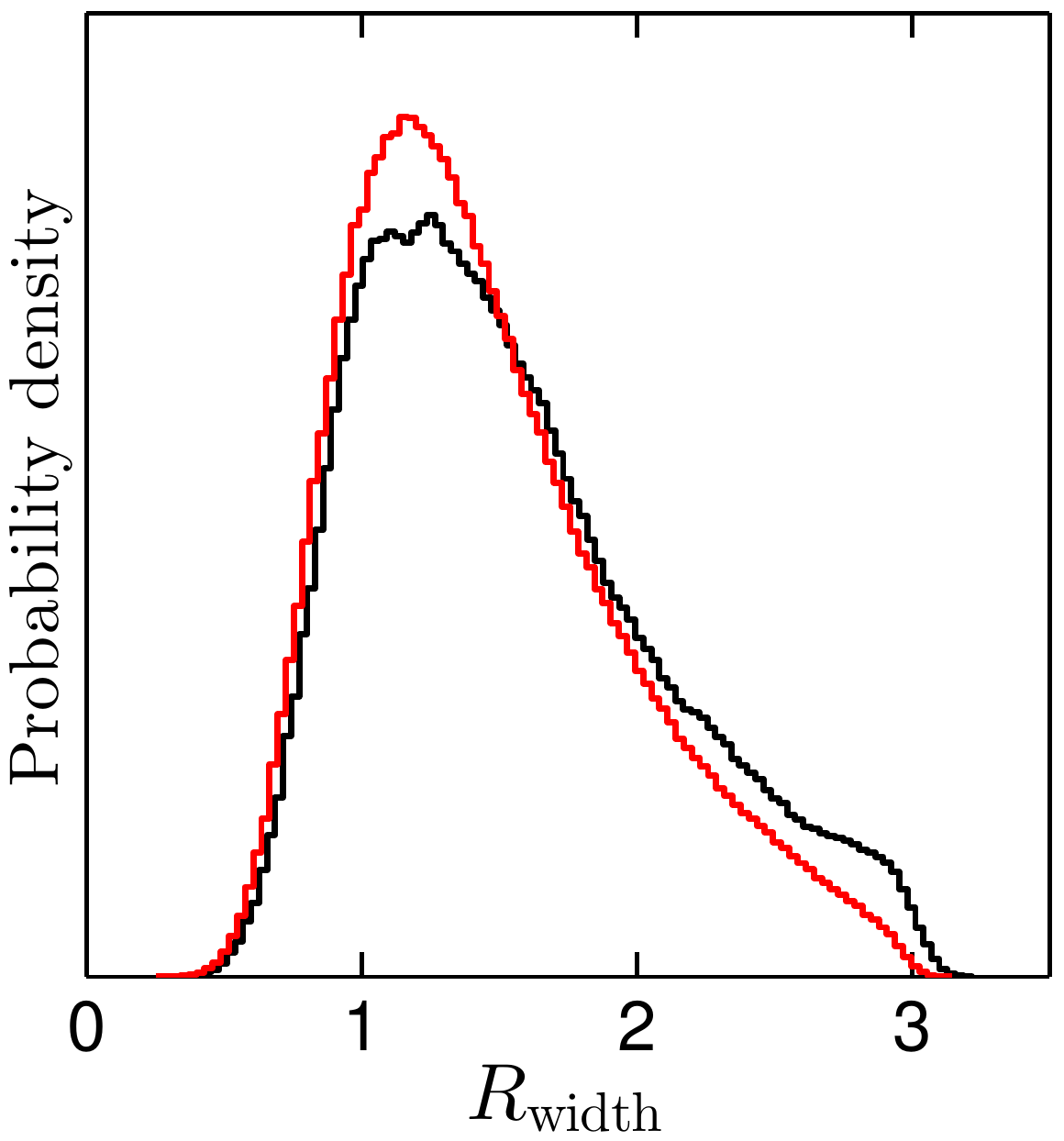}	
	\includegraphics[width=5.1cm]{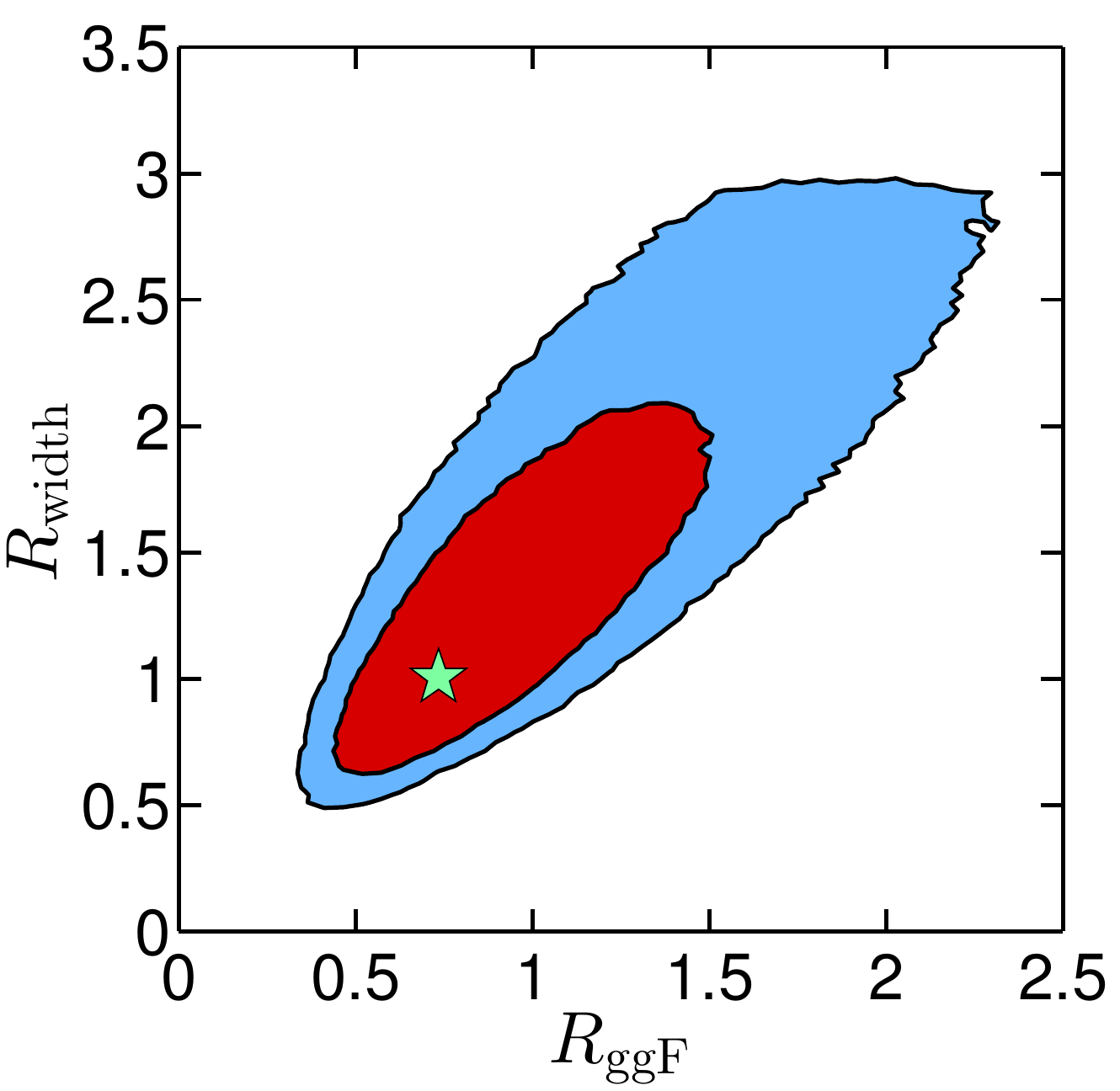}
	\includegraphics[width=5.1cm]{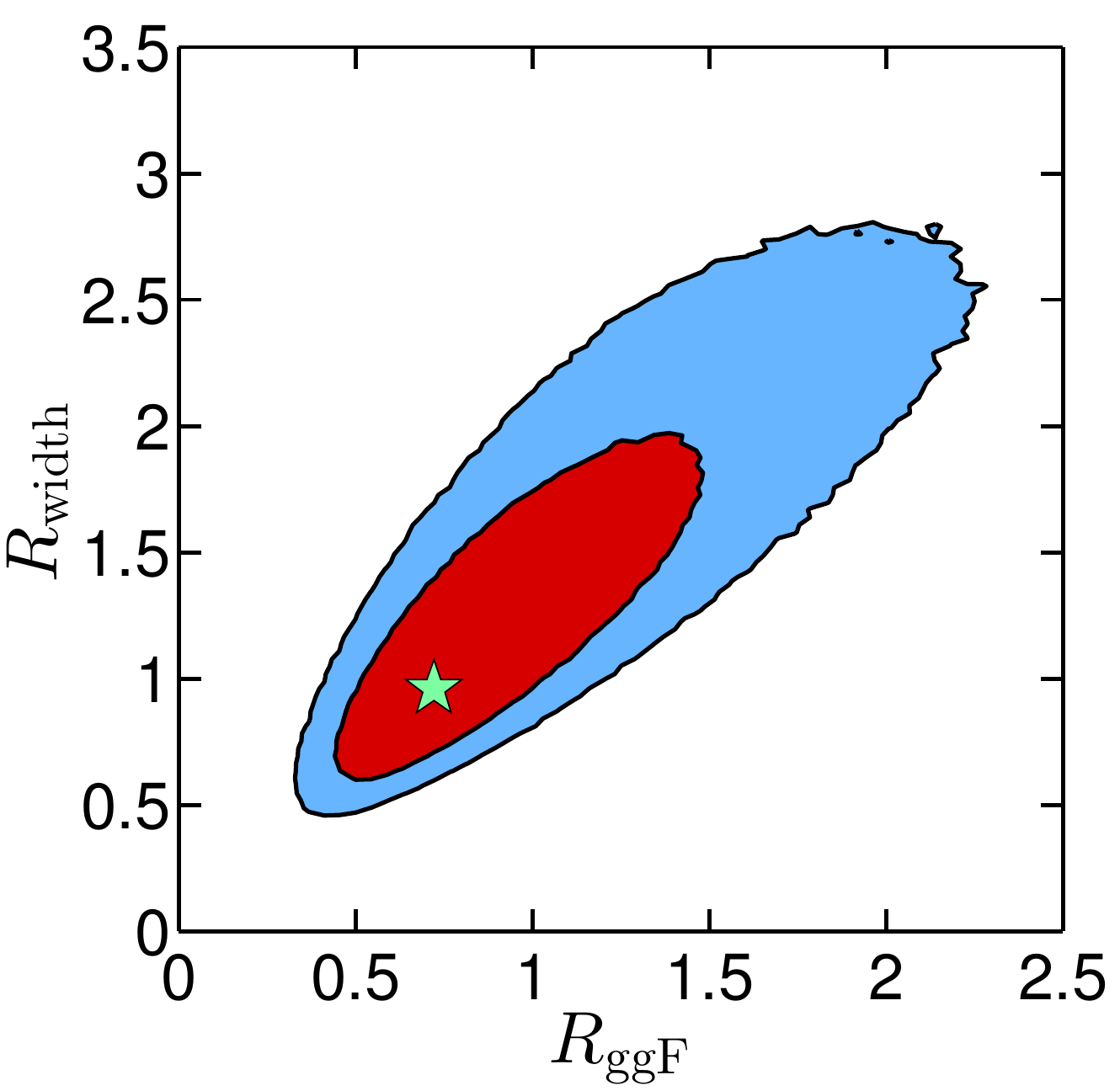}
	\caption{On the left, posterior PDF of $R_{\rm width} = \Gamma_h / \Gamma^{\rm SM}_h$ in scenario I (black) and scenario II (red). Also shown is the 2D posterior PDF of $R_{\rm width}$ versus $R_{\rm ggF}$ in scenario I (middle) and scenario II (right). Color code as in the previous figures. \label{fig:Rwidth}}
\end{figure}

The PDF of $\beta_b$ is asymmetric with a longer right tail. $\beta_b$ appears mainly in the $h \rightarrow b\bar{b}$ decay rate, \ie~in $R_{b\bar{b}}$.~\footnote{$b$ quark contributions in the loop-induced processes (ggF, $\gamma\gamma$, $Z\gamma$) are small and can be disregarded.} The reason of the asymmetry is the following: as the branching ratio ${\cal B}(h \rightarrow b\bar{b}) = 57\%$ in the SM, a deviation of $\beta_b$ from 0 (hence $c_b$ from 1) results in a sizeable modification of the total width of the Higgs.
Our signal strengths are expressed as $\hat\mu(X, b\bar{b}) = R_X R_{b\bar{b}} / R_{\rm width}$ and contain a strong correlation between $R_{b\bar{b}}$ and $R_{\rm width}$. As $R_{\rm width}$ significantly increases with $R_{b\bar{b}}$, the deviations from $\mu_{b\bar{b}} = 1$ are smaller than what we could naively expect, allowing large values of $R_{b\bar{b}}$, hence $\beta_b$. This explains the tails of the PDF of $\beta_b$.

The 1D and 2D PDFs of $R_{\rm width}$ are shown in the left pannel of Fig.~\ref{fig:Rwidth}. It turns out that a large increase of $R_{\rm width}$ is not forbidden by the measurements of other channels, in which this effect is compensated by an increase of the decay or production rates, in particular ggF. The upper bound on $R_{\rm width}$, $R_{\rm width} \lesssim 3$, comes from the requirement $\beta_b<1$.

\begin{figure}
	\centering
		\includegraphics[width=4.5cm]{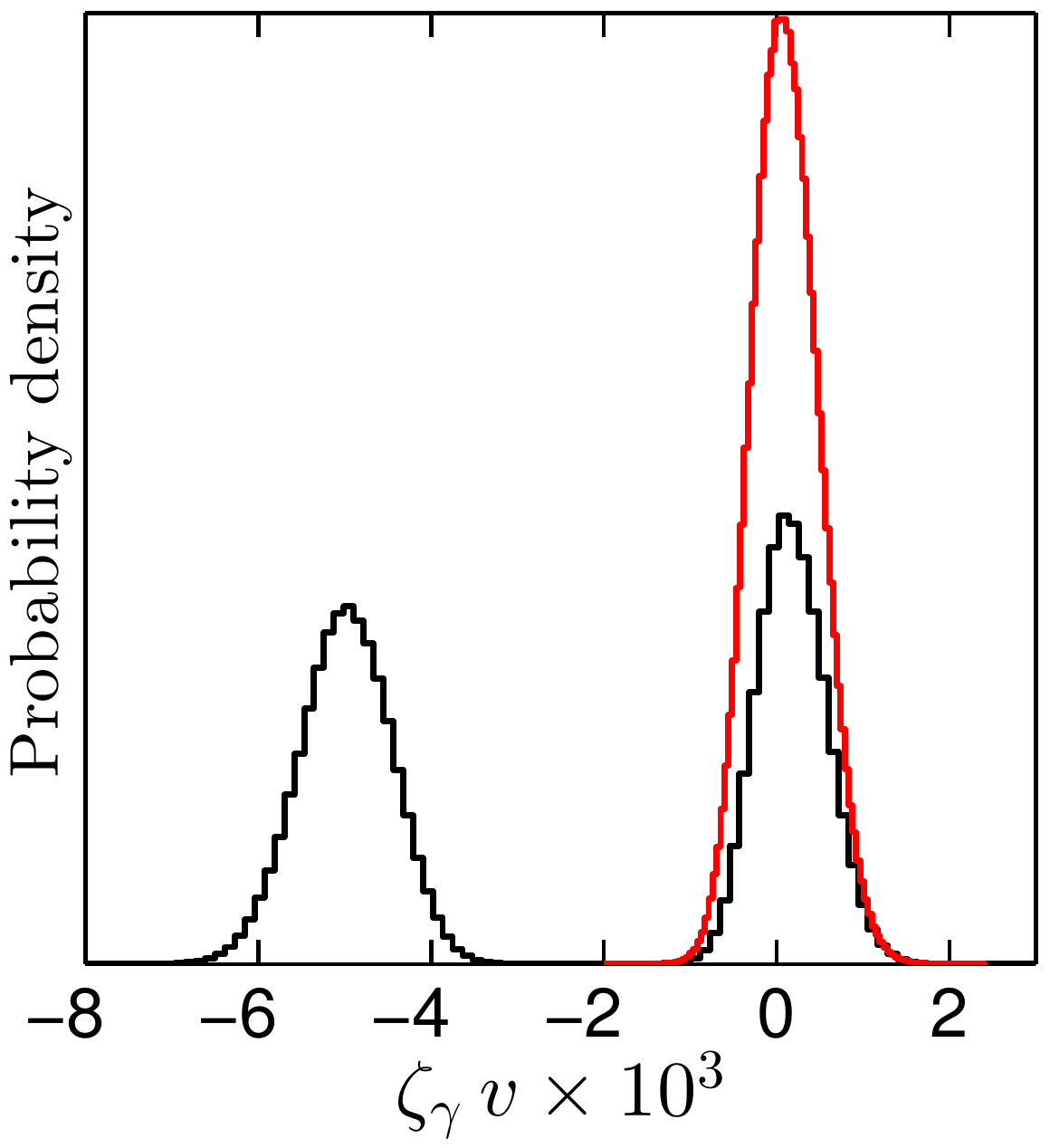}
        \includegraphics[width=4.5cm]{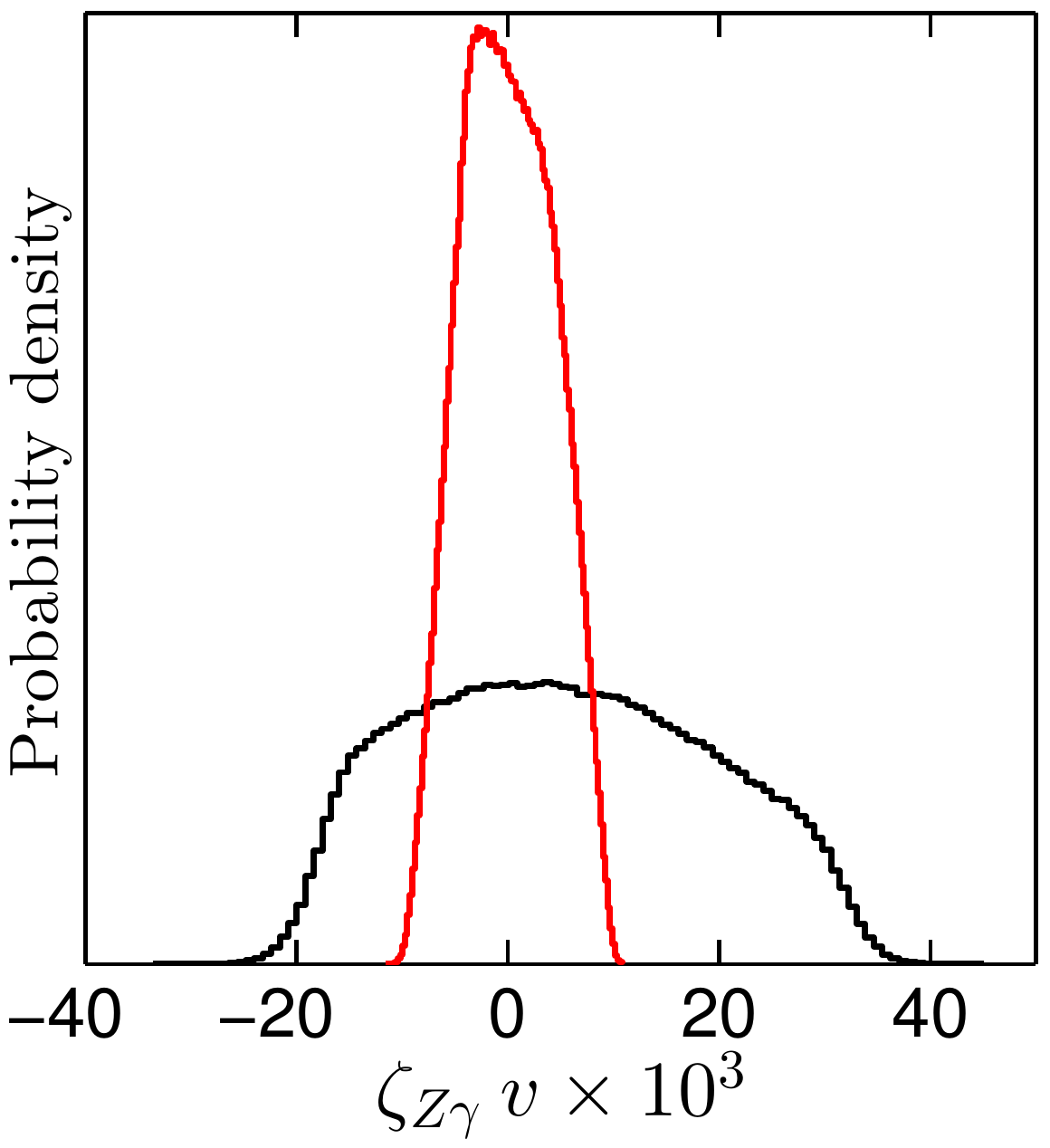}			
		\includegraphics[width=4.7cm]{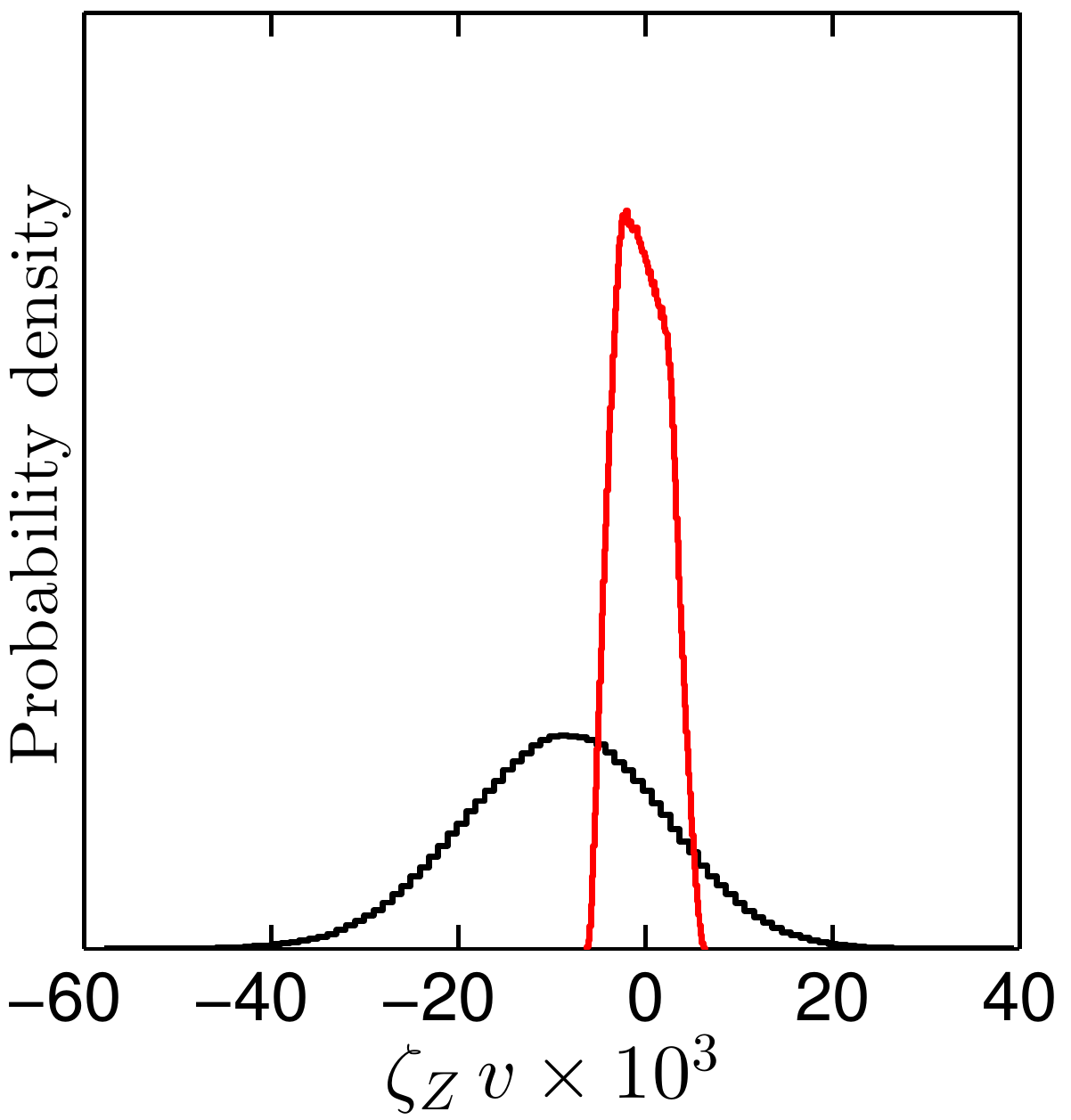}					
	\caption{Posterior PDFs of $\zeta_\gamma$, $\zeta_{Z\gamma}$, and $\zeta_Z$ in scenario I (black) and scenario II (red). \label{fig:zeta_g_Zg_Z}}
\end{figure}

\begin{figure}
	\centering
	\includegraphics[width=5cm]{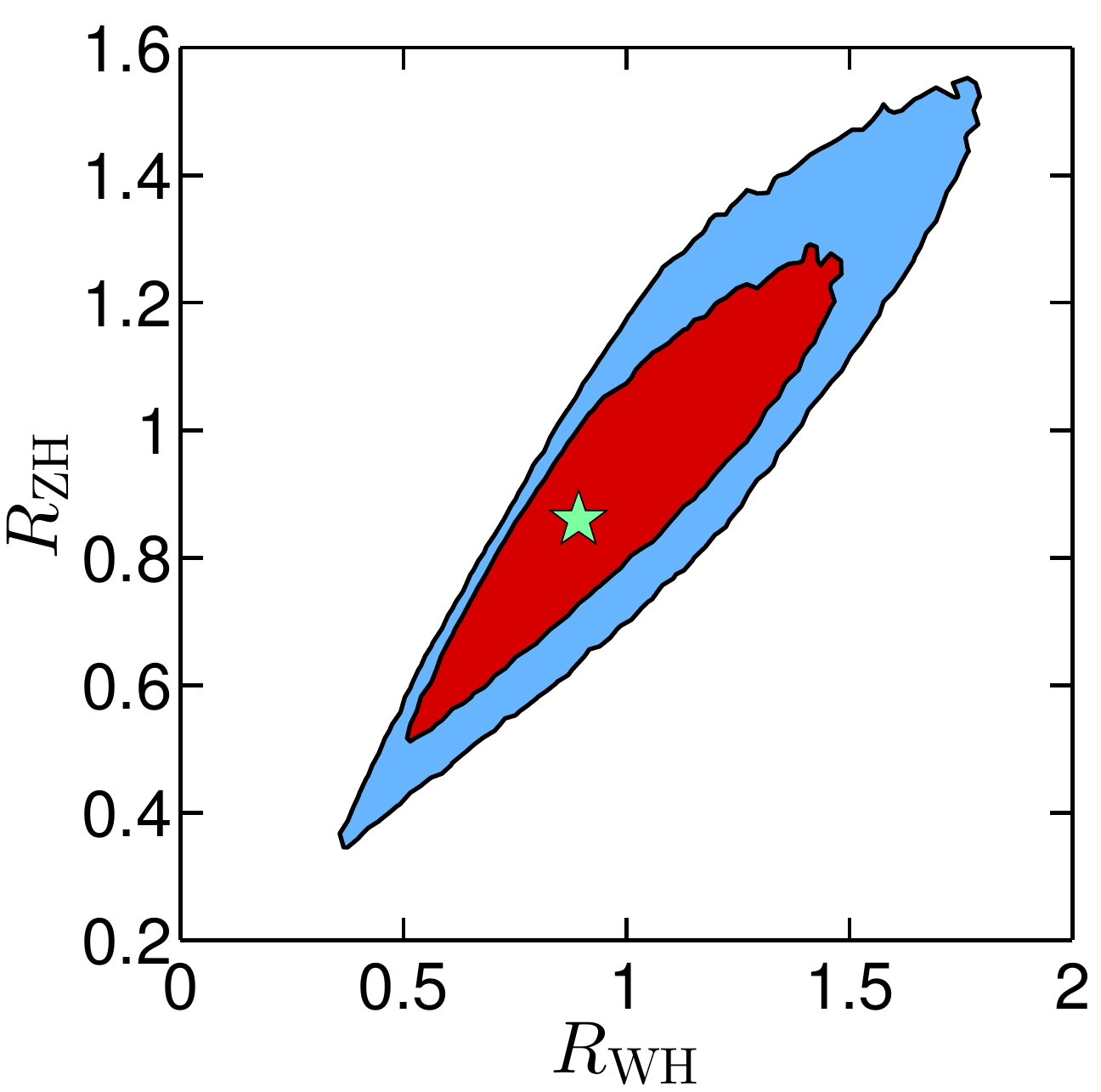}
	\includegraphics[width=5.1cm]{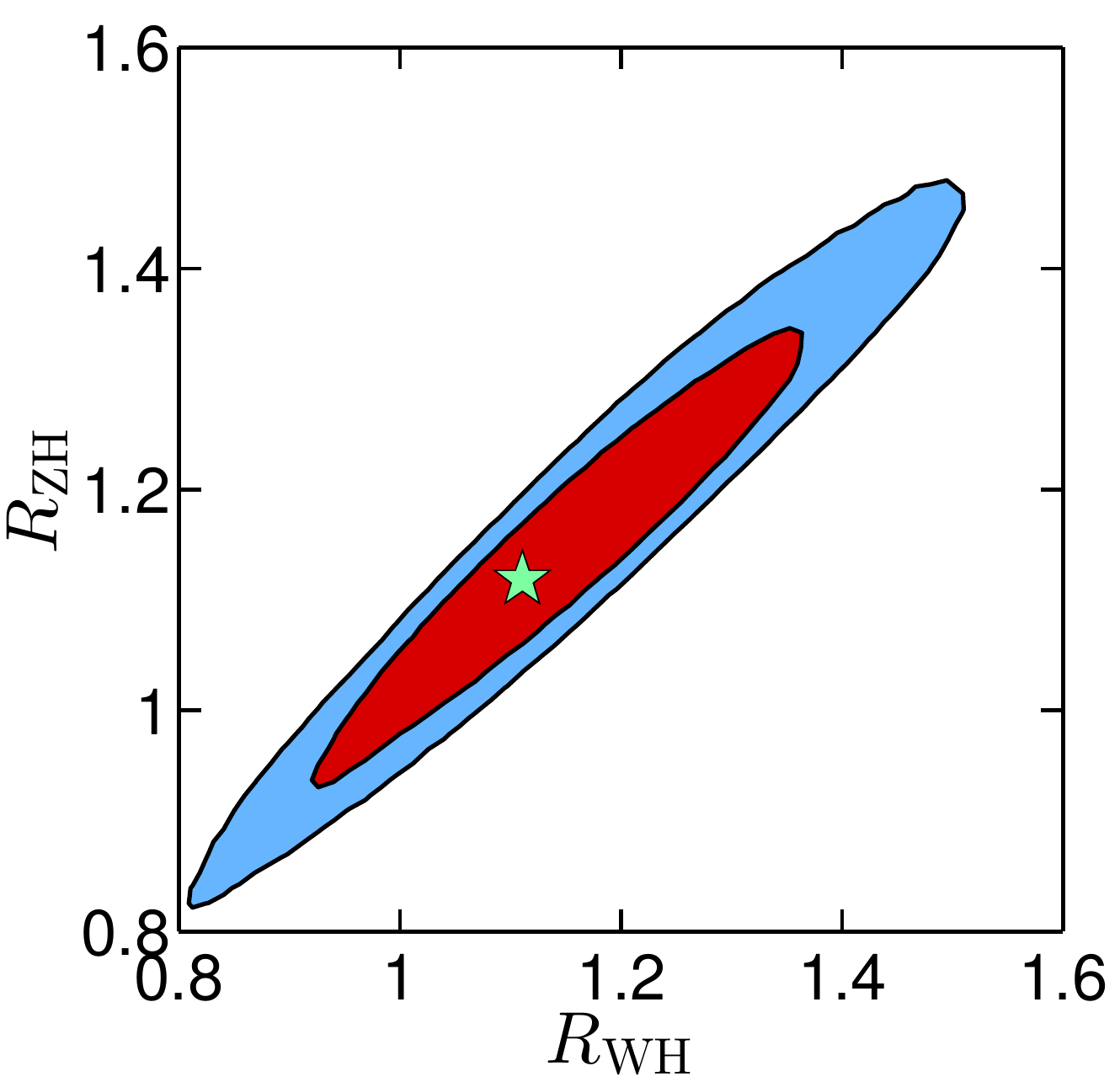}
	\caption{Posterior PDF of $R_{\rm ZH}$ versus $R_{\rm WH}$ in scenario I (left) and scenario II (right). Color code as in
the previous figures. \label{fig:RWH_RZH}}
\end{figure}

In Fig.~\ref{fig:zeta_g_Zg_Z}, we show  the PDFs of the tensorial couplings $\zeta_\gamma$, $\zeta_{Z\gamma}$, $\zeta_Z$ in Fig.~\ref{fig:zeta_g_Zg_Z} for scenario I.
The PDF of $\zeta_\gamma$ is constrained to small values in order to have the correct $H \rightarrow \gamma\gamma$ rate.
The PDF of $Z\gamma$ is much broader because of the weak experimental sensitivity to the $Z\gamma$ rate. The distribution for $\zeta_Z$ (and similarly $\zeta_W$) is mainly due to indirect effects on the fundamental parameters $\beta_{VV}$ ($\gamma\gamma$ and $Z\gamma$ rates, as well as TGVs) rather than because of direct experimental constraints. 
Notice that even with the assumption of custodial symmetry (which enforces $a_W=a_Z$), Eq.~(\ref{eq:VH}) allows the rates for associated production to be different for $Z$ and $W$ because of the contribution of the tensorial couplings. 
It turns out that $\zeta_W$ and $\zeta_Z$ can be large enough in scenario I to induce a substantial deviation from $R_{\rm WH}=R_{\rm ZH}$. This is shown in Fig.~\ref{fig:RWH_RZH}. 
This effect is also present in scenario II to a lesser extent. 

In scenario I, we observe two peaks of opposite signs for the tensorial couplings $\zeta_g$ and $\zeta_\gamma$. These features appear because of the competition between the tree-level $\zeta_{g,\gamma}$ and the loop-level SM couplings in the ggF and digamma amplitudes. In addition to the classical region  where $\zeta$ adds up to the SM coupling and cannot be very large, regions with $\zeta=\mathcal{O}(-2\lambda^{\rm SM})$ are also allowed. 
Note that  $\zeta_\gamma$ is a linear combination of $\beta_{\WW}$, $ \beta_{\WB}$, $ \beta_{\BB}$, but that the two $\zeta_\gamma$ peaks cannot be seen in the PDFs of these parameters.
These four regions do not show up in scenario II, because the $\zeta_i$ are loop-suppressed and thus cannot be large enough to cancel the SM couplings.

\begin{figure}
	\centering
	\includegraphics[width=4.7cm]{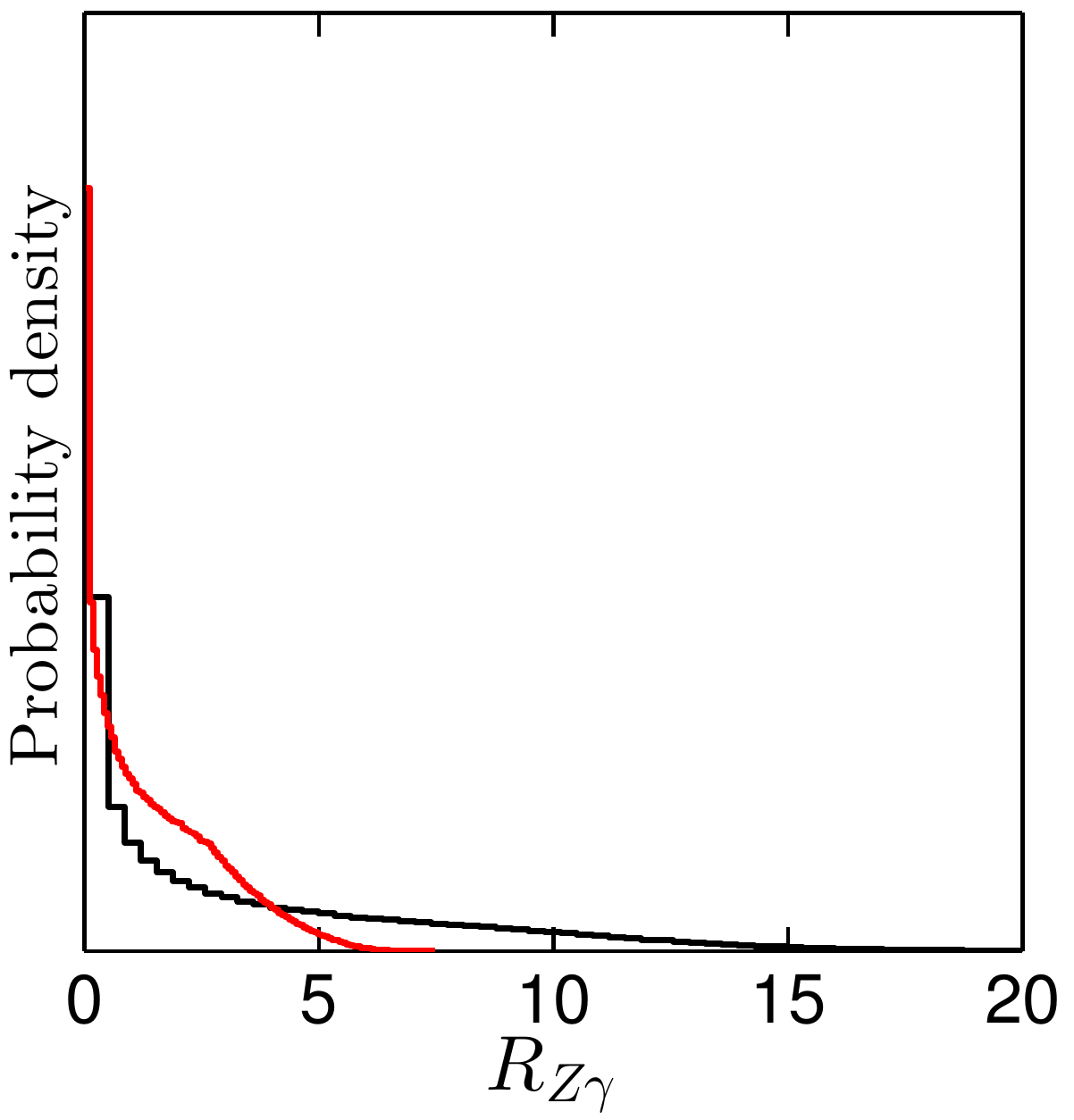}							
	\includegraphics[width=4.7cm]{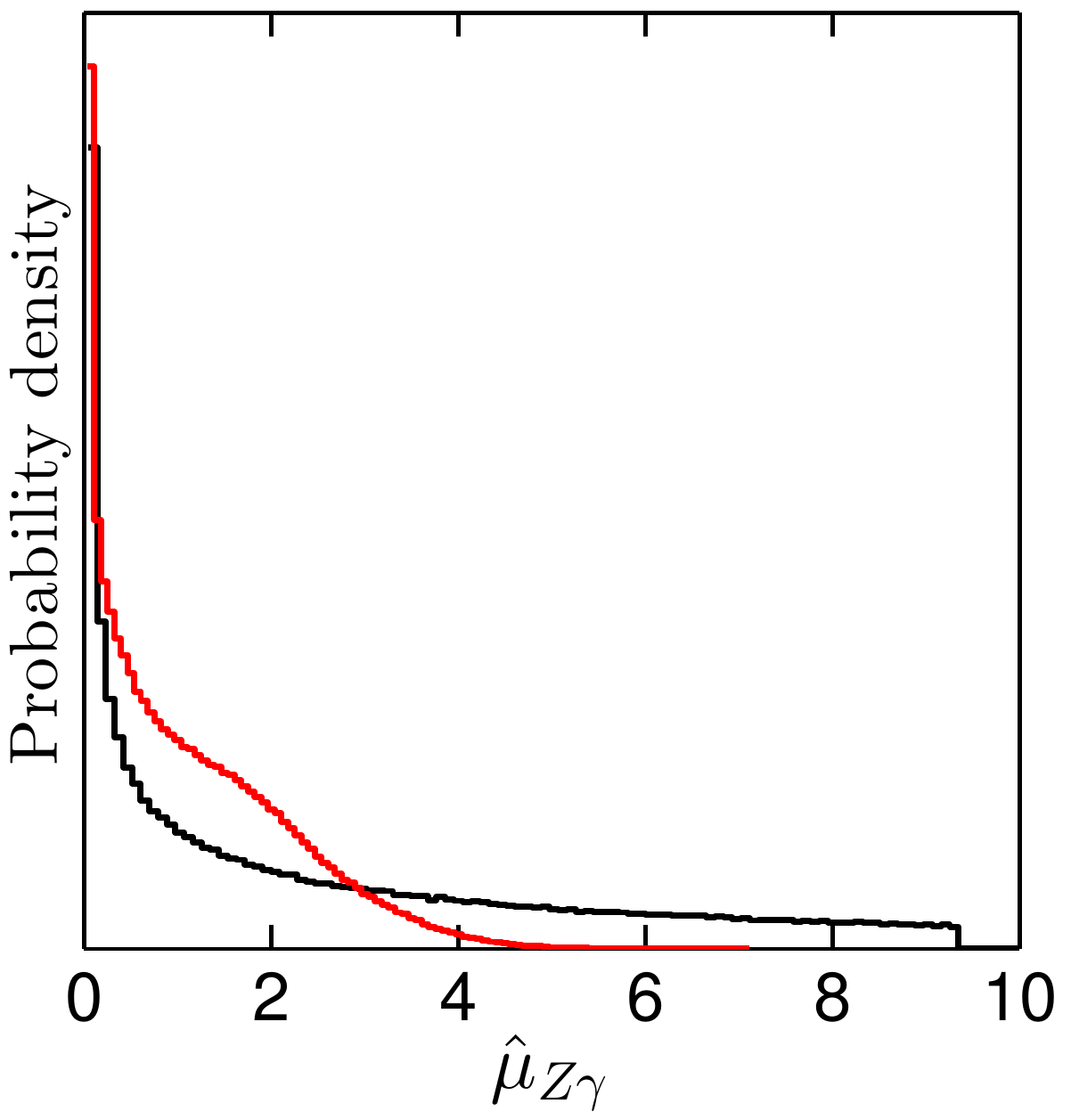}								
	\caption{Posterior PDFs of $R_{Z\gamma} \equiv \Gamma(h\rightarrow Z\gamma)/\Gamma^{\rm SM}(h\rightarrow Z\gamma)$ (left) and $\hat \mu_{Z\gamma}$ (right) in scenario I (black) and scenario II (red). \label{fig:RZgam_1D}}
\end{figure}

A feature of the PDF of $\zeta_{Z\gamma}$ is that, in spite of the various constraints on $\beta_{\WW}$, $\beta_{\WB}$, and $\beta_{\BB}$, the $Z\gamma$ rate is still likely to be considerably enhanced.
The shape of the $\zeta_{Z\gamma}$ PDF is mainly constrained by the CMS bound $\hat \mu_{Z\gamma}<9.3$ in scenario I, while indirect constraints from the $S$ parameter and trilinear gauge vertices dominate in scenario II. 
In the enhanced rate, the HDO contribution dominates, such that $R_{Z\gamma}$ is mostly proportional to $(\zeta_{Z\gamma})^2$.
This happens in scenario I, but also in scenario II although the $\zeta_i$ are smaller.
The PDF of the ratio $R_{Z\gamma}$ can be seen in Fig.~\ref{fig:RZgam_1D}. The $95\%$ Bayesian credible intervals are $[0, 12.0]$ for scenario I and $[ 0, 4.3]$ for scenario II. As large deviations are allowed in this channel within this framework, it is therefore particularly promising for the discovery of a NP signal.

In scenario I, $\zeta_{Z\gamma}$ is sufficiently large to cancel the SM coupling, such that enhancement with both signs of $\zeta_{Z\gamma}$ is realized. In contrast, for scenario II, only the branch with constructive interference $\zeta_{Z\gamma}<0$ can  enhance $R_{Z\gamma}$. In both scenarios, $\zeta_{Z\gamma}$ can cancel the SM coupling such that having a small or vanishing $R_{Z\gamma}$ is likely.~\footnote{We do not focus on this aspect as the direct searches at the LHC are still far from this level of precision.
The shape of the $R_{Z\gamma}$ PDF follows the distribution of $|\zeta_{Z\gamma}+\lambda_{Z\gamma}|^2$, which presents a peak in 0. Schematically, for a uniform distribution of $\zeta_{Z\gamma}$, the peak behaves as $\zeta_{Z\gamma}^{-1/2}$. One can observe a similar behaviour for $R_{\rm ttH}$.}

Finally, we compute the signal strength of $h \rightarrow Z\gamma$ in case of a fully inclusive analysis at the LHC. The PDFs are shown in the right panel of Fig.~\ref{fig:RZgam_1D} for both scenarios. In scenario I, the distribution reaches the CMS 95\% C.L.~bound $\hat\mu_{Z\gamma}<9.3$, while it vanishes before in scenario II. The $68\%$ and $95\%$ BCIs are $[0,3.6]$, $[0,8.1]$ in scenario I and $[0,1.6]$, $[0,3.2]$ in scenario II. 

\begin{figure}
	\centering
		\includegraphics[width=5.1cm]{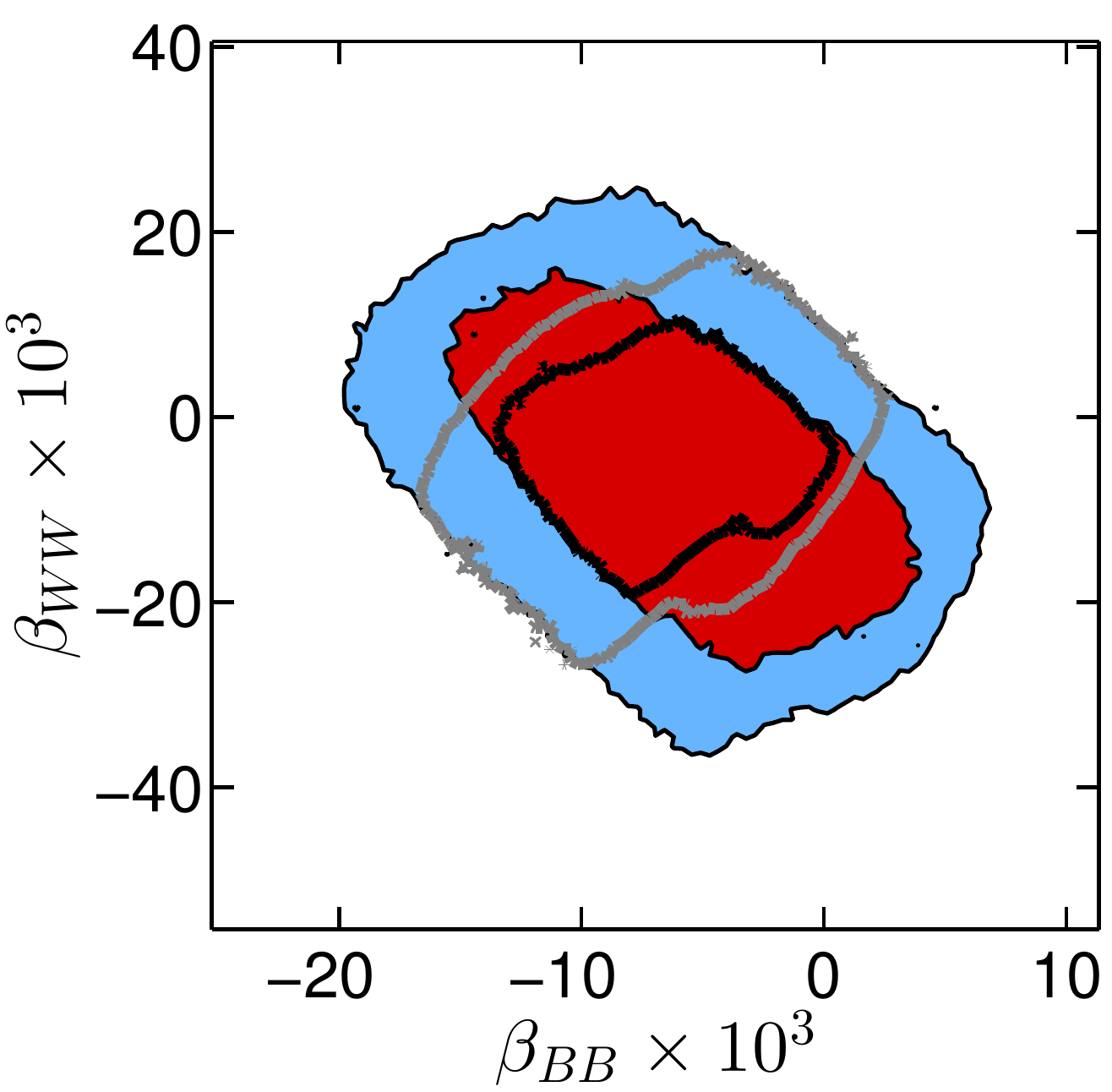}
		\includegraphics[width=4.95cm]{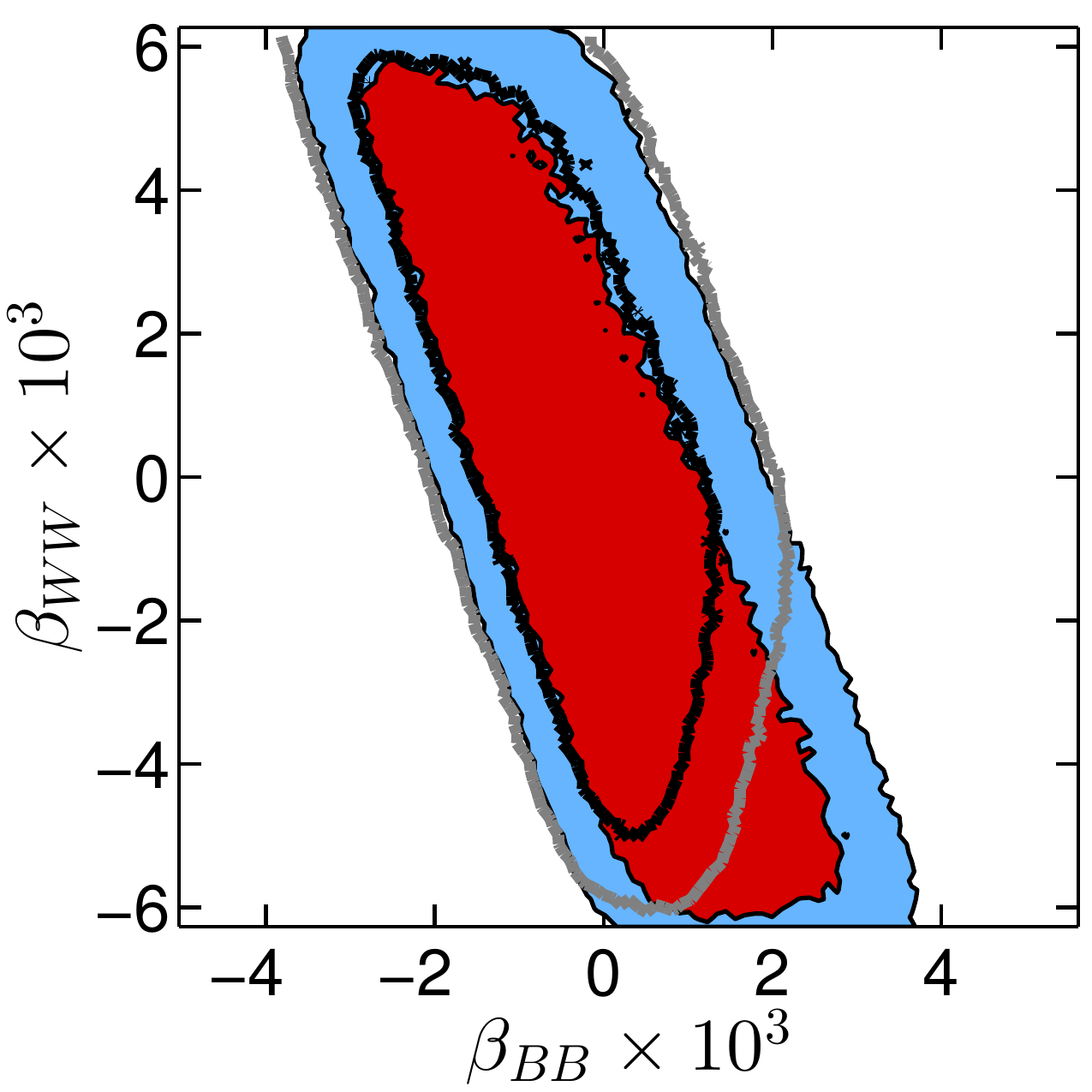}		
	\caption{Posterior PDF of $\beta_{\WW}$ versus $\beta_{\BB}$ in scenario I (left) and scenario II (right). The red and blue regions correspond to the 68\% and 95\% BCRs from the current measurements, while the black and grey contours correspond to the 68\% and 95\% BCRs assuming in addition that $\hat\mu_{Z\gamma}<2$.
\label{fig:Zgam}}
\end{figure}

Given that the 13/14 TeV LHC has a good potential to constrain the $h \rightarrow Z\gamma$ rate, one may wonder about the impact of a more precise measurement on our results. Therefore, we investigate the possibility of having $\hat\mu_{Z\gamma}<2$ at 95\% CL, and we implement this bound as a step function. \footnote{Note that the relative SM production rates $\sigma^{\rm SM}_X / \sum_X \sigma^{\rm SM}_X$ do not change significantly between 8 TeV and 14 TeV. Thus, assuming a fully inclusive analysis, we can take the decomposition into production channels as given in Sec.~\ref{se_data}, Table~\ref{CMSresults}.} 
It mainly results in a better determination of $\beta_{\BB}$ and $\beta_{\WW}$ in both scenarios, as can be seen in Fig.~\ref{fig:Zgam}.
This new limit has an effect on the Higgs phenomenology in scenario I only. It leads to a slightly better prediction of $R_{\rm WH}$: the 95\% BCI is $[0.7,1.5]$, instead of $[0.5,1.7]$ when we take into account the current limit on $h \rightarrow Z\gamma$.

\section{Conclusion} \label{se_conclusion}

Less than one year after the first announcement of a signal in LHC data, the existence of a Higgs boson is at present firmly established. Time has come to probe in detail the structure of the now complete electroweak sector, searching for indirect signs of high-energy New Physics.

We use a complete basis of dimension-six operators encoding NP effects in an effective Lagrangian in which all tensorial couplings are taken into account. The basis is chosen such that field-strength--Higgs operators ($\mathcal{O}_{FF}$) are exactly mapped into tensorial couplings. In this basis it is straightforward to study the well-motivated hypothesis of loop-suppression of these operators. 

The data taken into account in our analysis are the whole set of results from ATLAS and CMS, including all available correlations, as well as Tevatron data. Trilinear gauge vertices measurements and constraints on electroweak precision observables are also included in our study.

It turns out that weak bosons tensorial couplings can in principle modify non-trivially the kinematic structure of the  VBF, VH and $h\rightarrow VV$ processes, and thus the efficiency of the kinematic selections. Scrutinizing the experimental analyses, we find that one can consider unchanged kinematic cuts efficiencies for the rates of these processes to a good approximation.
The effect of tensorial couplings is crucial in the ggF, $h \rightarrow \gamma\gamma$, $h \rightarrow Z\gamma$ processes, where they compete with the one-loop SM couplings. In our predicted rates, leading loop corrections are taken into account, consistently  with respect to the loop-level HDO framework. 

In order to put constraints on the higher dimensional operators we consider, we carry out a global analysis in the framework of Bayesian inference. We find this approach particularly appropriate as it deals with weakly constrained problems and naturally takes into account fine-tuning, such that the results we present are free of improbable, \ie~fine-tuned, cancellations. Markov Chain Monte Carlo techniques are used to perform the numerical integrations.

Our analysis is centered on New Physics arising at $\lesssim$~3 TeV. 
It allows us to express our results in terms of $\beta_i \equiv \alpha_i\, v^2 / \Lambda^2$, {\it i.e.} to factorize the effect of the coefficients $\alpha_i$ and of the scale of New Physics $\Lambda$.
Among other things, this parameterization is $\Lambda$-independent up to a small $\log\Lambda$ dependence. Distributions of HDO coefficients at any low $\Lambda$ can be easily mapped from our results.
Two general scenarios are considered: I) democratic HDOs, where the coefficients of all the dimension-six operators are essentially unconstrained, and II) loop-suppressed $\mathcal{O}_{FF}$'s, where the field-strength--Higgs operators are loop suppressed with respect to the other HDOs.

We find overall a substantial amount of freedom in both of these scenarios. For instance, the coupling to the top quark still allows for $\mathcal O(1)$ deviations, while the couplings to bottom and tau are slightly more constrained. We report in both scenarios a large and natural region at small or vanishing $c_t$, favored by goodness of fit, which we trace back to the slight deficit observed in the $b\bar b$ channel with the ttH production mode, as reported by CMS.
Regarding the couplings of the vector bosons, we find that only small deviations are allowed, of the order of $10\%-20\%$. 
Despite the fact that the inclusive rate of the $h\to\gamma\gamma$ channel is close to one, we observe a slightly enhanced $R_{\gamma\gamma}=\Gamma_{\gamma\gamma}/\Gamma^{SM}_{\gamma\gamma}$, with the 95\% Bayesian credible interval (BCI) for $R_{\gamma\gamma}$ given by $[0.8,2.5]$ ($[0.8,2.3]$) in scenario I (II). Correlations with $R_{\rm width} = \Gamma_h / \Gamma^{\rm SM}_h$ as well as the deviations of the various production modes from the SM are then identified to be responsible for fitting the experimental signal strengths.
We also find that an enhanced coupling to the $b$ quark is likely, leading to a larger total width, with the 95\% BCI for $R_{\rm width}$ given by $[0.7,2.7]$ ($[0.6,2.5]$) in scenario I (II). A strong correlation with $R_{\rm ggF}$ exists and makes the predicted signal strengths compatible with the data.

Overall, it appears that the tensorial couplings play a crucial role in the SM loop-induced processes. In particular, after taking into account all the constraints we find that the Higgs boson decay width into $Z\gamma$ can be enhanced by up to a factor 12 (4) in scenario I (II), within 95\% BCI. Conversely, future measurements in the $Z\gamma$ channel will provide important bounds on the coefficients of the higher dimensional operators we consider.

\section*{Acknowledgments}
We would like to thank Christophe Grojean, Bill Murray, Michael Spira and Gauthier Alexandre for useful discussions. We are grateful to Sabine Kraml for helpful comments and for carefully reading the manuscript. GG would like to thank ICTP in Trieste for hospitality during part of this work.
SF acknowledges the Brazilian Ministry of Science, Technology and Innovation for financial support, and the LPSC and the Ecole Polytechnique for hospitality during part of this work. GG would like to thank the Funda\c{c}\~ao de Amparo \`a Pesquisa do Estado de S\~ao Paulo (FAPESP) for financial support.

\newpage

\noindent{\Large\bf Appendix}

\appendix

\section{Deriving the tree-level effective Lagrangian} \label{inputs}

Here we give some details on the derivation of the effective Lagrangian $\mathcal L_{v,f}+\mathcal L_t$.~\footnote{The derivation follows closely Ref.~\cite{Burgess:1993vc}.}
We note that the following field rescalings need to be performed in the SM Lagrangian in order to pass to canonically normalized kinetic terms for all fields:
\bea
h		&\to&	\left[1+\xi_h\right]h\,,\nn\\
G_\mu	&\to&	\left[1+\xi_g\right]G_\mu\,,\nn\\
A_\mu	&\to&	\left[1+\xi_\gamma\right]A_\mu	+\xi_{Z \gamma}Z_\mu\,,\nn\\
Z_\mu	&\to&	\left[1+\xi_Z\right]Z_\mu\,,\nn\\
W_\mu^\pm
		&\to& 	\left[1+\xi_W\right]W_\mu^\pm\,,
\eea
with
\bea
\xi_h		&=&	-\left(\frac{1}{4}\alpha_{D^2}+\frac{1}{4}\alpha_{D^2}'\right)\tilde v^2\,,\nn\\
\xi_g		&=&	\alpha_{GG}\,\tilde v^2\,, \nn\\
\xi_\gamma	&=&	\left(\alpha_{\WW}\tilde s_w^2+\alpha_{\BB}\tilde c_w^2-\frac{1}{2}\alpha_{\WB}\tilde s_w\tilde c_w\right)\tilde v^2\,,\nn\\
\xi_{Z \gamma}	
			&=&		\left(2\alpha_{\WW}\tilde c_w\tilde s_w-2\alpha_{\BB}\tilde c_w\tilde s_w-\frac{1}{2}\alpha_{\WB}(\tilde c_w^2-\tilde s_w^2)\right)\tilde v^2\,,\nn\\
\xi_{ Z}	&=&	\left(\alpha_{\WW}\tilde c_w^2+\alpha_{\BB}\tilde s_w^2+\frac{1}{2}\alpha_{\WB}\tilde s_w\tilde c_w\right)\tilde v^2\,, \nn\\
\xi_{W}		&=& 	\alpha_{\WW}\,\tilde v^2\,.
\eea

The quantities $\tilde g,\ \tilde g',\ \tilde v$ are replaced by the quantities  $ g,\  s_w,\  v$, which by definition are related to the fine-structure constant $\alpha$, the $Z$ mass $m_Z$ and the Fermi constant $G_F$ via 
\be
4\pi\alpha \equiv s_w^2 g^2\,,\qquad
m_Z^2\equiv\frac{ v^2\, g^2}{4\,  c_w^2}\,,\qquad 
G_F\equiv\frac{1}{\sqrt{2}\, v^2}.
\label{hat2}
\ee
Taking into account the HDOs we can compute in terms of the SM parameters
\bea
4 \pi \alpha&=&\tilde s_w^2\tilde g^2(1+2\tilde s_w^2\alpha_{\WW}\,\tilde v^2+2\tilde c_w^2\alpha_{\BB}\, \tilde v^2-\tilde c_w \tilde s_w\alpha_{\WB}\, \tilde v^2)\,,\nn\\
m_Z^2&=&\frac{\tilde v^2\tilde g^2}{4\tilde c_w^2}\left(1+\frac{1}{2}\alpha_{D^2}\tilde v^2+\frac{1}{2}\alpha'_{D^2}\tilde v^2+2\tilde c_w^2\alpha_{\WW}\tilde v^2+2\tilde s_w^2\alpha_{\BB}\tilde v^2+\tilde c_w\tilde s_w\alpha_{\WB}\tilde v^2\right)\,,\nn\\
G_F&=&\frac{1}{\sqrt{2}\,\tilde v^2}\left(1-\frac{1}{2}\alpha_{D^2}\,\tilde v^2+\frac{1}{2}\alpha_{D}\,\tilde v^2\right)\,.
\label{input}
\eea
For $\alpha$, all corrections come from wave function renormalization (WFR) of $A_\mu$. For $m_Z$, there are both WFR and direct mass corrections, while for  $G_F=g^2/(4\sqrt{2}m_W^2)$  the WFR for $W$ drops out and we are left with mass corrections as well as explicit vertex corrections.  
Combining Eq.~(\ref{hat2}) and (\ref{input}) and inverting we obtain
\bea
\tilde v^2&=& v^2-\frac{ v^4}{2}\left(\alpha_{D^2}-\alpha_{D}\right)\,,\nn\\
\tilde g^2&=& g^2+ g^2 v^2\left( -\frac{ c_w^2}{2( c_w^2- s_w^2)}\left[\alpha_{D^2}'+\alpha_D\right]
		-\frac{ c_w s_w}{ c_w^2- s_w^2}\alpha_{\WB}-2\alpha_{\WW}
	\right)\,,\nn\\
\tilde s_w^2&=& s_w^2+ s_w^2 c_w^2 v^2\left(\frac{1}{2( c_w^2- s_w^2)}\left[\alpha_{D^2}'+\alpha_D\right] 
				+\frac{2 c_w s_w}{ c_w^2- s_w^2}\alpha_{\WB}+2[\alpha_{\WW}-\alpha_{\BB}]\right)\,.\nn\\
\eea
Next we need to express the bare Yukawa couplings in terms of the physical masses. One finds:
\be
\tilde y_f=\frac{\sqrt{2}\, m_f}{ v}\left(1+\frac{1}{4}\,[\alpha_{D^2}-\alpha_D]\, v^2+\alpha_f\, v^2\right)\,.
\ee  

Finally we define a $ g_s$ by $4\pi\alpha_s\equiv g_s^2$ and write
\be
\tilde g_s^2= g_s^2(1-2\, \alpha_{GG}\, v^2)\,.
\ee 

\section{Loop-induced SM couplings} \label{App_SMloops}

Here we give the expressions of the form factors for SM loop-induced couplings Eqs.~\eqref{lambda_loop_gammaW}, \eqref{lambda_loop_gf}. 

\be
A_f(\tau)=\frac{3}{2}\tau(1+(1-\tau)f(\tau))\,,
\ee
\be
A_v(\tau)=\frac{1}{7}(2+3\tau+3\tau(2-\tau)f(\tau))\,,
\ee
\be
A_{Z \gamma}(\tau,\kappa)=-\frac{\tau\kappa}{(\tau-\kappa)}(f(\tau)-f(\kappa))\,,
\ee
\be
B_{Z \gamma}(\tau,\kappa)=\frac{\tau \kappa}{(\tau-\kappa)}+\frac{\tau^2 \kappa^2}{(\tau-\kappa)^2}(f(\tau)-f(\kappa))+ \frac{2\tau^2 \kappa}{(\tau-\kappa)^2}(g(\tau)-g(\kappa))\,.
\ee
\be
f(\tau)= \left\{\begin{array}{cc}
\arcsin^2(\tau^{-1/2}) & \textrm{if}\,\, \tau\geq1\,, \\
-\frac{1}{4} (\log\frac{1+\sqrt{1-\tau}}{1-\sqrt{1-\tau}}-i\pi)^2 & \textrm{if} \,\, \tau <1 \,.
 \end{array} \right.
\ee
\be
g(\tau)= \left\{\begin{array}{cc}
\sqrt{\tau-1}\arcsin(\tau^{-1/2}) & \textrm{if}\,\, \tau\geq1\,, \\
\frac{1}{2} \sqrt{1-\tau} (\log\frac{1+\sqrt{1-\tau}}{1-\sqrt{1-\tau}}-i\pi) & \textrm{if} \,\, \tau <1 \,.
 \end{array} \right.
\ee

\section{Tensorial Higgs decay to weak bosons}\label{App_tens}

In the vectorial $h\to VV$ decay we encounter the following integral \cite{Pocsik:1980ta}
\be
R_\delta(x)\equiv\int_{\sqrt{x}}^{\frac{x+1}{2}} dy\ \frac{\left(y^2-x\right)^\frac{1}{2}}{(1-2y)^2}\left(\frac{\delta }{x}\,(y^2-x)+1+x-2y\right)\,,
\ee
where $y=E_V/m_h$ is the energy of the on-shell gauge boson, $x=m_V^2/m_h^2$ and $\delta=0 (1)$ for transversely (longitudinally) polarized gauge boson.
In terms of the function 
\be
F(x)\equiv \frac{1}{\sqrt{4x-1}}\ \arccos \left(\frac{3x-1}{x^\frac{3}{2}}\right)
\ee
we can evaluate
\begin{multline}
R_1+2R_0=\frac{1}{32 x^2}\biggl((-1+x)(2-13x+47x^2)\biggr.\\
-3x(1-6x+4x^2)\log(x)\\
\biggl.+6x(1-8x+20x)F(x)\biggr)\,.
\end{multline}
This is indeed proportional to the total decay width given in Ref.~\cite{Djouadi:2005gi}. The translation of notation is
\be
R_T=16 (R_1+2R_0)\,,
\ee
with the decay width given by
\be
\Gamma(h\rightarrow VV^*)=\frac{3G_F^2 m_V^4}{16\pi^3}m_h \delta_V R_T(x)   \,,
\ee
and $\delta_W=1$, $\delta_Z=\frac{7}{12}-\frac{10}{9}s_W^2+\frac{40}{9}s_W^4$.

To compute the tensorial contributions, we need to multiply the integrand with $\nu_{VV}\equiv (y-x)$ and $\nu_{VV}^2$
\bea
Q_\delta(x)&\equiv&\int_{\sqrt{x}}^{\frac{x+1}{2}} dy\ \frac{\left(y^2-x\right)^\frac{1}{2}}{(1-2y)^2}\left(\frac{\delta }{x}\,(y^2-x)+1+x-2y\right)(y-x)\,,\nn\\
P_\delta(x)&\equiv&\int_{\sqrt{x}}^{\frac{x+1}{2}} dy\ \frac{\left(y^2-x\right)^\frac{1}{2}}{(1-2y)^2}\left(\frac{\delta }{x}\,(y^2-x)+1+x-2y\right)(y-x)^2\,.\nn
\eea
The expressions are of similar form but slightly more complicated
\begin{multline}
Q_1+2Q_0=-\frac{1}{96 x^2}\biggl((-1+x)(-3+31 x-140 x^2+160 x^3)\biggr.\\
+3x(2-15x+42x^2-12 x^3)\log(x)\\
\biggl.+6x(-2+19 x-68x^2+84 x^3)F(x)\biggr)\,,
\end{multline}
\begin{multline}
P_1+2P_0=\frac{1}{768 x^2}\biggl((-1+x)(12-173x+971x^2-2299x^3+1405x^4)\biggr.\\
-6x(5-46x+174x^2-288x^3+48x^4)\log(x)\\
\biggl.+12x(5-56 x+256 x^2-544 x^3+432 x^4)F(x)\biggr)\,.
\end{multline}
The final quantities $\bar \nu_{VV}$, $\bar \nu^2_{VV}$ introduced in Eq. \eqref{nu_VV} are 
\be
\langle \nu_{VV}\rangle=\frac{Q_1+2Q_0}{R_1+2R_0}\,,\,\,\,\, \langle \nu_{VV}^2\rangle=\frac{P_1+2P_0}{R_1+2R_0}\,.
\ee

\bibliographystyle{unsrt}

\end{document}